\documentclass[11pt]{article}

\usepackage[T1]{fontenc}
\usepackage[utf8]{inputenc}
\usepackage{lmodern}
\usepackage[margin=1in]{geometry}
\usepackage{microtype}
\usepackage{amsmath,amssymb,bm,mathtools}
\usepackage{graphicx}
\usepackage{float}
\usepackage{booktabs,longtable,tabularx,array,adjustbox,multirow}
\usepackage{etoolbox}
\usepackage{indentfirst}
\usepackage{enumitem}
\usepackage{caption}
\usepackage{xurl}
\usepackage{cite}
\usepackage{newunicodechar}
\usepackage{xcolor}
\definecolor{arxivblue}{HTML}{1F5AA6}
\usepackage{hyperref}

\hypersetup{
  pdftitle={A Hierarchical Validity-Audit Framework for Neural Mass Models in Simulation-Based Inference: From Observational Coverage to Mechanistic Interpretation},
  pdfauthor={Tianming Cai; Ning Jiang; Yuan Yang; Jiayuan He},
  pdfsubject={Neural mass models and simulation-based inference},
  pdfdisplaydoctitle=true,
  colorlinks=true,
  linkcolor=black,
  citecolor=arxivblue,
  urlcolor=arxivblue,
  filecolor=arxivblue
}

\allowdisplaybreaks[2]
\AtBeginEnvironment{longtable}{\setlength{\tabcolsep}{3pt}}

\makeatletter
\def\citepunct{,\penalty\@m}
\newcommand{\spacedcite}[1]{%
  \begingroup
  \def\citepunct{,\penalty\@m\space}%
  \cite{#1}%
  \endgroup
}
\makeatother

\newunicodechar{​}{}
\newunicodechar{−}{\ensuremath{-}}

\providecommand{\real}[1]{#1}

\begin{document}
\begin{center}
{\LARGE\bfseries A Hierarchical Validity-Audit Framework for Neural Mass Models in Simulation-Based Inference: From Observational Coverage to Mechanistic Interpretation\par}
\vspace{1.2em}
{\large Tianming Cai\textsuperscript{1,2}, Ning Jiang\textsuperscript{1,2}, Yuan Yang\textsuperscript{3}, and Jiayuan He\textsuperscript{1,2,*}\par}
\vspace{0.8em}
\begin{minipage}{0.96\textwidth}
\centering\small
\textsuperscript{1}National Clinical Research Center for Geriatric, West China Hospital, Sichuan University, Chengdu, Sichuan 610017, China.\par
\textsuperscript{2}Med-X Center for Manufacturing, Sichuan University, Chengdu, Sichuan 610017, China.\par
\textsuperscript{3}Department of Neurosurgery, West China Hospital, Sichuan University, No. 37 Guo Xue Xiang, Chengdu 610041, China.\par
\vspace{0.35em}
\textsuperscript{*}Corresponding author.\par
\end{minipage}
\end{center}
\vspace{0.8em}

\hypertarget{sec-abstract}{%
\section{Abstract}\label{sec-abstract}}

\noindent
Neural mass models describe neural population activity using low-dimensional dynamical parameters and estimate parameter posteriors through simulation-based inference when an explicit likelihood is difficult to construct. However, simulation-internal posterior recovery does not guarantee that a model fits real observations or that its parameters have physiological interpretations. We therefore propose NMM-SBI Audit, a hierarchical audit framework that sequentially evaluates the capacity of a model configuration to cover the data, trains separate posteriors for multilevel parameter coordinates to comprehensively detect summary information loss, and finally examines joint parameter compensation and cross-track consistency. Experiments with known ground truth showed that the framework controlled the empirical error rate and identified prespecified failures. In real data, the single-source Epileptor model failed to cover the core seizure statistics of SOZ-local iEEG; consequently, its simulation-internally recoverable targets were not eligible for patient-specific mechanistic interpretation. No systematic mismatch across representations was detected for the CMC-inspired auditory network model, which supported conditional recovery of some superficial-layer and inhibitory gains while also revealing parameter compensation, summary information loss, and instability of the active structure. These results indicate that observation fit, target recoverability, and joint interpretability constitute non-interchangeable levels of evidence. NMM-SBI Audit provides a scalable methodological tool for constraining excessive mechanistic interpretation in simulation-based inference of neural dynamics.

\vspace{0.8em}
\noindent\textbf{Keywords:} neural mass model; simulation-based inference; parameter posterior; summary statistics; identifiability

\hypertarget{sec-1}{%
\section{1. Introduction}\label{sec-1}}

Linking macroscopic neurophysiological signals, such as electroencephalography (EEG) or electrocorticography (ECoG), to underlying cortical microcircuit mechanisms is a long-standing central problem in computational neuroscience and neural engineering \cite{ref1,ref2}. Neural mass models (NMMs) provide a generative modelling framework with mechanistic interpretability for addressing this problem \cite{ref4}. These models use a small number of state variables to characterize mean-field interactions between excitatory and inhibitory neuronal populations and can reproduce normal rhythms, evoked responses, and various forms of pathological electrophysiological activity within a unified dynamical framework \cite{ref5,ref6}. NMM parameter inversion is therefore often regarded as a potential means of inferring local physiological states from macroscopic signals. This objective, however, is subject to a fundamental limitation: the mapping from macroscopic observations to microscopic mechanisms is generally not one-to-one. Because EEG/ECoG signals are affected by spatial mixing, measurement noise, finite sampling, feature compression, and other factors \cite{ref3}, different parameter combinations may produce highly similar observed dynamics. This parameter degeneracy means that observable changes in the data can often constrain only low-dimensional directions or manifolds within the full parameter space \cite{ref7,ref8}, rather than uniquely determining every candidate physiological mechanism. NMM inversion thus constitutes a typical ill-posed inverse problem. Even if an inversion algorithm yields a concentrated and apparently credible posterior distribution, that posterior may reflect constraints imposed by prior assumptions or the feature representation rather than genuine support in the data for the target mechanism.

Dynamic causal modelling (DCM) \cite{ref9,ref10} and simulation-based inference (SBI) \cite{ref11,ref12,ref13} have substantially expanded the capacity to invert complex neural dynamical models. DCM provides a classical framework based on generative models and Bayesian inversion, whereas SBI enables posterior estimation for nonlinear simulators whose likelihood functions are difficult to express explicitly. Corresponding diagnostic tools, including predictive checks \cite{ref14,ref15}, simulation-based calibration (SBC) \cite{ref16,ref17}, expected coverage analysis, and local classifier two-sample tests (C2STs) \cite{ref18,ref19,ref20}, provide important evidence for evaluating model coverage, posterior calibration, predictive consistency, and approximation error. These diagnostics primarily assess whether posterior parameters obtained under a given dynamical model are credible \cite{ref21}.

However, reliable posterior estimation for a given target in NMM inversion does not imply that other interpretable targets in the same model are also supported by the current data representation \cite{ref22}. Interpretive targets in neural mass models typically occur at different levels, including raw model parameters, derived coordinates defined by parameter combinations, and active coordinates characterized by data-sensitive directions. These three types of target are not equivalent substitutes for one another but correspond to distinct languages of interpretation. The theory of sloppy models holds that the parameter-sensitivity spectra of complex models are generally distributed approximately uniformly on a logarithmic scale: data constrain only a few stiff directions composed of parameter combinations while providing almost no information about many sloppy directions \spacedcite{ref23,ref24}. Active subspace methods offer a related but independent tool from the perspective of data-driven dimensionality reduction and have been used to alleviate non-identifiability in biological dynamical systems \spacedcite{ref25,ref26}. Non-identifiability and degeneracy in DCM are central issues in the development of this framework \cite{ref27}, and reporting posteriors for linear parameter combinations (contrasts), rather than for raw coupling parameters, has increasingly been used to improve inferential specificity \cite{ref28}.

These studies show that identifiability at different levels is not equivalent. However, most related discussions remain confined to a single target level and have not incorporated raw parameters, candidate derived coordinates, and data-driven sensitivity directions as parallel candidate targets for systematic comparison within a common audit procedure. Evaluating posterior quality for only a few prespecified parameters is therefore insufficient to determine which level of mechanistic interpretation is supported by the current observations. NMM inversion must also examine the consistency and coupling among different interpretive targets, thereby delimiting the conclusions that can be drawn when microcircuit mechanisms are inferred from macroscopic observations \cite{ref29}. This issue becomes more pronounced when high-dimensional neural time series are compressed into low-dimensional features. To make SBI or other Bayesian inversion procedures computationally tractable for EEG/ECoG data, researchers commonly transform raw waveforms into manually designed summary statistics or low-dimensional representations learned by neural networks. Previous studies have shown that the choice of feature representation can markedly affect parameter recovery and posterior estimation quality \cite{ref30,ref31,ref32,ref62}, while systematic discrepancies between the simulator and real data can similarly weaken the credibility of inferential results \cite{ref33}. For NMM inversion, macroscopic predictive performance is not equivalent to mechanistic identifiability. A representation capable of reconstructing waveforms or predicting clinical labels well may nevertheless discard information required to distinguish changes in specific synaptic connections, time constants, or excitation--inhibition balance \cite{ref34}. Conversely, a target that appears unrecoverable under the current procedure may not be physiologically unobservable; its dynamical signature may instead have been attenuated or erased in the selected feature space \cite{ref35}. Evaluation of a summary representation therefore cannot be limited to overall prediction or condition-discrimination performance but must also examine its informational sufficiency for different inversion targets. The resulting failure of target recovery may arise either because the target itself lacks identifiability under the current observations or because information is lost during summary compression. Without distinguishing these two causes, it is difficult to determine whether the failure should be attributed to the model mechanism or the data representation.

\begin{figure}[p]
\centering
\includegraphics[keepaspectratio,width=1\textwidth,height=0.80\textheight]{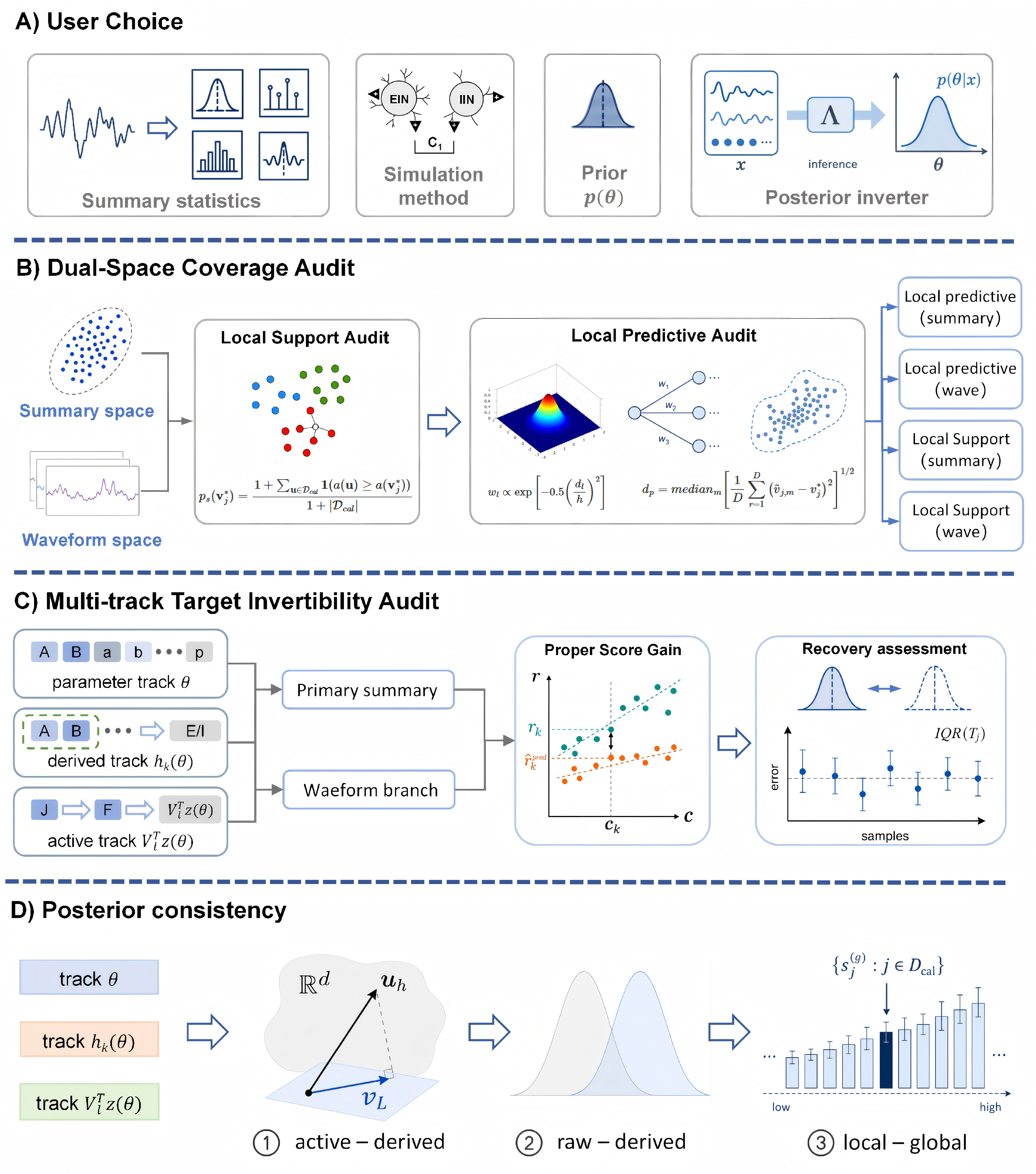}
\captionsetup{font=scriptsize,labelfont=bf,textfont=normalfont,labelsep=period,justification=justified,singlelinecheck=true,skip=2pt}
\caption{Overview of the target-audit framework. Given a simulator, prior, observation operator, summary representation and posterior estimator, the framework sequentially audits observation compatibility, simulation-internal recoverability and calibration across raw, derived and active targets, followed by joint and cross-track consistency. Its outputs delimit evidence and failure modes under the specified configuration.}\label{fig:main1}
\end{figure}

To address these issues, we propose a multi-track NMM-SBI target-audit framework (\hyperref[fig:main1]{Fig. 1}). Raw parameters, preregistered derived coordinates, and data-driven active coordinates are treated as three classes of parallel candidate targets. The framework sequentially performs three audits: after shared observation calibration, it examines observation compatibility in the summary and waveform diagnostic spaces; it trains target-specific posteriors and uses a controlled Wave+ branch to localize summary information loss; and it compares single-target, joint, and cross-track posteriors to identify changes in marginal performance, parameter dependence, and representational inconsistency.

The framework does not treat simulation-internal posterior estimation as validation of real mechanisms \cite{ref36}. Instead, it embeds posterior estimation within a multi-target audit oriented toward mechanistic interpretation. It does not assume that all candidate targets have equal interpretive eligibility. It integrates observation coverage, target recovery, summary information retention, and predictive consistency to distinguish the support status of different targets and the sources of their failure. We first evaluated the error rate and power of the audit statistics under known ground truth and then applied the framework to two real-data settings: SOZ-local iEEG--Epileptor and MMN--CMC-inspired. We show that the framework can identify not only whether a candidate NMM configuration sufficiently covers real observations, but also which targets have interpretable posterior support under a given summary. Our framework may provide an optional methodological constraint for subsequent NMM-SBI applications. The complete implementation, configuration files, and instructions for standalone execution are available at \url{https://github.com/wutuobangC/SBI-NMM-Audit}

\clearpage
\clearpage

\hypertarget{sec-2}{%
\section{2. Methods}\label{sec-2}}

\hypertarget{sec-2-1}{%
\subsection{2.1 Observation-layer calibration of signals}\label{sec-2-1}}

This study used simulation-based inference as the foundational framework for posterior estimation and integrated multi-target track diagnostics with a dual-space audit. Under this audit framework, the user specifies a neural dynamical simulator, a parameter prior distribution, a summary-statistics method, and a posterior inverter. A candidate inference configuration can be expressed as:\[\mathcal{S}=\psi\left(M_{\theta},p(\theta),g_{s},q_{\eta}\right)\tag{1}\label{eq:main-1}\]where \(M_{\theta}\) denotes the neural dynamical simulator, \(p(\theta)\) the parameter prior distribution, \(g_{s}\) the summary statistics, \(q_{\eta}\) the posterior inverter, and \(\psi\) the observation-layer calibration.

Distributional shift between real empirical data and data simulated by the generative model may arise not only from mismatch in the dynamical model \(M_{\theta}\) or parameter prior \(p(\theta)\), but also from differences in the macroscopic observation layer \(\psi\). To prevent data leakage, the framework partitions the real data by patient or participant into a non-overlapping observation calibration set \(\mathcal D^{\mathrm{obs}}_{\mathrm{cal}}\) (70\%) and observation test set \(\mathcal D^{\mathrm{obs}}_{\mathrm{test}}\) (30\%). Candidate observation parameters \(\Psi = \{\psi^{(1)},\psi^{(2)},\dots,\psi^{(B)}\}\) are searched only in \(\mathcal D^{\mathrm{obs}}_{\mathrm{cal}}\), and the following is estimated:\[\hat{\psi} = \arg\min_{\psi^{(b)}\in\Psi} \mathcal{L}_{\mathrm{obs}} \left( D(\mathbf{Y}_{\mathrm{obs}}^{\mathrm{train}}), D(\psi^{(b)}(M_{\theta})) \right)\tag{2}\label{eq:main-2}\]where \(D(\cdot)\) denotes user-selected summary statistics, \(\mathbf{Y}_{\mathrm{obs}}^{\mathrm{train}}\) the real data in \(\mathcal D^{\mathrm{obs}}_{\mathrm{cal}}\), and \(\psi^{(b)}(M_{\theta})\) the observation layer of the candidate simulated data. \(\mathcal L_{\mathrm{obs}}\) compares differences in the medians, interquartile widths, and tail coverage of diagnostic features after standardization by the simulated IQR, and is defined as \(\mathcal L_{\mathrm{obs}}=\mathcal L_{\mathrm{median}}+0.20\mathcal L_{\mathrm{IQR}}+0.35\mathcal L_{\mathrm{tail}}\).

After \(\widehat{\boldsymbol\psi}\) is selected, it is frozen and shared across all candidate summaries. Step 1 and subsequent diagnostics involving real observations are evaluated only on the held-out \(\mathcal D^{\mathrm{obs}}_{\mathrm{test}}\); the training and calibration of target posteriors and known-truth recovery metrics use independently partitioned simulated data. In this study, observation-layer calibration only reduces confounding from bounded observation nuisance factors and does not involve the dynamical model or subsequent diagnostics.

\hypertarget{sec-2-2}{%
\subsection{2.2 Dual-space coverage audit}\label{sec-2-2}}

After automatic calibration, the framework first evaluates whether a given neural mass model, parameter prior, and summary statistics can locally cover the data to be analysed, thereby avoiding direct SBI inversion when the observed data lie outside the joint support of the simulator and prior. This step provides a prerequisite compatibility audit for subsequent target-specific inference. The framework maps the reference simulated data \(\mathbf{Y}_{ref}\) and actual observed data \(\mathbf{Y}_{obs}\) into two complementary representation spaces: the inference Summary Space \(\mathbf{R} = \mathcal{E}_{p}(\mathbf{Y})\) and the Waveform Space \(\mathbf{Z} = \mathcal{D}(\mathbf{Y})\). The Summary Space is the feature space that directly drives SBI. \(\mathcal{E}_{p}\) may consist of hand-crafted statistics or learnable summary representations such as a CNN--LSTM or masked self-attention network (Masked Transformer) \spacedcite{ref37,ref38}. The Waveform Space is extracted by a fixed diagnostic function \(\mathcal{D}\) that is not involved in posterior training; \(\mathcal{D}\) is predefined before the experiment according to the observed phenomena that the candidate dynamical model purports to explain. The dual-space diagnostic is intended to reduce the risk that excessive compression by learned summaries masks waveform-level mismatch.

Within each representation space, the framework independently performs two complementary diagnostics: a Local Support Audit and a Local Predictive Audit. The former quantifies the local proximity of observed data within the domain generated by the reference model, whereas the latter further assesses whether this proximity results from stochastic flukes caused by random inputs or observation noise.

Within each experimental condition, Step 1 partitions the simulated samples into three non-overlapping subsets: \(\mathcal D_{\mathrm{ref}}^{\mathrm{sup}}\), \(\mathcal D_{\mathrm{cal}}^{\mathrm{sup}}\), and \(\mathcal D_{\mathrm{null}}^{\mathrm{sup}}\). \(\mathcal D_{\mathrm{ref}}^{\mathrm{sup}}\) is used to fit the standardization transform and construct the local \(k\)-nearest-neighbour space; \(\mathcal D_{\mathrm{cal}}^{\mathrm{sup}}\) is used to estimate the distribution of nonconformity scores for valid samples and single-observation conformal \(p\) values; and \(\mathcal D_{\mathrm{null}}^{\mathrm{sup}}\) provides a condition-matched pool of valid \(p\) values for generating the batch-level pseudo-observed null. The real observation test set \(\mathcal D^{\mathrm{obs}}_{\mathrm{test}}\) is not involved in this fitting or null-distribution construction.

\hypertarget{local-support-audit}{%
\subsubsection{Local Support Audit}\label{local-support-audit}}

The Local Support Audit follows the inductive conformal anomaly detection (ICAD) framework \cite{ref39}. For any sample \(\mathbf{v}^*_j\) to be audited in the real observation set \(\mathcal D^{\mathrm{obs}}_{\mathrm{test}}\) (where \(\mathbf v\) generically denotes a representation in either the \(\mathbf R\) or \(\mathbf Z\) space), the algorithm retrieves the \(k\) nearest neighbours \(\mathcal{N}_k(\mathbf{v}^*_j)\) from \(\mathcal D_{\mathrm{ref}}^{\mathrm{sup}}\) and defines the nonconformity score as:\[a(\mathbf{v}^*_j) = \frac{1}{k} \sum_{\mathbf{u} \in \mathcal{N}_k(\mathbf{v}^*_j)} \|\mathbf{v}^*_j - \mathbf{u}\|_2\tag{3}\label{eq:main-3}\]The framework treats each valid simulated sample in \(\mathcal D_{\mathrm{cal}}^{\mathrm{sup}}\) in turn as a pseudo-query, performs the same calculation, and obtains the condition-specific split-conformal upper-tail fraction:\[p_{s}(\mathbf{v}^*_j) = \frac{1 + \sum_{\mathbf{u} \in \mathcal D_{\mathrm{cal}}^{\mathrm{sup}}} \mathbf{1}[a(\mathbf{u}) \geq a(\mathbf{v}^*_j)]}{1 + |\mathcal D_{\mathrm{cal}}^{\mathrm{sup}}|}\tag{4}\label{eq:main-4}\]where \(\mathbf{1} (\cdot)\) is the indicator function, equal to 1 when \((\cdot)\) is true and 0 otherwise. \(p_{s}(\mathbf{v}^*_j)\) ranges from \([0, 1]\) and measures the upper-tail rank of the observation's nonconformity score in the condition-specific valid calibration distribution. A lower conformal \(p\) value indicates that the observation's nonconformity score exceeds that of most valid simulated samples. Its local neighbourhood structure may therefore extend beyond the reference support provided by the current model--prior configuration, suggesting potentially insufficient coverage. For samples at the edge of the distribution, the posterior in subsequent SBI training is more susceptible to local sample sparsity and model extrapolation. Equation (4), and the subsequent \(p\) metrics, are used in this study as calibrated scores of distributional or structural matching and should be distinguished from \(p\) values in independent hypothesis tests.

\hypertarget{local-predictive-audit}{%
\subsubsection{Local Predictive Audit}\label{local-predictive-audit}}

The preceding Local Support Audit can establish only that the simulation database contains waveforms similar to the observed data; it cannot rule out the possibility that this similarity is a stochastic fluke caused by random inputs or observation noise \spacedcite{ref40,ref41}. The Local Predictive Audit therefore conducts a further assessment by independently restarting the physical simulation in the local parameter region of the observed sample and examining whether the similarity remains stable. For \(\mathbf v_j^*\), the framework identifies the \(k\) nearest neighbours in the condition-matched \(\mathcal D_{\mathrm{ref}}^{\mathrm{sup}}\) and constructs kernel weights according to distance \(d_l\):\[w_l \propto \exp\left[-0.5 \left(\frac{d_l}{h}\right)^2\right]\tag{5}\label{eq:main-5}\]where \(h\) is the median neighbour distance. Based on the weights \(w_l\), the framework samples multiple sets of local parameters from the parameter space \(\{\boldsymbol{\theta}_l\}\) of the corresponding samples. Each sampled parameter set is then assigned a new random seed for an independent forward dynamical simulation, and its output is remapped to the corresponding feature space. We define the local predictive discrepancy \(d_p\) of sample \(\mathbf{v}^*_j\) as the median RMSE between the representations \(\tilde v_{j,m}\) of multiple independent predictive draws and the original representation:\[d_p = median_m \left[\frac{1}{D}\sum_{r=1}^{D}\left(\tilde v_{j,m}-v^{*}_{j}\right)^2\right]^{1/2}\tag{6}\label{eq:main-6}\]Using the calibration-set samples \(\mathcal D_{\mathrm{cal}}^{\mathrm{sup}}\), we then replace the nonconformity score \(a(\mathbf{v}^*_j)\) in Equation (3) with \(d_p\) to calculate the predictive conformal \(p\) value \(p_{r}(\mathbf{v}^*_j)\). \(d_p\)\hspace{0pt} is a random quantity generated by kernel-weighted resampling and independent resimulation. Thus, \(p_r(\mathbf{v}^*_j)\) evaluates the representational stability of the local parameter region under repeated simulations rather than the geometric distance between the observation and the reference samples.

\hypertarget{batch-statistics}{%
\subsubsection{Batch statistics}\label{batch-statistics}}

Single-observation conformal \(p\) values are used to localize anomalies but cannot be simply averaged to form a dataset-level conclusion. Let \(\mathcal D^{\mathrm{obs}}_{\mathrm{test}}\) and \(\mathcal D_{\mathrm{null}}^{\mathrm{sup}}\), after calibration using the same \(\mathcal D_{\mathrm{cal}}^{\mathrm{sup}}\), yield \(\{p_i^{\mathrm{obs}}\}_{i=1}^{n}\) and \(\{p_j^{\mathrm{null}}\}_{j=1}^{m}\), respectively. The framework uses a Welch-studentized lower-tail difference as the batch statistic:\[T_{\mathrm{obs}}=\frac{\overline{-\log p^{\mathrm{obs}}}-\overline{-\log p^{\mathrm{null}}}}{\sqrt{s_{\mathrm{obs}}^2/n+s_{\mathrm{null}}^2/m}} \tag{7}\label{eq:main-7}\]A larger \(T_{\mathrm{obs}}\) indicates more numerous or more extreme low \(p\) values in the real batch. To construct the null distribution, within each experimental condition the framework samples with replacement from \(\mathcal D_{\mathrm{null}}^{\mathrm{sup}}\) a pseudo-observed batch with the same condition composition and sample size as \(\mathcal D^{\mathrm{obs}}_{\mathrm{test}}\), and independently samples a pseudo-null baseline with the same composition as the finite null pool. \(T_b^{\mathrm{null}}\) is recalculated in every replicate. We used \(B=2{,}000\) Monte Carlo pseudo experiments and defined:\[p_{\mathrm{global}}=\frac{1+\sum_{b=1}^{B}\mathbf 1\left(T_b^{\mathrm{null}}\geq T_{\mathrm{obs}}\right)}{B+1} \tag{8}\label{eq:main-8}\]This construction holds \(\mathcal D_{\mathrm{ref}}^{\mathrm{sup}}\), \(\mathcal D_{\mathrm{cal}}^{\mathrm{sup}}\), and the observation-layer calibration parameters fixed, while matching the condition composition of the real batch and propagating sampling variation from the finite \(\mathcal D_{\mathrm{null}}^{\mathrm{sup}}\). A lower \(p_{\mathrm{global}}\) indicates that the concentration of low-tail values in the real observation batch exceeds normal variation under the valid batch null; a value that is not low indicates only that this type of batch-level mismatch was not detected.

The framework calculates global \(p\) values for summary support, summary predictive, waveform support, and waveform predictive, and applies a Bonferroni combination across all non-duplicate audits:\[p_{\mathrm{B}}=\min\left(1,m*\min_{\ell=1,\ldots,m}p_\ell\right)\tag{9}\label{eq:main-9}\]where \(m\) is the number of non-duplicate audits. When the summary space and waveform diagnostic space are identical, waveform-space statistics are not included a second time, such that \(m=2\); otherwise, \(m=4\).

In the dual-space diagnostic, the framework reports four batch-level global \(p\) values for summary support, summary predictive, waveform support, and waveform predictive, namely \(p_{\mathrm{sup}}^R,p_{\mathrm{pred}}^R,p_{\mathrm{sup}}^Z,p_{\mathrm{pred}}^Z\), together with the Bonferroni omnibus \(p_{\mathrm{B}}\) value. If the p-value distribution in the summary space shows no evident lower-tail shift whereas the p values in the waveform diagnostic space are systematically reduced, the current summary representation may have compressed or attenuated waveform-level evidence of model misspecification. If the p values in the summary space are themselves systematically low, subsequent posterior inference may face insufficient local support or predictive instability in the actual SBI input space, and its mechanistic interpretation should be treated cautiously.

\hypertarget{sec-2-3}{%
\subsection{2.3 Multi-track target recoverability audit}\label{sec-2-3}}

After determining whether the model configuration is mismatched to the actual observed data, the framework proceeds to the second stage. This stage identifies which specific targets can be supported by the observed data and provides a statistically credible assessment of recoverability within the simulation configuration. Unlike conventional methods that directly train a single joint posterior distribution over the full high-dimensional parameter space \(\theta\), which often obscures low-dimensional degeneracy caused by locally non-identifiable parameters \cite{ref27}, our framework constructs multilevel Target Tracks and independently trains Target-specific Posteriors to audit the interpretive boundaries of all specified targets.

To comprehensively probe the information-carrying capacity of the system, the framework divides all potential inference targets \(t_k \in \mathcal{T}\) into three interpretive tracks:

\begin{enumerate}
\def\labelenumi{\arabic{enumi}.}
\item
  Raw-parameter track (Raw Track, \(t_k = \theta_k\)): This track directly examines the recoverability of individual underlying physiological parameters, such as excitatory synaptic gains or inhibitory time constants. It tests whether the current macroscopic summary features contain sufficiently specific information to disentangle a single microscopic physical quantity.
\item
  Derived-coordinate track (Derived-coordinate Track, \(t_k = h_k(\theta)\)): This track examines predefined nonlinear parameter combinations or derived mechanistic quantities, such as the E/I balance ratio or timescale ratio \cite{ref42}. In neural mass models, parameter compensation often makes individual parameters difficult to constrain independently, whereas specific parameter combinations may be highly identifiable.
\item
  Active track (Active Track, \(t_k = \mathbf{v}_l^\top \mathbf{z}(\theta)\)): This is a fully data-driven exploratory track. The framework selects local centres in the standardized parameter space, fits local ridge regressions under controlled experimental conditions to estimate the Jacobian \(\widehat{\mathbf J}_j\), and constructs the globally averaged sensitivity matrix \(\widehat{\mathbf F}_{\mathrm{global}} = \frac{1}{N_{\mathrm{center}}} \sum_{j=1}^{N_{\mathrm{center}}} \widehat{\mathbf J}_j^\top \widehat{\mathbf J}_j\). Its principal eigen-directions define candidate Active Targets \cite{ref43}. The active dimensionality \(L\) is determined automatically according to cumulative sensitivity, the relative eigengap, and bootstrap subspace stability (see \hyperref[app-3]{Appendix 3}). This track is intended to reveal the low-dimensional parameter manifold most strongly constrained by the observed signals, serving as a data-driven reference for sensitivity directions.
\end{enumerate}

For Steps 2--3, the simulated data are stratified by experimental condition into four non-overlapping subsets: a training set \(\mathcal D_{\mathrm{train}}^{\mathrm{post}}\) (60\%), validation set \(\mathcal D_{\mathrm{val}}^{\mathrm{post}}\) (10\%), calibration set \(\mathcal D_{\mathrm{cal}}^{\mathrm{post}}\) (15\%), and test set \(\mathcal D_{\mathrm{test}}^{\mathrm{post}}\) (15\%). For each candidate target \(t_k\), the framework uses only \(\mathcal D_{\mathrm{train}}^{\mathrm{post}}\) to train a target-specific low-dimensional posterior estimator \(q_{\phi,k}(t_k \mid \widetilde{\mathbf{R}}, c)\) based on the prespecified inverter, where \(c\) denotes the experimental condition and \(\mathbf{R}\) the summary representation of the simulated signal. To eliminate scale differences among summaries of different dimensionalities, the input features are standardized as \(\widetilde{\mathbf{R}}=\frac{\mathbf{R}-\mu_{\mathrm{train}}}{\sigma_{\mathrm{train}}+\epsilon}\). Importantly, the standardization parameters (mean and standard deviation) are estimated independently from \(\mathcal D_{\mathrm{train}}^{\mathrm{post}}\) alone. During this procedure, \(\mathcal D_{\mathrm{val}}^{\mathrm{post}}\) is used for model selection and early stopping, whereas \(\mathcal D_{\mathrm{cal}}^{\mathrm{post}}\) is strictly excluded from all posterior training and is used only for posterior-coverage calibration and construction of subsequent empirical reference distributions. Finally, on the independent test set \(\mathcal D_{\mathrm{test}}^{\mathrm{post}}\), the framework evaluates the recoverability of each target-specific posterior in terms of distributional information gain, point-estimation accuracy, and summary information loss.

\hypertarget{sec-2-3-1}{%
\subsubsection{2.3.1 Gain evaluation (Proper Score Gain, PSG)}\label{sec-2-3-1}}

To test whether the summary genuinely conveys information, the framework introduces the Proper Score Gain (PSG). This metric is based on the skill-score framework, which uses the ratio of a strictly proper scoring rule for model and reference predictions to quantify relative improvement \cite{ref44}. Here, we use the Continuous Ranked Probability Score (CRPS) as the proper scoring rule \cite{ref45} and replace the reference prediction with the condition-specific prior baseline defined below.

The framework uses an empirical Monte Carlo estimator to calculate the CRPS of the posterior distribution. For the \(i\)th sample in the test set \(\mathcal D_{\mathrm{test}}^{\mathrm{post}}\), given its point-valued true target \(t_i^{true}\), \(M\) predictive samples \(\{t_i^{(m)}\}_{m=1}^M \sim q_{\phi,k}(t_k \mid \tilde{\mathbf{R}}_i)\) are drawn without bias from the posterior, where \(\mathbf R_i\) denotes the feature vector obtained by applying the selected summary extractor to the waveform \(\mathbf Y_i\) of the \(i\)th sample. The empirical estimator of the sample-specific \(\mathrm{CRPS}_i\) is defined as:\[\mathrm{CRPS}_i = \frac{1}{M} \sum_{m=1}^{M} \left|t_i^{(m)} - t_i^{\mathrm{true}}\right| - \frac{1}{2M^2} \sum_{m=1}^{M} \sum_{m'=1}^{M} \left|t_i^{(m)} - t_i^{(m')}\right| \tag{10}\label{eq:main-10}\]The first term represents the mean absolute error between the posterior samples and the true value \(t_i^{true}\), whereas the second corrects for dispersion among the posterior samples themselves, allowing CRPS to evaluate both the location and uncertainty of the distribution. Further, let \(N_{\mathrm{test}}^{\mathrm{post}}=|\mathcal D_{\mathrm{test}}^{\mathrm{post}}|\); then:\[\mathbb E[\mathrm{CRPS}_{\mathrm{post}}]=\frac{1}{N_{\mathrm{test}}^{\mathrm{post}}}\sum_{i\in\mathcal D_{\mathrm{test}}^{\mathrm{post}}}\mathrm{CRPS}_i \tag{11}\label{eq:main-11}\]To evaluate the actual information gain after the model extracts the summary, the framework then constructs an uninformative baseline distribution, the condition-specific prior baseline \(p(t_k)\). For each real sample \(i\) in the test set, \(M\) baseline samples are drawn from the empirical target distribution for the corresponding condition \(c_i\) in the simulated training set: \(\{\widetilde t_i^{(m)}\}_{m=1}^M \sim \widehat p(t_k\mid c_i)\). The baseline score \(\mathbb{E}[\mathrm{CRPS}_{base}]\) is calculated in exactly the same manner. This baseline does not use \(\mathbf R_i\), and its target-value distribution is constructed solely from \(\mathcal D_{\mathrm{train}}^{\mathrm{post}}\); it therefore introduces no information from the test set. Finally, Proper Score Gain (PSG) is defined as the normalized reduction in the posterior score relative to the condition-specific baseline score:\[\mathrm{PSG}=1-\frac{\mathbb E[\mathrm{CRPS}_{\mathrm{post}}]}{\mathbb E[\mathrm{CRPS}_{\mathrm{base}}]} \tag{12}\label{eq:main-12}\]Because a lower CRPS indicates a predictive distribution closer to the true target value, PSG measures the improvement in distributional quality obtained by conditioning on the summary relative to the baseline prediction. \(\mathrm{PSG}>0\), \(\approx0\), and \(<0\) indicate, respectively, that the posterior achieves a positive gain relative to the condition-specific prior baseline, provides approximately no additional information, or performs worse than the baseline.

\hypertarget{sec-2-3-2}{%
\subsubsection{2.3.2 Evaluation of point recoverability}\label{sec-2-3-2}}

In addition to gain evaluation, the framework assesses whether the Posterior Mean can serve as a Point Estimator of target \(t_k\) and accurately track its true trajectory of variation. For the \(i\)th sample in \(\mathcal D_{\mathrm{test}}^{\mathrm{post}}\), \(M\) samples are drawn from \(q_{\phi,k}(t_k\mid\widetilde{\mathbf R}_i,c_i)\), and \(\widehat t_{k,i}=M^{-1}\sum_m t_{k,i}^{(m)}\) is used as the point estimate. The coefficient of determination is defined as:\[R_k^2=1-\frac{\sum_{i\in\mathcal D_{\mathrm{test}}^{\mathrm{post}}}(\widehat t_{k,i}-t_{k,i}^{\mathrm{true}})^2}{\sum_{i\in\mathcal D_{\mathrm{test}}^{\mathrm{post}}}(t_{k,i}^{\mathrm{true}}-\bar t_k^{\mathrm{true}})^2} \tag{13}\label{eq:main-13}\]A higher \(R_k^2\) indicates that the posterior mean more accurately recovers the true target value.

Posterior uncertainty is evaluated solely from the raw samples output by the posterior estimator on \(\mathcal D_{\mathrm{test}}^{\mathrm{post}}\). For the \(i\)-th test sample, its 0.05 and 0.95 quantiles are defined as \(L_{k,i}^{90} = Q_{0.05}\left(\left\{t_{k,i}^{(m)}\right\}_{m=1}^{M}\right)\) and \(U_{k,i}^{90} = Q_{0.95}\left(\left\{t_{k,i}^{(m)}\right\}_{m=1}^{M}\right)\), respectively, and the empirical coverage on the blind test set is defined as:\[\widehat{\operatorname{Cov}}_{90,k} = \frac{1}{N} \sum_{i\in\mathcal D_{\mathrm{test}}} \mathbf 1 \left(L_{k,i}^{90} \leq t_{k,i}^{\mathrm{true}} \leq U_{k,i}^{90}\right) \tag{14}\label{eq:main-14}\]The resulting coverage error is \(E_{90,k} = \left\vert{} \widehat{\operatorname{Cov}}_{90,k}-0.90 \right\vert{}\). Ideally, \(E_{90,k}\approx 0.0\). To evaluate calibration of the full posterior distribution, let \(r_{k,i}\) denote the number of posterior draws smaller than the true value; if ties are present, their ranks are determined at random. For finite \(M\), we further introduce \(\xi_{k,i}\sim U(0,1)\) to construct the randomized PIT:\[u_{k,i}^{\star}=\frac{r_{k,i}+\xi_{k,i}}{M+1} \tag{15}\label{eq:main-15}\]If the raw posterior is calibrated, \(u_{k,i}^{\star}\) should approximately follow \(U(0,1)\). Let \(\widehat F_k\) denote its empirical distribution function. The raw SBC deviation is defined as:\[D_{\mathrm{SBC},k}^{\mathrm{raw}}=\sup_{u\in[0,1]}|\widehat F_k(u)-u| \tag{16}\label{eq:main-16}\]The closer \(D_{\mathrm{SBC},k}^{\mathrm{raw}}\) is to 0, the closer the randomized-PIT distribution is to the uniform distribution. In this study, only \(E_{90,k}^{\mathrm{raw}}\) and \(D_{\mathrm{SBC},k}^{\mathrm{raw}}\) are used to evaluate uncertainty calibration of the raw posterior. In brief, coverage error is an intuitive check at one specific credible level, whereas SBC is a global check across all posterior quantiles; the two are related but not entirely redundant \cite{ref13}.

\hypertarget{sec-2-3-3}{%
\subsubsection{2.3.3 Summary information-loss diagnostic}\label{sec-2-3-3}}

In the target-specific recoverability audit, a low or non-positive intrinsic scoring gain for a target \(t_k\) does not by itself distinguish whether the target lacks stable, usable information under the current macroscopic-signal observation conditions, or whether relevant information may be present in the raw waveform but was not adequately retained during compression by the current summary representation. Moreover, previous work has shown that incidental performance inflation caused solely by increased model capacity can also affect the final outcome. Information gain must therefore be evaluated against appropriate control conditions rather than by inspecting performance values alone \cite{ref46}.

To further distinguish these two possibilities, the framework compares three input paths while holding the simulator, parameter prior, target definition, and posterior inverter fixed. The Original path uses only the current summary, \(\mathbf X_{\mathrm p}=\mathbf R\). The Wave+ path concatenates a branch obtained by standardizing fixed waveform descriptors using \(\mathcal D_{\mathrm{train}}^{\mathrm{post}}\) and applying PCA, \(\mathbf X_{\mathrm r}=[\mathbf R;\mathbf W_{\mathrm{PCA}}]\). The primary negative control, PermWave, retains exactly the same dimensionality, marginal distribution, scale, and network architecture but separately permutes the rows of \(\mathbf W_{\mathrm{PCA}}\) by condition within \(\mathcal D_{\mathrm{train}}^{\mathrm{post}}\), \(\mathcal D_{\mathrm{val}}^{\mathrm{post}}\), \(\mathcal D_{\mathrm{cal}}^{\mathrm{post}}\), and \(\mathcal D_{\mathrm{test}}^{\mathrm{post}}\). This removes sample-level waveform--target associations without exchanging samples across splits \cite{ref47}.

The three paths separately report PSG, \(R^2\), \(E_{90}^{\mathrm{raw}}\), and \(D_{\mathrm{SBC}}^{\mathrm{raw}}\). Only when the PSG for Wave+ exceeds those for both Original and PermWave, without an evident deterioration in the coverage error or SBC distance of its raw posterior, is the result interpreted as indicating that the additional waveform descriptors contain target-relevant information not retained by the current summary. If Wave+ and PermWave improve to a similar extent, the change is more likely to arise from the input structure, optimization, or finite-sample variation. If neither improves, no additional information is detected within the current waveform-description system.

\hypertarget{sec-2-4}{%
\subsection{2.4 Posterior consistency and mechanistic alignment}\label{sec-2-4}}

After establishing the local coverage of the model's observation manifold and the single-target marginal recoverability of each candidate Target, the framework proceeds to the third stage. The preceding single-target posterior estimation can establish only that a target is recoverable in isolation; it cannot reveal joint coupling among multiple physical parameters or cross-level inconsistency. This step does not rescreen targets. Instead, it conducts a structured audit of parameter sets that already exhibit single-target recoverability, primarily distinguishing three classes of higher-order features: the effect of joint modelling on the stability of single-target inference, synchrony and coupling among targets within the same track, and mechanistic alignment across the different interpretive tracks (Raw, Derived, and Active).

\hypertarget{sec-2-4-1}{%
\subsubsection{2.4.1 Single-track joint posterior and coupling}\label{sec-2-4-1}}

If \hyperref[sec-2-3]{Section 2.3} has identified a reasonable target set \(\mathcal{T}_{\ell} = \{t_{\ell,1}, t_{\ell,2}, \dots, t_{\ell,K_{\ell}}\}\) in a track \(\ell\), the within-track targets for the \(i\)th simulated sample are vectorized as \(\mathbf{t}_{\ell,i} = (t_{\ell,1,i}, \dots, t_{\ell,K_{\ell},i})^\top\). The framework then trains a track-specific joint posterior \(q_{\phi,\ell}(\mathbf{t}_{\ell} \mid \mathbf{R})\) using the selected posterior estimator. For each test sample \(i\), Monte Carlo sampling from this joint posterior generates \(M\) multidimensional joint sample vectors \(\mathbf{t}_{\ell,i}^{(m)} \sim q_{\phi,\ell}(\mathbf{t}_{\ell} \mid \mathbf{R}_i), m=1,\dots,M\). Each sample vector simultaneously contains values for all targets in that track. Thus, for target \(t_{\ell,k}\), its joint marginal posterior can be approximated directly by the \(k\)th component of all joint samples:\[t_{\ell, k,i}^{(m)} = \left[ \mathbf{t}_{\ell,i}^{(m)} \right]_k, \quad m=1,\dots,M\tag{17}\label{eq:main-17}\]The framework recalculates PSG, \(R^2\), \(E_{90}^{\mathrm{raw}}\), and \(D_{\mathrm{SBC}}^{\mathrm{raw}}\) from the joint marginal distribution and compares them with the single-target posterior from Step 2. To retain the same scale as the primary metric in Step 2, the change in joint performance is defined as:\[\Delta\mathrm{PSG}_k=\mathrm{PSG}_k^{\mathrm{joint}}-\mathrm{PSG}_k^{\mathrm{single}} \tag{18}\label{eq:main-18}\]The single and joint models use the same \(\mathcal D_{\mathrm{train}}^{\mathrm{post}}\) to construct the condition-specific prior baseline. Accordingly, \(\Delta\mathrm{PSG}_k>0\) indicates that the joint marginal outperforms the single-target posterior, whereas \(\Delta\mathrm{PSG}_k<0\) indicates deterioration. The framework bootstraps the sample indices in \(\mathcal D_{\mathrm{test}}^{\mathrm{post}}\) with replacement and evaluates directional stability using the 2.5\%--97.5\% percentile interval \cite{ref48}. An interval entirely above zero, entirely below zero, or spanning zero is recorded as improvement, deterioration, or no detected stable change, respectively. \(E_{90}^{\mathrm{raw}}\) and \(D_{\mathrm{SBC}}^{\mathrm{raw}}\) are always based on the unscaled raw posterior.

In addition, to identify potential association structures among targets within the same track, the framework evaluates two levels: posterior-mean synchrony across observations and posterior-sample coupling within a single observation. For any two targets \(t_a\) and \(t_b\), their posterior means \(\hat{t}_{a,i} = \mathbb{E}[t_a \mid \mathbf{R}_i]\) and \(\hat{t}_{b,i} = \mathbb{E}[t_b \mid \mathbf{R}_i]\) are first extracted for each sample \(i\) in the full test set \(\mathcal D_{\mathrm{test}}^{\mathrm{post}}\). The Pearson correlation coefficient \(\rho_{ab}^{\mathrm{mean}}\) is then calculated across the test population, and its uncertainty is evaluated by bootstrap resampling of the test samples.

For a fixed test sample \(i\), the component sequences \(\{ t_{\ell,a,i}^{(m)} \}_{m=1}^{M}\) and \(\{ t_{\ell,b,i}^{(m)} \}_{m=1}^{M}\) corresponding to targets \(a\) and \(b\) are further extracted from its joint sample set \(\{ \mathbf{t}_{\ell,i}^{(m)} \}_{m=1}^{M}\). The correlation \(\rho_{ab,i}^{\mathrm{joint}}\) between the two variables is calculated within that observation using the \(M\) samples. The framework summarizes its distribution across observations and reports both the median and the proportion of observations satisfying the prespecified practical coupling threshold \(|\rho_{ab,i}^{\mathrm{joint}}|>\tau\). Here, \(\tau\) denotes the prespecified minimum practical coupling strength used to distinguish interpretable posterior dependence from numerical correlations near zero. The primary analysis uses \(\tau=0.3\), following classifications of correlation coefficients in the statistical literature, in which \(|\rho|\approx0.3\) is generally regarded as the lower bound of a moderate effect size \cite{ref49}. Corresponding confidence intervals are finally obtained by observation-level bootstrapping.

These two metrics are complementary. The former reflects macroscopic covariation under external driving, whereas the latter indicates whether two parameters within the dynamical model are statistically coupled when the observation is held fixed.

\hypertarget{sec-2-4-2}{%
\subsubsection{2.4.2 Cross-track mechanistic consistency}\label{sec-2-4-2}}

After auditing targets within each track, the framework further examines whether the Raw, Derived, and Active tracks have consistent interpretive structures. This analysis comprises three levels: directional consistency, distributional consistency, and global stability. Importantly, these analyses evaluate only the mathematical self-consistency of the different tracks under the current model and data representation; they do not constitute causal evidence that a mechanism actually exists.

\hypertarget{sec-2-4-2-a}{%
\paragraph{(a) Active--derived direction consistency}\label{sec-2-4-2-a}}

First, active--derived direction consistency assesses whether the global active subspace contains the globally optimal linear proxy direction for a predefined derived coordinate under the specified prior and parameterization. For the \(h\)-th predefined derived coordinate \(m_h=T_h(\boldsymbol\theta)\), we first standardize the \(d\) free raw parameters and the derived coordinate separately on \(\mathcal D_{\mathrm{train}}^{\mathrm{post}}\):\[\widetilde{\theta}_{ik}=\frac{\theta_{ik}-\mu_k}{\sigma_k},\widetilde m_{ih}=\frac{m_{ih}-\mu_{m_h}}{\sigma_{m_h}}\tag{19}\label{eq:main-19}\]The optimal global linear direction of this derived coordinate over the current prior range is then obtained by least-squares fitting:\[\widehat{\boldsymbol\beta}_h=\arg\min_{\boldsymbol\beta}\sum_{i\in\mathcal D_{\mathrm{train}}}\left(\widetilde m_{ih}-\boldsymbol\beta^\top\widetilde{\boldsymbol\theta}_i\right)^2\tag{20}\label{eq:main-20}\]For raw parameters absent from the definition of the derived coordinate \(T_h\), the corresponding coefficients are set to zero using the structural mask \(\mathbf M_h\); fixed parameters and observation-layer nuisance parameters are likewise excluded from the derived direction. The effective coefficients are then normalized to obtain the unit derived direction:\[\mathbf u_h
=\frac{\mathbf M_h\odot\widehat{\boldsymbol\beta}_h}{\left\|\mathbf M_h\odot\widehat{\boldsymbol\beta}_h\right\|_2}\tag{21}\label{eq:main-21}\]For random seed \(s\), let \(\mathbf V_{\mathrm{global}}\in\mathbb R^{d\times L}\) be the orthonormal basis formed by the first \(L\) global active directions independently selected for that seed, and let \(\mathbf u_h\in\mathbb R^d\) be the normalized linear proxy direction for the \(h\)th derived coordinate estimated from the simulation-training split for that seed. Because the sign and order of individual active directions, as well as rotations within the subspace, are not unique, we use the active subspace spanned by the \(L\) active directions as the formal object of comparison. We define:\[\begin{aligned}A_h^{\mathrm{AD}}=\left\|\mathbf V_{\mathrm{global}}^\top \mathbf u_h \right\|_2^2\\
s_h^{\mathrm{AD}}=1-A_h^{\mathrm{AD}}\end{aligned}\tag{22}\label{eq:main-22}\]where \(A_h^{\mathrm{AD}}\in[0,1]\) is the proportion of the derived direction projected into the active subspace, and \(s_h^{\mathrm{AD}}\) is the degree of directional misalignment. Because the absolute projection proportion is affected by the total number of parameters and the dimensionality of the active subspace, the framework further constructs a structure-matched finite geometric reference set. Let \(\mathcal I_{\mathrm{free}}\) denote the set of parameter coordinates that actually vary freely under the current prior. The framework rearranges the components of \(\mathbf u_h\) only among these free coordinates, while strictly preserving the number of nonzero coefficients, their values, sign relationships, and unit norm. Duplicate permutations are removed, and two directions with globally opposite signs are treated as the same geometric direction, yielding the permutation set:\[\mathcal U_h^{\mathrm{perm}}=\left\{\mathbf u_{h,1}^{\mathrm{perm}},\mathbf u_{h,2}^{\mathrm{perm}},\dots,\mathbf u_{h,B_h}^{\mathrm{perm}}\right\}\tag{23}\label{eq:main-23}\]where \(B_h=|\mathcal U_h^{\mathrm{perm}}|\) is the total number of unique permuted directions. For each direction in the set, its alignment with the same global active subspace is calculated as \(A_{h,b}^{\mathrm{perm}}=\left\lVert\mathbf V_{\mathrm{global}}^\top\mathbf u_{h,b}^{\mathrm{perm}}\right\rVert_2^2, b=1,\ldots,B_h\). Because the complete finite reference set is enumerated here, the empirical upper-tail fraction of the derived direction within this complete finite reference set is defined as:\[p_{h}^{\mathrm{AD}} = \frac{1}{B_h} \sum_{b=1}^{B_h} \mathbf 1 \left( A_{h,b}^{\mathrm{perm}} \ge A_h^{\mathrm{AD}} \right)\tag{24}\label{eq:main-24}\]A lower \(p_h^{\mathrm{AD}}\) indicates that the predefined derived direction is more closely aligned with the global active subspace than most other parameter combinations with the same sparsity, coefficient magnitudes, and sign structure. In addition, for a one-dimensional active subspace with \(L=1\), \(A_h^{\mathrm{AD}}\) reduces to the squared absolute cosine similarity between the derived direction and that active direction.

\hypertarget{sec-2-4-2-b}{%
\paragraph{(b) Raw--derived distributional consistency}\label{sec-2-4-2-b}}

Second, for a derived target \(t_{\mathrm{d}}=h(\boldsymbol{\theta}_{\mathrm{raw}})\) that can be explicitly defined from the raw parameters, the framework estimates its posterior distribution through two paths. In the first path, samples are drawn for sample \(i\) from the joint posterior of the Raw track in \(\mathcal D_{\mathrm{test}}^{\mathrm{post}}\) and are pushed forward through the derived function \(h\):\[\begin{aligned}\boldsymbol{\theta}_{\mathrm{raw},i}^{(m)}\sim q_{\phi,\mathrm{raw}}\left(\boldsymbol{\theta}_{\mathrm{raw}}\mid\mathbf{R}_i\right), \\ \widetilde{t}_{\mathrm{d},i}^{(m)}=h\left(\boldsymbol{\theta}_{\mathrm{raw},i}^{(m)}\right) \end{aligned} \tag{25}\label{eq:main-25}\]This yields the push-forward distribution of the raw posterior in derived space:\[h_{\#} q_{\phi,\mathrm{raw}}(\boldsymbol{\theta}_{\mathrm{raw}} \mid \mathbf{R}_i)\tag{26}\label{eq:main-26}\]where \(h_{\#}\) is the push-forward operator \cite{ref51}.

The second path estimates the target directly in the Derived track for that sample:\[t_{\mathrm{d},i}^{(m)}\sim q_{\phi,\mathrm{d}}(t_{\mathrm{d}}\mid\mathbf R_i) \tag{27}\label{eq:main-27}\]The framework uses the 1-Wasserstein distance to compare the empirical distributions obtained through the two paths:\[s_i^{\mathrm{test}}=W_1\!\left(\{\widetilde t_{\mathrm{d},i}^{(m)}\}_{m=1}^{M},\{t_{\mathrm{d},i}^{(m)}\}_{m=1}^{M}\right)\tag{28}\label{eq:main-28}\]Because finite posterior sampling, density-estimation error, and training stochasticity all produce nonzero Wasserstein distances \cite{ref52}, the posterior estimator is fitted using only \(\mathcal D_{\mathrm{train}}^{\mathrm{post}}\), with \(\mathcal D_{\mathrm{val}}^{\mathrm{post}}\) used for inverter selection. The framework performs the identical two-path calculation on the simulated calibration set \(\mathcal D_{\mathrm{cal}}^{\mathrm{post}}\), which is independent of posterior fitting, to obtain the reference distance set \(\{s_j^{\mathrm{cal}}\}_{j=1}^{N_{\mathrm{cal}}}\). It then calculates the empirical upper-tail fraction of the two-path discrepancy distance \(s_i^{\mathrm{test}}\) in the calibration reference distribution. Let \(N_{\mathrm{cal}}^{\mathrm{post}}=|\mathcal D_{\mathrm{cal}}^{\mathrm{post}}|\); then:\[p_i^{\mathrm{RD}}=\frac{1+\sum_{j\in\mathcal D_{\mathrm{cal}}^{\mathrm{post}}}\mathbf 1(s_j^{\mathrm{cal}}\ge s_i^{\mathrm{test}})}{N_{\mathrm{cal}}^{\mathrm{post}}+1} \tag{29}\label{eq:main-29}\]A lower \(p_{i}^{\mathrm{RD}}\)\hspace{0pt} indicates that, for this observation, the difference between the raw push-forward distribution and the directly estimated derived posterior substantially exceeds the normal error level of the complete reference inference procedure; the two tracks therefore lack self-consistency for this derived target.

\hypertarget{sec-2-4-2-c}{%
\paragraph{(c) Local--global active consistency audit}\label{sec-2-4-2-c}}

Finally, local--global active consistency assesses whether the global active subspace estimated from \(\mathcal D_{\mathrm{train}}^{\mathrm{post}}\) can represent the local sensitivity structure near each real observation in \(\mathcal D^{\mathrm{test}}_{\mathrm{obs}}\). For the \(s\)th random seed, let the independently selected Active dimensionality be \(L_s\) and the global Active basis be \(\mathbf V_{\mathrm{global}} \in \mathbb R^{d \times L_s}\). For the \(i\)th real-observation summary \(\mathbf R_i^{\mathrm{obs}}\) in \(\mathcal D^{\mathrm{test}}_{\mathrm{obs}}\), the framework first identifies the \(k\) nearest training centres in the standardized summary space. Here, \(k=36\). Using the outer products of local gradients already estimated at these centres, a weighted local sensitivity matrix is constructed:\[\begin{aligned}\mathbf F_i^{\mathrm{local}}=\sum_{j\in\mathcal N_k(i)}\omega_{ij}\mathbf J_j^\top\mathbf J_j \\
\sum_{j\in\mathcal N_k(i)}\omega_{ij}=1\end{aligned}\tag{30}\label{eq:main-30}\]where \(\omega_{ij}\) is a kernel weight defined according to neighbourhood distance \(d_{ij}\):\[\omega_{ij} = \frac{ \exp\left[-\frac{1}{2}(d_{ij}/h_i)^2\right] }{ \sum_{r\in\mathcal{N}_k(i)} \exp\left[-\frac{1}{2}(d_{ir}/h_i)^2\right]}\tag{31}\label{eq:main-31}\]The local basis \(\mathbf V_i\in\mathbb R^{d\times L_s}\) consists of the first \(L_s\) eigenvectors of \(\mathbf F_i^{\mathrm{local}}\), and we define:\[\begin{aligned}A_i^{\mathrm{LG}}=\frac{1}{L_s}\left\|\mathbf V_{\mathrm{global}}^\top\mathbf V_i\right\|_F^2 \\ 
s_i^{\mathrm{LG}}=1-A_i^{\mathrm{LG}}\end{aligned}\tag{32}\label{eq:main-32}\]This statistic is equivalent to the mean complement of the squared cosines of the principal angles {[}50,53{]}. \(\mathcal D_{\mathrm{cal}}^{\mathrm{post}}\) provides a dimension-matched reference distribution of simulated misalignment. For a real sample in \(\mathcal D^{\mathrm{test}}_{\mathrm{obs}}\), we define:\[p_i^{\mathrm{LG}}=\frac{1+\sum_{j\in\mathcal D_{\mathrm{cal}}^{\mathrm{post}}}\mathbf 1(s_j^{\mathrm{cal}}\ge s_i^{\mathrm{LG}})}{1+|\mathcal D_{\mathrm{cal}}^{\mathrm{post}}|} \tag{33}\label{eq:main-33}\]A lower \(p_i^{LG}\) indicates that the local--global shift in the real observations substantially exceeds normal variation in the reference simulation procedure.

\hypertarget{sec-3}{%
\section{3. Experiments}\label{sec-3}}

\hypertarget{sec-3-1}{%
\subsection{3.1 Experimental data and dynamical simulations}\label{sec-3-1}}

To evaluate the applicability of the proposed framework across different neural dynamical modelling settings, we designed one set of controlled experiments and two sets of real-data experiments. The controlled experiments used statistical simulations with analytically controllable ground truth to estimate the type I error rate, specificity, and statistical power of each audit module. The two real-data experiments covered two representative problems: spontaneous pathological dynamics and stimulus-evoked response dynamics. In the first experiment, the Epileptor model was applied to real SOZ-local iEEG seizure data to assess whether the framework could evaluate the coverage, parameter recoverability, and stability of derived interpretation of a candidate epileptic dynamical model in real ictal intracranial electrophysiological signals. In the second experiment, the Canonical Microcircuit model was applied to real ERP data to evaluate the framework's audit capability in a stimulus-response task.

\hypertarget{sec-3-1-1}{%
\subsubsection{3.1.1 Controlled simulation experiments}\label{sec-3-1-1}}

Real data generally lack observable ground truth for dynamical parameters, precluding direct quantification of the framework's own error rate. We therefore first evaluated the operating characteristics of each audit module in a set of synthetic tasks with analytically controllable structures. All data-generating processes were specified in advance and did not depend on the outputs of the audit statistics. The purpose of this experiment was to determine whether the framework controlled false positives when no mismatch was present and produced the corresponding responses when prespecified mismatch, information removal, or posterior distortion was present. The formal analysis comprised 200 independent Monte Carlo replicates. Random seeds represented computational replicates only and were not treated as biological samples.

For Step 1, we constructed an 8-dimensional Gaussian generative model with four conditions, in which samples under the \(c\)th condition followed \(\mathbf z \mid c \sim \mathcal N(\boldsymbol\mu_c,\mathbf I_8)\). Condition differences were encoded only in the first two dimensions, whereas dimensions 3--8 contained background noise across conditions. Each replicate generated 1,200 training samples and 320 audit samples. In addition to a matched condition drawn from the same generative distribution, we introduced two mean shifts. The first uniformly shifted dimensions 1--3 of the audit samples by \(1.6\). Because it affected only the first three coordinates, it appeared in both the inference summary and the complete diagnostic space. The second uniformly shifted dimensions 4--6 of the audit samples by \(1.6\). Because the compressed summary retained only the first two coordinates (the summary in this controlled experiment was a fixed two-dimensional projection), whereas the waveform diagnostic space retained all eight coordinates, the latter shift was invisible in the summary space but fully preserved in the waveform diagnostic space. This construction provided an explicit representation-dependent mismatch: its presence was determined by the generative equation, whereas whether it could be observed depended on the representation used. The Local Support Audit directly compared these conditional distributions, whereas the Local Predictive Audit regenerated local predictions to test whether neighbouring samples could be reproduced under independent random realizations.

The controlled experiment for Step 2 further separated target information, the summary-loss control, and posterior calibration. The target variable was generated from \(\mathcal U(-1,1)\). Each replicate contained 2,400 \(\mathcal D_{\mathrm{train}}^{\mathrm{post}}\) samples and 600 \(\mathcal D_{\mathrm{test}}^{\mathrm{post}}\) samples, and 96 posterior draws were generated for each posterior-test sample. Because this experiment directly constructed posteriors of known quality rather than training a neural estimator, no separate validation split was used. Under the information-sufficient Primary condition, both the posterior-mean error and posterior width were set to \(0.20\). The PermWave condition retained the same centre error, width, and input size but removed the sample-level association between the supplementary branch and the target. In the Waveform augmented condition, both the centre error and width were reduced to \(0.10\), representing a true supplementary-information improvement in distributional prediction accuracy. The No-temporal ablation sampled from the empirical prior to remove the summary--target association, whereas the Unsupported target condition made the true values independent of the predictive distribution.

To distinguish point recovery from uncertainty calibration, we started from the Primary posterior and reduced the deviations of all draws from their mean to \(25\%\) of the original values while holding the posterior mean fixed, thereby constructing an explicitly underdispersed posterior. This condition should retain a high \(R^2\), while \(E_{90}^{\mathrm{raw}}\) and \(D_{\mathrm{SBC}}^{\mathrm{raw}}\) should detect uncertainty distortion in the raw posterior.

The controlled experiment for Step 3 was designed to separate synchrony across observations from joint compensation under a fixed observation. For each blind-test observation, the posterior means of the two targets were set to \((z_i,z_i)\), giving them the same mean-synchrony structure across observations. Under a fixed observation, the joint posterior used Gaussian covariance with \(\rho=0\) and \(\rho=-0.85\), respectively. Thus, the two settings had the same posterior-mean synchrony, but only the latter contained explicit negative sample-shape coupling, allowing assessment of whether the framework correctly distinguished population covariation from single-observation posterior compensation.

Cross-track consistency used the known derived relation \(t_{\mathrm{derived}} = \theta_1 + \theta_2\). In the matched condition, the direct-derived posterior and the push-forward distribution of the raw posterior had the same centre and scale; in the mismatched condition, the entire direct-derived posterior was shifted by two target standard deviations. Finally, in a 7-dimensional parameter space, \(\mathbf u = \frac{1}{\sqrt 2}(1,-1,0,0,0,0,0)^\top\) was prespecified as the derived direction, and a slightly perturbed version of this direction and a random unit direction were used as the matched and unmatched conditions, respectively, for Active--derived alignment.

For each audit, at a fixed nominal level of \(\alpha=0.05\), we estimated the type I error rate (false-positive rate) from the rejection proportion under the matched condition and power (Statistical Power) from the detection proportion under the prespecified deviation condition; Wilson intervals were used to quantify Monte Carlo uncertainty. Because these deviations were deliberately constructed with fixed effect sizes, the resulting operating characteristics apply only to the current dimensionality and sample size. They confirm that the audit statistics respond as expected to the target failure modes but do not constitute a guarantee of universal performance across arbitrary neural dynamical problems.

\hypertarget{sec-3-1-2}{%
\subsubsection{3.1.2 SOZ-local iEEG and the Epileptor model}\label{sec-3-1-2}}

The first real-data experiment examined intracranial electrophysiological signals during epileptic seizures. We used Epileptor as the candidate neural dynamical model (see \hyperref[app-1]{Appendix 1}); this model can simulate transitions from the preictal baseline state to ictal activity and then to postictal recovery within a unified dynamical framework \cite{ref54}. We used the HUP iEEG epilepsy dataset \cite{ref55}, including 56 patients, with only the first available seizure file from each patient used as an independent observation. Clinically annotated SOZ channels were extracted. After channels flagged by automated quality control were removed, the channel median was taken, and a continuous window from 5 s before to 25 s after clinical onset was extracted. Signals were resampled to 256 Hz and sequentially processed using short-duration artefact interpolation, a 60 Hz notch filter (\(Q=30\)), a fourth-order 0.5--80 Hz Butterworth bandpass filter, and robust scaling based on the median and MAD from \(-5\) to \(-1\) s before seizure onset; the same operators were applied to the simulated signals. Each patient constituted an independent observational unit and was assigned by patient to \(\mathcal D^{\mathrm{obs}}_{\mathrm{cal}}\) (70\%) or \(\mathcal D^{\mathrm{obs}}_{\mathrm{test}}\) (30\%); no patient appeared in both sets.

Each simulated trajectory was integrated autonomously under fixed parameters. The endogenous onset was detected from the state transition of the first fast subsystem, after which the same \(-5\) to \(+25\) s window was extracted. Given the limited number of observed samples in the real SOZ-local iEEG experiment, and because our aim was not unconstrained inversion of the complete Epileptor parameter space but an assessment of the recoverability of key seizure-related mechanisms under real observation conditions, we adopted a restricted parameterization strategy and inverted only three subparameters important to epileptic dynamics: \(\theta = \{x_0,I_1,I_2\}\). Detailed parameter values are provided in \hyperref[tab:table1]{Table 1}.

{\footnotesize
\begin{longtable}[]{@{}
  >{\raggedright\arraybackslash}p{(\columnwidth - 4\tabcolsep) * \real{0.1700}}
  >{\raggedright\arraybackslash}p{(\columnwidth - 4\tabcolsep) * \real{0.2500}}
  >{\raggedright\arraybackslash}p{(\columnwidth - 4\tabcolsep) * \real{0.5800}}@{}}
\caption{Prior values and ranges of the parameters used in the Epileptor simulations. \(U(\cdot)\) and \(N(\cdot)\) denote uniform and normal distributions, respectively, and \(\operatorname{clip}(f,a,b)\) indicates that the range of the result obtained from \(f\) is forcibly truncated at \((a,b)\) \label{tab:table1}}\tabularnewline
\toprule
Parameter & Values & Meaning \\
\midrule
\endfirsthead
\toprule
Parameter & Values & Meaning \\
\midrule
\endhead
\(x_0\)\hspace{0pt} & \(U(-1.75,-1.00)\) & Epileptogenicity \\
\(y_0\) & 1.0 & Offset constant of the first fast-subsystem variable \\
\(\tau_0\) \hspace{0pt} & 61.25 & Time scale of slow variable \\
\(\tau_1\) & 1.0 & Time scale of the first fast subsystem \\
\(\tau_2\) \hspace{0pt} & 10.0 & Time scale of the second spike-and-wave subsystem \\
\(I_1\)\hspace{0pt} & \(U(3.00,4.20)\) & Drive to the first fast discharge subsystem \\
\(I_2\)\hspace{0pt} & \(U(0.05,1.40)\) & Drive to the second spike-and-wave subsystem \\
\(b_g\) & 2.0 & Gain from the historical low-pass state to the second subsystem \\
\(\lambda_g\) & 0.01 & Decay rate of the historical state \\
\(c_g\) & 0.10 & Drive coefficient from the first fast variable to the historical state \\
\(s_{\mathrm{fast}}\) & 12.0 & Fixed conversion factor of the fast variable relative to physical seconds \\
\(w_{x_1}\) & -1.0 & Fixed weight of \(x_1\) in the source-signal readout \\
\(G_{\mathrm{SW}}\) & 1.05 & Fixed weight of \(x_2\) in the source-signal readout \\
\(\sigma_{\mathrm{dyn}}\) & 0.111 & Standard deviation of dynamical noise \\
\(z(0)\) & \(\operatorname{clip}(I_1+0.80+\eta_z,0.5,7.5)\) & Conditioned initial state of the preictal slow variable \\
\(\eta_z\) & \(N(0,0.04^2)\) & Initial perturbation of the slow variable \\
\bottomrule
\end{longtable}
}

The latent source signal was defined as \(r(t)=w_{x_1}x_1(t)+G_{\mathrm{SW}}x_2(t)\). Source gain, the stationary background component, and observation noise were treated as bounded observation nuisance factors, estimated only from \(\mathcal D^{\mathrm{obs}}_{\mathrm{cal}}\) and then frozen. Notably, we restricted the priors for \(x_0\), \(I_1\), and \(I_2\) to ranges near the classical Epileptor seizure operating point and rescaled the slow-variable timescale to suit a 30 s clinical window.

To provide greater scientific interpretability, we derived two candidate derived coordinates from the parameters above. 1. Effective Fast Drive:\[t_{\mathrm{eff}}=I_1+4x_0\tag{34}\label{eq:main-34}\]\(t_{\mathrm{eff}}\) represents the effective additive bias applied to the first fast subsystem under a quasi-steady-state approximation of the slow variable. A higher \(t_{\mathrm{eff}}\) corresponds to a stronger net drive to the fast subsystem. 2. Fast-spike Drive Balance:\[ t_{\mathrm{F/S}}=\log\left[\frac{1+I_1/3.1}{1+I_2/0.45}\right]\tag{35}\label{eq:main-35}\]Because the nominal scales of \(I_1\) and \(I_2\) differ, directly calculating \(I_1/I_2\) would make the result strongly affected by dimensional differences and by the near-zero region of \(I_2\). This derived coordinate is therefore defined with reference to the classical operating points \(I_1^{\mathrm{ref}}=3.1\) and \(I_2^{\mathrm{ref}}=0.45\). When \(I_1=3.1\) and \(I_2=0.45\), \(t_{\mathrm{F/S}}=0\), indicating that both subsystems are at the classical reference operating point. \(t_{\mathrm{F/S}}>0\) indicates that the normalized drive to the first fast discharge subsystem predominates over that to the second spike-and-wave subsystem.

\hypertarget{sec-3-1-3}{%
\subsubsection{3.1.3 ERP CORE MMN and the CMC model}\label{sec-3-1-3}}

The second real-data experiment examined the mismatch negativity (MMN) response in event-related potentials. We used a CMC-inspired fixed five-node auditory neural mass network model as the candidate neural dynamical model. Its local population organization was based on the canonical microcircuit architecture \cite{ref56} and incorporated the hierarchical feedforward--feedback structure of the auditory mismatch response. Real EEG data were obtained from the MMN task in the ERP CORE dataset. This dataset contains auditory oddball EEG recordings from healthy participants, in which deviant stimuli are used to evoke the MMN response \cite{ref57}. We ultimately included 40 healthy participants with complete preprocessing results. Deviant events were defined as stimuli coded 70, whereas the standard control comprised standard-after-standard events coded 80 for which the preceding stimulus was also coded 80. Each participant constituted an independent observational unit and was assigned by patient to \(\mathcal D^{\mathrm{obs}}_{\mathrm{cal}}\) (70\%) or \(\mathcal D^{\mathrm{obs}}_{\mathrm{test}}\) (30\%); no patient appeared in both sets.

For each participant, EEG data were downsampled to 256 Hz, initially referenced to the mean of P9/P10, and used to construct horizontal and vertical electro-oculogram channels. A 0.1 Hz high-pass filter was applied, and ICA was used to remove artefacts when ocular components were detectable \cite{ref58}. Events were epoched from \(-200\) to \(800\) ms and baseline-corrected using the prestimulus interval. Robust within-participant artefact rejection and a whole-head average reference were then applied, and deviant and standard-after-standard ERPs were calculated separately, followed by a 20 Hz low-pass filter.

We averaged the three frontocentral electrodes Fz, FCz, and Cz to construct a virtual frontocentral channel. Each participant was ultimately represented as:\[\begin{aligned} \mathbf Y_i=\begin{bmatrix}S_i(t)\\ \Delta_i(t)\end{bmatrix}\\ \Delta_i(t)=D_i(t)-S_i(t) \end{aligned}\tag{36}\label{eq:main-36}\]where \(S_i(t)\) is the standard-after-standard ERP and \(D_i(t)\) the deviant ERP. The input retained the complete -200 -- 800 ms time window and microvolt units, without participant-wise normalization of the poststimulus waveform, thereby preserving differences in response amplitude relative to the prestimulus baseline.

For computational convenience, we did not use the complete, full-parameter CMC model, but instead used a simplified fixed five-node form (see \hyperref[app-2]{Appendix 2}). As in the iEEG experiment, we assessed the recoverability of key ERP/ERF-generating mechanisms under real observation conditions. We again adopted a restricted parameterization strategy and inverted only the five subparameters most directly related to evoked-response morphology: \(\theta_{\mathrm{CMC}}=\{g_{\mathrm{ss}},g_{\mathrm{sp}},g_{\mathrm{dp}},g_{\mathrm{ii}},\tau_i\}\). Detailed parameter values are provided in \hyperref[tab:table2]{Table 2}.

{\footnotesize
\begin{longtable}[]{@{}
  >{\raggedright\arraybackslash}p{(\columnwidth - 4\tabcolsep) * \real{0.3100}}
  >{\raggedright\arraybackslash}p{(\columnwidth - 4\tabcolsep) * \real{0.2800}}
  >{\raggedright\arraybackslash}p{(\columnwidth - 4\tabcolsep) * \real{0.4100}}@{}}
\caption{Prior values and ranges of the parameters used in the CMC simulations. \(CN(\mu,\sigma; a,b)\) denotes a truncated normal distribution with mean \(\mu\), standard deviation \(\sigma\), and truncation range \([a,b]\) \label{tab:table2}}\tabularnewline
\toprule
Parameter & Values & Meaning \\
\midrule
\endfirsthead
\toprule
Parameter & Values & Meaning \\
\midrule
\endhead
\(g_{\mathrm{ss}}\)\hspace{0pt} & \(CN(1.00,0.25;0.50,1.60)\) & Effective gain of external stimulus input \\
\(g_{\mathrm{sp}}\) & \(CN(1.00,0.25;0.50,1.60)\) & Superficial pyramidal-cell gain \\
\(g_{\mathrm{dp}}\) & \(CN(1.00,0.25;0.50,1.60)\) & Deep pyramidal-cell gain \\
\(g_{\mathrm{ii}}\) & \(CN(1.00,0.25;0.50,1.60)\) & Inhibitory interneuron gain \\
\(g_{\mathrm{F}}\) & 0.85 & Feedforward connection gain from superficial to deep pyramidal cells \\
\(g_{\mathrm{B}}\) & 0.75 & Feedback connection gain from deep pyramidal cells to the input layer \\
\(\tau_e\) & 12 ms & Excitatory-population timescale \\
\(\tau_i\) & \(CN(42,10;20,65)\) ms & Inhibitory-population timescale \\
\(\tau_{\mathrm{dp}}\)\hspace{0pt} & \(55+15\tanh[0.7(g_{\mathrm{dp}}-1)]\) ms & Effective deep-pyramidal timescale \\
\(\tau_{\mathrm{slow}}\) & 180 ms & Auxiliary slow-state timescale \\
\(\sigma_{\mathrm{ss}},\sigma_{\mathrm{sp}},\sigma_{\mathrm{ii}}, \sigma_{\mathrm{dp}}\) & 0.035, 0.040, 0.035, 0.030 & State-noise coefficients of the four local populations \\
\bottomrule
\end{longtable}
}

To construct more scientifically interpretable analysis axes, we derived three coordinates from the parameters above. 1. Local recurrent excitation/inhibition gain ratio:\[t_{\mathrm{E/I}}=\log\frac{g_{\mathrm{sp}}}{g_{\mathrm{ii}}} \tag{37}\label{eq:main-37}\]This coordinate describes the magnitude of the superficial pyramidal gain relative to the local inhibitory gain and is an interpretive proxy for local recurrent E/I balance. 2. Superficial/deep pyramidal gain ratio:\[t_{\mathrm{SP/DP}}=\log\frac{g_{\mathrm{sp}}}{g_{\mathrm{dp}}} \tag{38}\label{eq:main-38}\]This coordinate describes the relative gain structure of the superficial and deep pyramidal populations in the generation of evoked responses. 3. Sensory input/inhibitory integration ratio:\[t_{\mathrm{S/I}}=\log \frac{g_{\mathrm{ss}}\tau_e}{g_{\mathrm{ii}}\tau_i} \tag{39}\label{eq:main-39}\]This coordinate provides an integrated description of the relative relationship between sensory-input drive and local inhibitory integration. These derived coordinates are interpretive variables designed specifically for this study. Their role is to serve as candidate targets in the Derived track and undergo posterior audit and cross-track consistency testing within the framework.

\hypertarget{sec-3-2}{%
\subsection{3.2 Experimental settings}\label{sec-3-2}}

To evaluate the effects of different summary choices on dynamical-parameter recoverability, we defined three classes of summary. For each random seed, a single set of observation-calibration parameters based on the preregistered diagnostic space was estimated only from \(\mathcal D^{\mathrm{obs}}_{\mathrm{cal}}\), with patients/participants held separate, and was frozen and shared across all summaries. \(\mathcal D^{\mathrm{obs}}_{\mathrm{test}}\) did not participate in calibration, summary training, or model selection.

The Epileptor experiment used three summaries. The first was a 15-dimensional set of hand-crafted features of epileptic dynamics describing the baseline-to-ictal transition, slow waves, response envelope, and slow autocorrelation structure (see \hyperref[app-4]{Appendix 4} for the complete list). The second was a 64-dimensional waveform-PCA representation obtained by applying PCA to the standardized raw waveform samples, which was used to assess whether general linear compression retained parameter information. The third was a 64-dimensional CNN--LSTM representation \cite{ref59}. The CNN--LSTM used an Encoder--Decoder architecture and was trained for 200 epochs on reference simulations without real data. It extracted local and long-range dynamics through convolution and LSTM components and predicted 136-dimensional descriptors deterministically extracted from the same waveforms. The CMC experiment used the same waveform PCA and CNN--LSTM, but the hand-crafted summary was replaced with 39-dimensional ERP/ERF features covering peaks, latencies, N1/P2/MMN windows, late recovery, and other phenomena; its CNN--LSTM prediction target was 56-dimensional. The different summaries were not forced to have the same dimensionality because the dimensionality of the hand-crafted features was determined by the scientific question.

The Step 1 waveform diagnostic space \(Z_{\mathrm{diag}}\) comprised the preregistered hand-crafted features described above and was used to assess whether the model covered the key observed phenomena it purported to explain. When the formal summary consisted of the hand-crafted features, \(R\) and \(Z_{\mathrm{diag}}\) coincided and were therefore not reported twice. The Step 2 summary-loss diagnostic additionally extracted 136-dimensional descriptors of the time domain, frequency domain, autocorrelation, waveform morphology, and other properties, which were compressed by PCA to 32 dimensions as the waveform branch. It therefore differed from the 64-dimensional waveform-PCA summary obtained by applying PCA directly to raw waveform samples.

For simplicity of presentation, we selected only sequential neural posterior estimation (SNPE) \cite{ref60} as the default SBI method in all experiments. SNPE directly outputs posterior samples and is therefore well suited to the audit procedure. Importantly, the framework itself does not depend on SNPE. In principle, any inverter that outputs target posterior samples given a summary, or provides posterior predictive samples suitable for proper scoring, can be incorporated into the same procedure, including SNRE, NLE, and other methods.

\hypertarget{sec-4}{%
\section{4 Results}\label{sec-4}}

\hypertarget{sec-4-1}{%
\subsection{4.1 Controlled simulation results}\label{sec-4-1}}

To evaluate the error-rate control and mismatch-detection capability of Step 1, we constructed three known-truth conditions across 200 independent algorithmic seeds: matched-null (no mean shift), visible mismatch (mean shift in dimensions 1--3), and summary-masked mismatch (mean shift in dimensions 4--6), as detailed in \hyperref[sec-3-1-1]{Section 3.1.1}. A batch-level global upper-tail fraction of \(p_{\mathrm{global}}\leq0.05\) was used as the criterion for mismatch detection. The results are shown in \hyperref[tab:table3]{Table 3}.

\clearpage
{\scriptsize
\begin{longtable}[]{@{}
  >{\raggedright\arraybackslash}p{(\columnwidth - 8\tabcolsep) * \real{0.2323}}
  >{\raggedright\arraybackslash}p{(\columnwidth - 8\tabcolsep) * \real{0.1919}}
  >{\raggedright\arraybackslash}p{(\columnwidth - 8\tabcolsep) * \real{0.1818}}
  >{\raggedright\arraybackslash}p{(\columnwidth - 8\tabcolsep) * \real{0.2020}}
  >{\raggedright\arraybackslash}p{(\columnwidth - 8\tabcolsep) * \real{0.1920}}@{}}
\caption{Error rates and detection power of the Step 1 dual-space audit under controlled mismatch conditions. Values are proportions over 200 independent algorithmic replicates; brackets indicate 95\% Wilson confidence intervals. \label{tab:table3}}\tabularnewline
\toprule
Audit & Type I error & Specificity & Visible-mismatch power & Masked-mismatch power \\
\midrule
\endfirsthead
\toprule
Audit & Type I error & Specificity & Visible-mismatch power & Masked-mismatch power \\
\midrule
\endhead
Summary local support \hspace{0pt} & 0.050, {[}0.027, 0.090{]} & 0.950 {[}0.910, 0.973{]} & 1.000 {[}0.981, 1.000{]} & 0.075 {[}0.046, 0.120{]} \\
Summary local predictive & 0.065, {[}0.038, 0.108{]} & 0.935 {[}0.892, 0.962{]} & 1.000 {[}0.981, 1.000{]} & 0.065 {[}0.038, 0.108{]} \\
Waveform local support & 0.060, {[}0.035, 0.102{]} & 0.940 {[}0.898, 0.965{]} & 1.000 {[}0.981, 1.000{]} & 1.000 {[}0.981, 1.000{]} \\
Waveform local predictive & 0.070, {[}0.042, 0.114{]} & 0.930 {[}0.886, 0.958{]} & 1.000 {[}0.981, 1.000{]} & 1.000 {[}0.981, 1.000{]} \\
\bottomrule
\end{longtable}
}

Under the matched-null condition, the type I error rates of the four audits ranged from 0.050 to 0.070, close to \(\alpha=0.05\), with corresponding specificities of 0.930--0.950. The distributions of all four global \(p\) values across replicates also did not deviate significantly from \(U(0,1)\). Detection power for visible mismatch was 1.000 for all four audits. Under summary-masked mismatch, the detection rates of summary support and summary predictive were only 0.075 and 0.065, close to their respective type I error rates, whereas both waveform support and waveform predictive reached 1.000. These results indicate that, at the current controlled scale, the condition-matched pseudo-observed null maintained the expected error rate and enabled the fixed diagnostic space to identify mismatch removed by the summary.

Thus, Step 1 identified not only configuration mismatch directly visible in the current summary, but also mismatch masked by summary compression through the independent waveform diagnostic space.

In the Step 2 tests, we constructed six different conditions. The corresponding results are shown in \hyperref[tab:table4]{Table 4}.

{\scriptsize
\begin{longtable}[]{@{}
  >{\raggedright\arraybackslash}p{(\columnwidth - 10\tabcolsep) * \real{0.1800}}
  >{\raggedright\arraybackslash}p{(\columnwidth - 10\tabcolsep) * \real{0.2500}}
  >{\raggedright\arraybackslash}p{(\columnwidth - 10\tabcolsep) * \real{0.1400}}
  >{\raggedright\arraybackslash}p{(\columnwidth - 10\tabcolsep) * \real{0.1400}}
  >{\raggedright\arraybackslash}p{(\columnwidth - 10\tabcolsep) * \real{0.1400}}
  >{\raggedright\arraybackslash}p{(\columnwidth - 10\tabcolsep) * \real{0.1500}}@{}}
\caption{Target information and posterior-calibration results in the Step 2 known-truth experiment. Values are means \(\pm\) standard deviations over 200 independent algorithmic replicates. \label{tab:table4}}\tabularnewline
\toprule
Controlled condition & Condition status & PSG & \(R^2\) & \(E_{90}^{\mathrm{raw}}\) & \(D_{\mathrm{SBC}}^{\mathrm{raw}}\) \\
\midrule
\endfirsthead
\toprule
Controlled condition & Condition status & PSG & \(R^2\) & \(E_{90}^{\mathrm{raw}}\) & \(D_{\mathrm{SBC}}^{\mathrm{raw}}\) \\
\midrule
\endhead
\textbf{Primary temporal summary} & Information-sufficient; raw posterior essentially calibrated & \textbf{\(0.660\pm0.012\)} & \textbf{\(0.878\pm0.008\)} & \textbf{\(0.020\pm0.011\)} & \textbf{\(0.034\pm0.010\)} \\
\textbf{PermWave control} & Branch distribution and structure retained; waveform--target association removed & \textbf{\(0.661\pm0.011\)} & \textbf{\(0.878\pm0.008\)} & \textbf{\(0.019\pm0.012\)} & \textbf{\(0.034\pm0.010\)} \\
\textbf{Waveform augmented} & Effective waveform information added and accuracy improved & \textbf{\(0.831\pm0.006\)} & \textbf{\(0.970\pm0.002\)} & \textbf{\(0.019\pm0.011\)} & \textbf{\(0.035\pm0.011\)} \\
\textbf{No-temporal ablation} & Target-relevant temporal information removed & \textbf{\(0.000\pm0.007\)} & \textbf{\(-0.012\pm0.009\)} & \textbf{\(0.020\pm0.013\)} & \textbf{\(0.037\pm0.012\)} \\
\textbf{Unsupported target} & Target independent of the observation & \textbf{\(-0.001\pm0.007\)} & \textbf{\(-0.013\pm0.008\)} & \textbf{\(0.021\pm0.012\)} & \textbf{\(0.037\pm0.011\)} \\
\textbf{Underdispersed posterior} & Posterior mean retained; raw width reduced to 25\% & \textbf{\(0.590\pm0.016\)} & \textbf{\(0.878\pm0.008\)} & \textbf{\(0.593\pm0.018\)} & \textbf{\(0.307\pm0.014\)} \\
\bottomrule
\end{longtable}
}

Using \(\mathrm{PSG}>0.05\) as the information-detection criterion yielded a type I error rate of \(0.000\) and power of \(1.000\); the same results were obtained when additional waveform information was identified using \(\Delta\mathrm{PSG}>0.05\) for Wave+ relative to PermWave. When \(E_{90}^{\mathrm{raw}}>0.10\) was used to identify coverage anomalies, the type I error rate was \(0.000\) and power was \(1.000\). At \(\alpha=0.05\), randomized-PIT SBC had a type I error rate of \(0.030\) and power of \(1.000\). Although the underdispersed condition retained the same \(R^2\), \(E_{90}^{\mathrm{raw}}\) and \(D_{\mathrm{SBC}}^{\mathrm{raw}}\) increased to \(0.593\pm0.018\) and \(0.307\pm0.014\), respectively, indicating that both raw-posterior metrics identified uncertainty distortion.

In the controlled evaluation of Step 3, we prespecified three scenarios: negative compensation in the joint posterior under a fixed observation, inconsistency between the push-forward distribution of the raw posterior and the direct-derived posterior, and alignment between the active direction and a predefined derived direction. Each scenario comprised a valid null condition and a known anomalous condition and was evaluated using 200 independent algorithmic replicates. The proportion classified as anomalous under the null condition was defined as the type I error rate, and the proportion correctly classified under the anomalous condition was defined as statistical power. Detailed results are shown in \hyperref[tab:table5]{Table 5}.

{\scriptsize
\begin{longtable}[]{@{}
  >{\raggedright\arraybackslash}p{(\columnwidth - 8\tabcolsep) * \real{0.2350}}
  >{\raggedright\arraybackslash}p{(\columnwidth - 8\tabcolsep) * \real{0.1900}}
  >{\raggedright\arraybackslash}p{(\columnwidth - 8\tabcolsep) * \real{0.2450}}
  >{\raggedright\arraybackslash}p{(\columnwidth - 8\tabcolsep) * \real{0.1650}}
  >{\raggedright\arraybackslash}p{(\columnwidth - 8\tabcolsep) * \real{0.1650}}@{}}
\caption{Error rates and statistical power in the Step 3 known-truth experiment. Values are classification proportions over 200 independent algorithmic replicates; brackets indicate 95\% Wilson confidence intervals. \label{tab:table5}}\tabularnewline
\toprule
Audit & Null condition & Injected condition & Type I error & Statistical power \\
\midrule
\endfirsthead
\toprule
Audit & Null condition & Injected condition & Type I error & Statistical power \\
\midrule
\endhead
Joint posterior compensation & Independent joint posterior & Strong negative posterior compensation & 0.000, {[}0.000,0.019{]} & 1.000,{[}0.981,1.000{]} \\
Raw--derived consistency & Correct raw push-forward & Shifted direct-derived posterior & 0.000, {[}0.000,0.019{]} & 1.000,{[}0.981,1.000{]} \\
Active--derived alignment & Random active direction & Active--derived alignment & 0.030,{[}0.014,0.064{]} & 1.000,{[}0.981,1.000{]} \\
\bottomrule
\end{longtable}
}

All three audits detected every known anomaly. The Joint posterior coupling audit produced no false positives under the independent condition and achieved 100\% statistical power under the negative-compensation condition. As the basis for classification, within-observation joint-posterior correlations were \(0.000\pm0.006\) under the independent condition and \(-0.851\pm0.001\) under the strong negative-compensation condition. The two conditions were deliberately constructed to have the same posterior-mean synchrony across observations, indicating that only joint-sample correlation under a fixed observation could identify this posterior compensation structure.

The Raw--derived consistency audit likewise achieved zero false positives and complete detection. Under the correct push-forward condition, finite posterior sampling produced a nonzero Wasserstein distance of \(W_1=0.202\pm0.004\), but its calibrated score was \(p^{\mathrm{RD}}=0.498\pm0.040\), indicating that the discrepancy lay within the normal range of the reference inference procedure. After the direct-derived posterior was artificially shifted, \(W_1\) increased to \(2.830\pm0.017\) and the calibrated score decreased to \(0.003\); inconsistency was identified in all 200 replicates. This finding indicates that the calibration procedure distinguished normal finite-sampling error from a systematic cross-track shift.

The Active--derived alignment audit had a type I error rate of \(3.0\%\) under the random-direction condition, below the prespecified 5\% level, whereas the injected direction was identified in every replicate. The mean projection proportion for the random direction was \(A^{\mathrm{AD}}=0.141\pm0.157\), with a structure-matched upper-tail fraction of \(p^{\mathrm{AD}}=0.509\pm0.272\); under the injected condition, the projection proportion increased to \(0.994\pm0.003\). Because the structure-matched reference set contained only 21 unique directions, the minimum resolution of its empirical upper-tail fraction was \(1/21=0.0476\). The injected condition therefore reached the strongest alignment evidence available from this finite reference set.

\hypertarget{sec-4-2}{%
\subsection{4.2 Real-data results}\label{sec-4-2}}

\hypertarget{sec-4-2-1}{%
\subsubsection{4.2.1 The dual-space audit distinguished representation-dependent observation compatibility}\label{sec-4-2-1}}

\begin{figure}[H]
\hypertarget{fig:main2}{%
\centering
\includegraphics[keepaspectratio,width=1.00\textwidth]{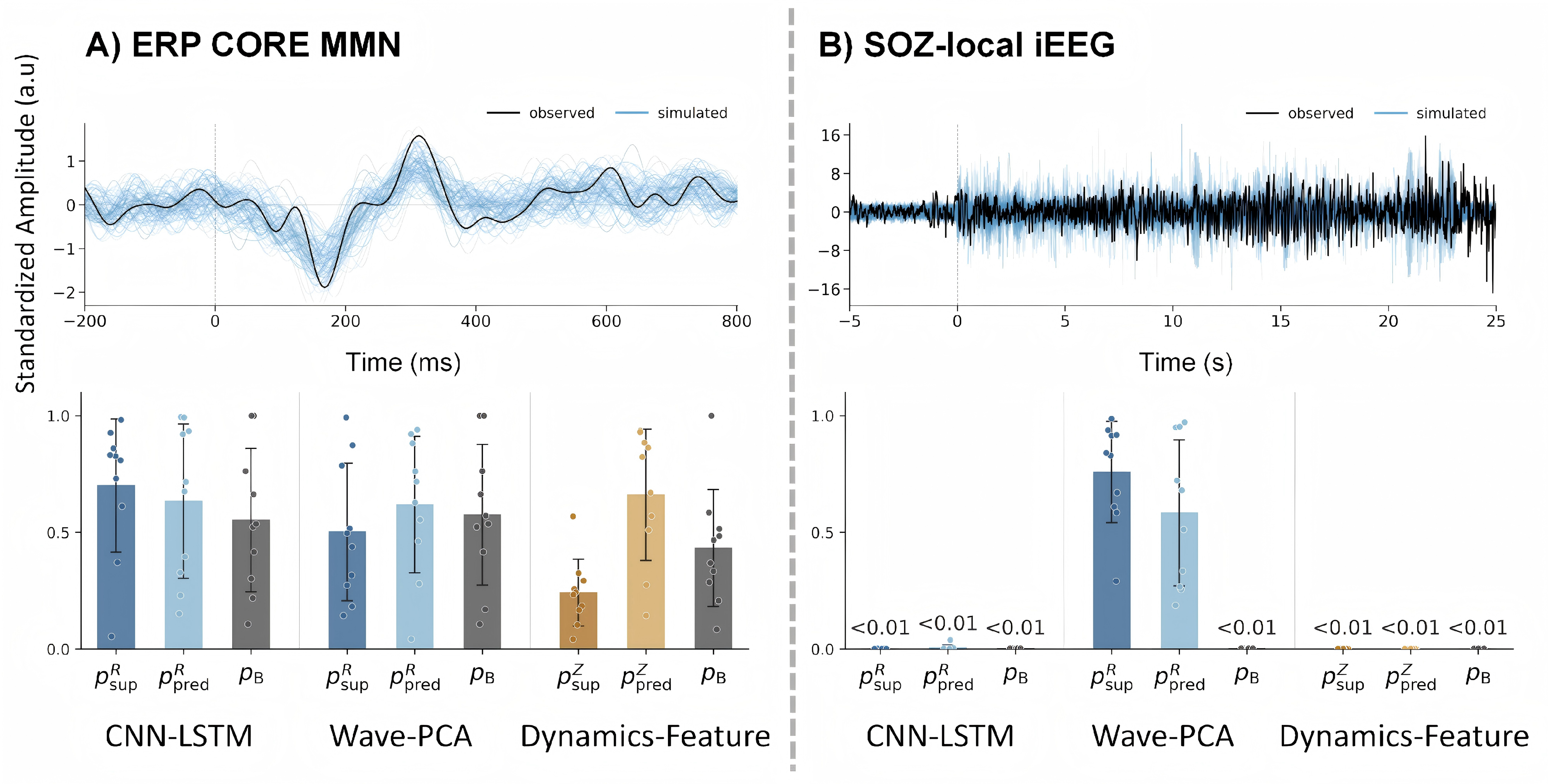}
\caption{Dual-space coverage audit for Epileptor/iEEG and CMC/MMN. (A). CMC/MMN audit. Top left: real EEG waveforms of example evoked MMN responses (black) and simulated data from the CMC dynamical model (blue; only 50 are shown as examples). Bottom left: means and standard deviations across 10 random seeds of the support, predictive global, and Bonferroni-combined \(p\) values for different summaries. Because the dynamical features and waveform-space features selected in this study were identical, waveform-space results are not reported separately. (B). Epileptor/iEEG audit. Top right: a real iEEG waveform from an example seizure (black) and simulated data from the Epileptor dynamical model (blue; only 50 are shown as examples). Bottom right: the same statistical representation as in (A).}\label{fig:main2}
}
\end{figure}

We first assessed whether the complete configuration comprising the candidate dynamical model, parameter prior, observation mapping, and summary could cover the real-data observations. Step 1 compared the real observations with the simulation configuration across three summaries and a fixed waveform diagnostic space. For each seed, 3,200 Epileptor and 4,000 CMC simulations were generated, and 10 algorithmic random seeds were run. The results are shown in \hyperref[fig:main2]{Fig. 2}.

The three summaries shared observation-calibration parameters estimated from \(\mathcal D^{\mathrm{obs}}_{\mathrm{cal}}\) and then frozen; their differences therefore primarily reflected the representation rather than changes in the observation model. The Summary Space \(R\) changed with the summary, whereas the preregistered waveform diagnostic space \(Z\) was fixed within each experiment. When the hand-crafted summary used the same features as \(Z\), it was audited only once and was not counted twice in the Bonferroni-combined \(p_{\mathrm B}\). Means and standard deviations across seeds were used only to describe stability and did not constitute additional significance tests.

For Epileptor/iEEG, the different representations yielded clearly different conclusions regarding coverage. The support and predictive global \(p\) values for waveform PCA were \(0.758\pm0.217\) and \(0.583\pm0.313\), respectively, indicating that no evident mismatch was detected in its compressed space. However, in the preregistered 15-dimensional space of core seizure dynamics, both global \(p\) values were approximately \(5\times10^{-4}\), corresponding to \(p_{\mathrm B}\approx0.002\) for the waveform-PCA configuration. The support, predictive, and \(p_{\mathrm B}\) values for the CNN--LSTM were likewise low, at \(0.0008\pm0.0008\), \(0.005\pm0.012\), and approximately \(0.002\), respectively. When the core dynamical features were used directly as the summary, the values were also near the Monte Carlo resolution limit. Thus, the overall waveform overlap in waveform PCA masked persistent mismatch in key seizure dynamics, and the current single-source Epileptor configuration cannot be regarded as covering real SOZ-local iEEG.

In contrast, CMC/MMN showed no systematic concentration in the lower tail. The support, predictive, and \(p_{\mathrm B}\) values for the CNN--LSTM were \(0.700\pm0.286\), \(0.634\pm0.331\), and \(0.553\pm0.307\), respectively; those for waveform PCA were \(0.502\pm0.295\), \(0.619\pm0.292\), and \(0.575\pm0.301\); and those for the 39-dimensional ERP diagnostic features were \(0.241\pm0.143\), \(0.660\pm0.281\), and \(0.433\pm0.250\). These results indicate that, under the current audit statistics, sample size, and specified model configuration, no systematic coverage mismatch was detected for CMC/MMN across the representation spaces.

Step 1 therefore distinguished two types of configuration. CMC/MMN provided stronger grounds for the subsequent target-specific audit, whereas Epileptor/iEEG overlapped with the real data only in some representation spaces and still failed the core dynamical diagnostic. Importantly, if the conditional null described above holds and the pseudo-observed batches match the condition composition and sampling units of the real test batch, the Monte Carlo null distribution of \(p_{\mathrm{global}}\) should remain approximately calibrated. Thus, \(p\approx0.3\) is a common non-extreme value rather than an anomalously low-tail value. On this basis, Steps 2--3 for CMC can support cautious interpretation conditional on the model; subsequent Epileptor results are interpreted only as recoverability within the current simulation configuration, although we report the complete analysis.

\hypertarget{sec-4-2-2}{%
\subsubsection{4.2.2 Target recoverability was summary-specific}\label{sec-4-2-2}}

\begin{figure}[p]
\centering
\includegraphics[keepaspectratio,width=1\textwidth,height=0.88\textheight]{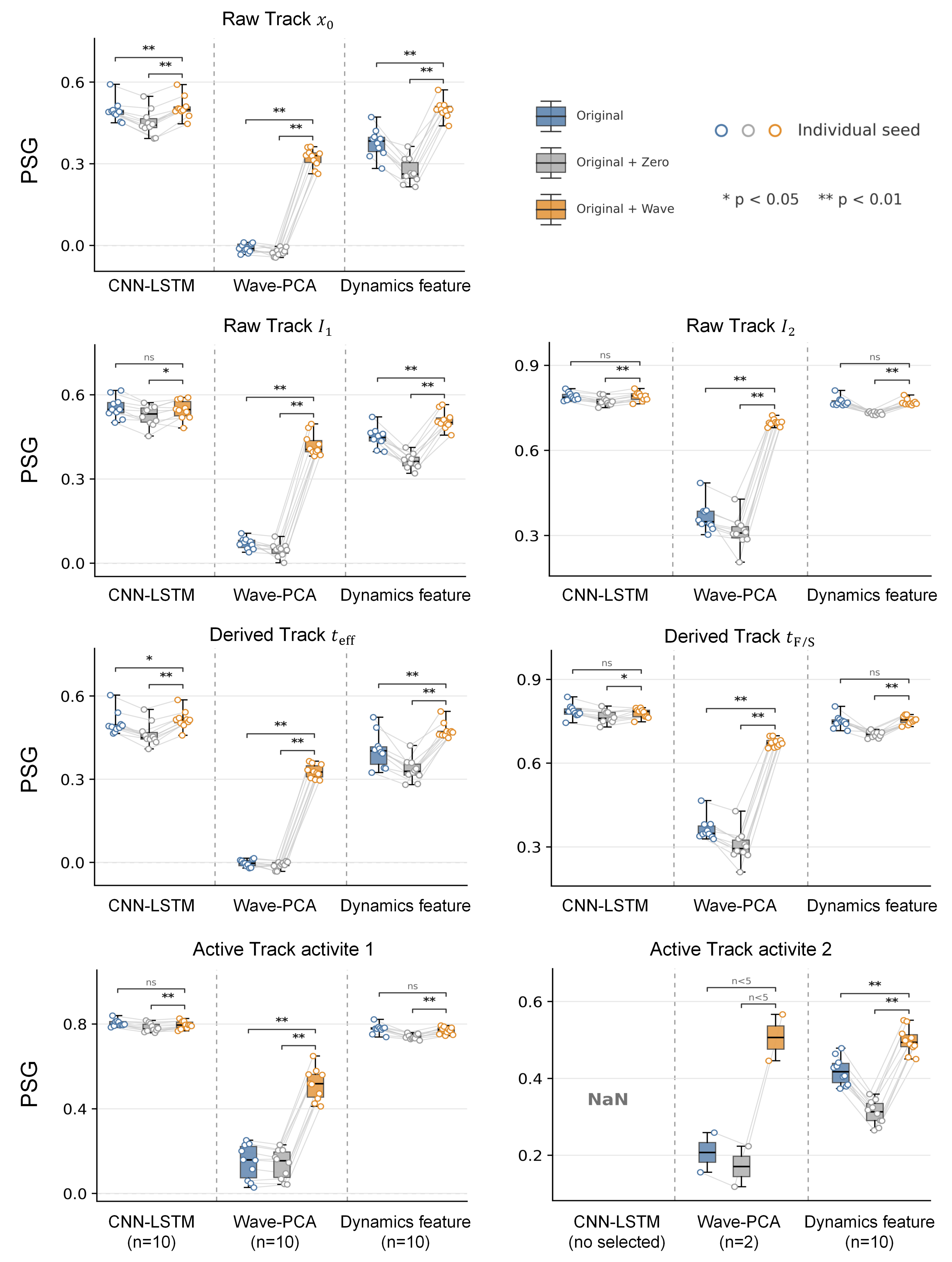}
\captionsetup{font=scriptsize,labelfont=bf,textfont=normalfont,labelsep=period,justification=justified,singlelinecheck=true,skip=2pt}
\caption{PSG of targets in each Epileptor/iEEG track under three summary input paths. Points indicate algorithmic random seeds, and grey lines connect paired results from the same seed. A one-sided exact sign-flip randomization test was used, with Holm correction applied within each target--summary combination to the Wave+--Original and Wave+--PermWave comparisons. For \(n=10\), the minimum attainable corrected value is \(2/1024=0.001953\), and therefore the highest significance marker is \(^{**}\); this test evaluates paired stability across algorithmic replicates.}\label{fig:main3}
\end{figure}

Following Step 1, we separately trained a target-specific SNPE posterior for each scalar target in the Raw, Derived, and Active tracks. Information and point recovery were evaluated using PSG and posterior-mean \(R^2\), respectively. Uncertainty in the raw posterior was audited only using \(E_{90}^{\mathrm{raw}}\) and randomized-PIT \(D_{\mathrm{SBC}}^{\mathrm{raw}}\), both calculated on \(\mathcal D_{\mathrm{test}}^{\mathrm{post}}\). The results are shown in \hyperref[fig:main3]{Figs. 3}--\hyperref[fig:main5]{5}.

\clearpage
\clearpage

PSG for the Epileptor targets showed clear summary dependence (\hyperref[fig:main3]{Fig. 3}). Under the CNN--LSTM, active 1, \(I_2\), and \(t_{\mathrm{F/S}}\) had the highest PSG values, at \(0.803\pm0.016\), \(0.790\pm0.014\), and \(0.786\pm0.024\), respectively. The dynamical features yielded a similar ranking, with corresponding values of \(0.777\pm0.022\), \(0.771\pm0.016\), and \(0.750\pm0.024\). Overall information gain was weaker for waveform PCA, with only \(I_2\) and \(t_{\mathrm{F/S}}\) reaching approximately \(0.36\). The Active dimensionality was selected independently within each seed: all 10 CNN--LSTM seeds selected \(L=1\), all dynamical-feature seeds selected \(L=2\), and waveform PCA selected \(L=1\) for 8 seeds and \(L=2\) for 2 seeds. Thus, the numbers of valid seeds for active 2 were 0, 10, and 2, respectively.

The summary-loss audit further showed that the information provided by additional waveform descriptors depended on the primary summary. Relative to PermWave, Wave+ increased PSG under the dynamical features by \(0.227\), \(0.187\), \(0.145\), and \(0.143\) for \(x_0\), active 2, \(I_1\), and \(t_{\mathrm{eff}}\), respectively. Under waveform PCA, the gains for the six fixed targets and active 1 ranged from \(0.338\) to \(0.386\) and reached \(p<0.01\) for all targets with 10 valid seeds, representing the most pronounced effect. Improvements for the CNN--LSTM were smaller, with increases of \(0.055\) and \(0.053\) for \(x_0\) and \(t_{\mathrm{eff}}\), respectively. These results localized potential information loss from compression in the different summaries and reflect recoverability within the current simulation configuration.

\clearpage
\begin{figure}[p]
\centering
\includegraphics[keepaspectratio,width=1\textwidth,height=0.92\textheight]{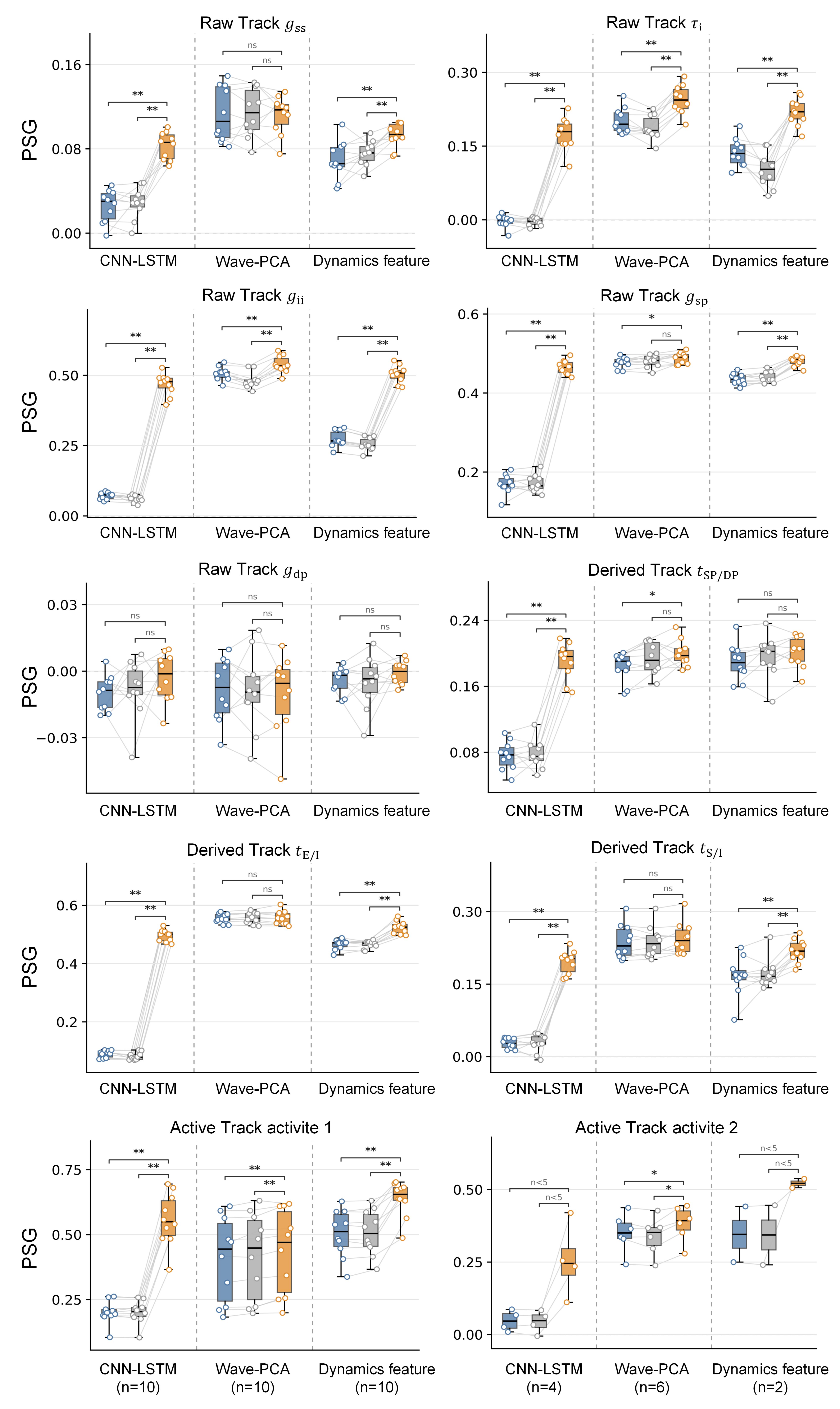}
\captionsetup{font=scriptsize,labelfont=bf,textfont=normalfont,labelsep=period,justification=justified,singlelinecheck=true,skip=2pt}
\caption{PSG gain for different parameters in different tracks under the three summaries in the CMC/MMN experiment. Colours, points, and paired within-seed lines have the same meanings as in \hyperref[fig:main3]{Fig. 3}. The sign-flip randomization test was Holm-corrected; the significance thresholds and resolution limit of the exact test are the same as in \hyperref[fig:main3]{Fig. 3}.}\label{fig:main4}
\end{figure}
\clearpage

CMC targets likewise showed summary dependence (\hyperref[fig:main4]{Fig. 4}). Under waveform PCA, \(t_{\mathrm{E/I}}\), \(g_{\mathrm{ii}}\), and \(g_{\mathrm{sp}}\) had the highest PSG values, at \(0.556\pm0.016\), \(0.507\pm0.024\), and \(0.477\pm0.014\), respectively. Under the ERP features, active 1, \(t_{\mathrm{E/I}}\), and \(g_{\mathrm{sp}}\) had values of \(0.506\pm0.084\), \(0.466\pm0.017\), and \(0.435\pm0.014\), respectively. Overall gain for the CNN--LSTM was weaker, with values of \(0.188\pm0.032\) and \(0.158\pm0.025\) for active 1 and \(g_{\mathrm{sp}}\), respectively. \(g_{\mathrm{dp}}\) was near the prior level under all three summaries. CNN--LSTM, ERP features, and waveform PCA selected \(L=2\) in 4, 2, and 6 seeds, respectively, giving corresponding valid-seed counts of 4, 2, and 6 for active 2.

Under the PermWave control, summary loss was most evident for the CNN--LSTM: the Wave+ PSG gains for \(t_{\mathrm{E/I}}\), \(g_{\mathrm{ii}}\), active 1, and \(g_{\mathrm{sp}}\) were \(0.421\), \(0.407\), \(0.384\), and \(0.305\), respectively. Under the ERP features, \(g_{\mathrm{ii}}\), active 1, and \(\tau_i\) improved by \(0.251\), \(0.156\), and \(0.118\), respectively; the largest gain for waveform PCA was only approximately \(0.057\). \(g_{\mathrm{dp}}\) showed no stable improvement in any comparison, suggesting that its weak recovery was primarily not caused by compression in the current summary.

\begin{figure}[H]
\hypertarget{fig:main5}{%
\centering
\includegraphics[keepaspectratio,width=0.88\textwidth]{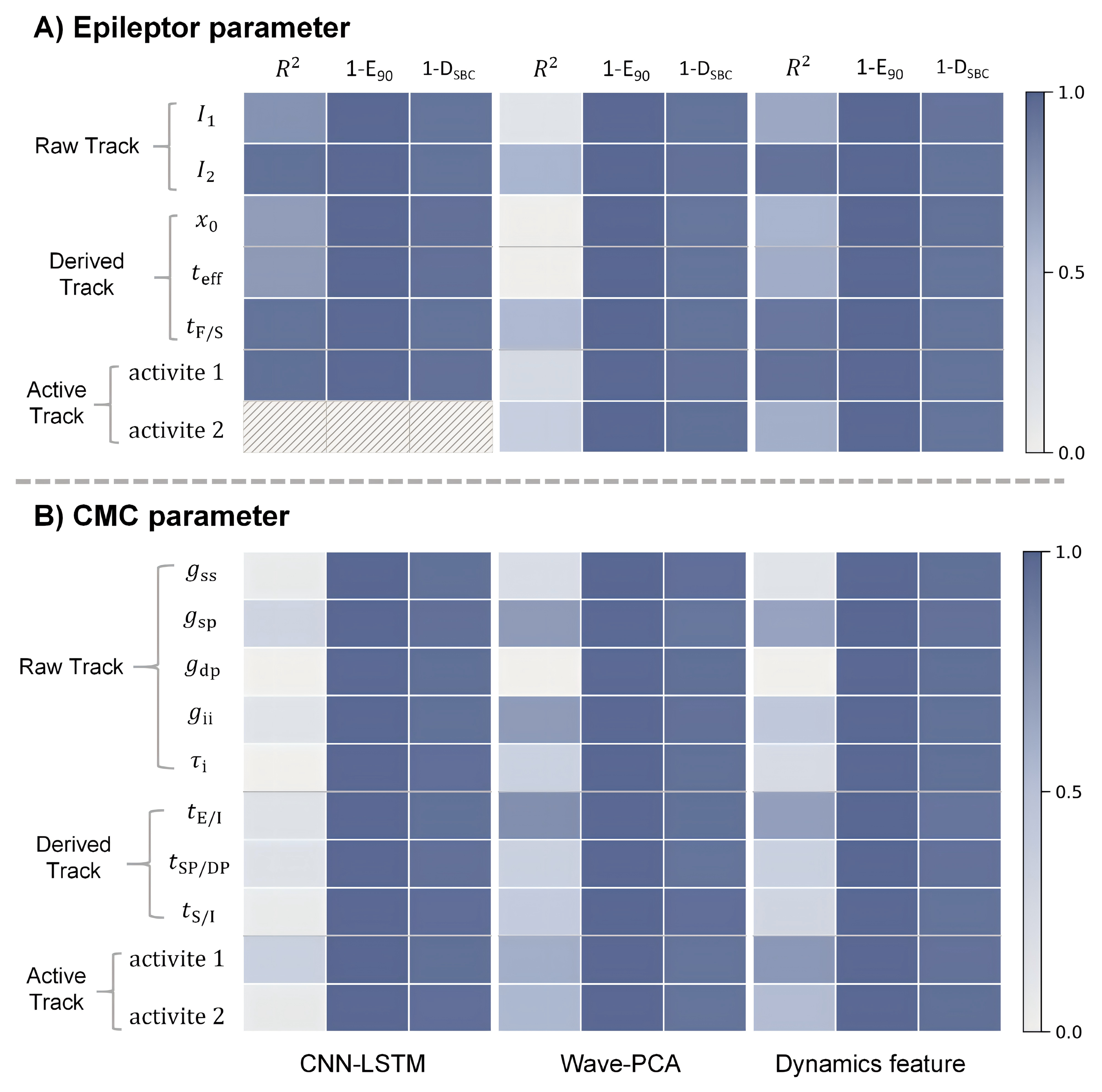}
\caption{Point recovery and raw-posterior calibration for candidate targets in the three tracks for Epileptor/iEEG and CMC/MMN. (A) Raw, Derived, and Active targets for Epileptor; (B) Raw, Derived, and Active targets for CMC. Columns are ordered by CNN--LSTM, waveform PCA, and model-specific hand-crafted features and show \(R^2\), \(1-E_{90}^{\mathrm{raw}}\), and \(1-D_{\mathrm{SBC}}^{\mathrm{raw}}\) averaged across algorithmic seeds. Active 2 summarizes only valid seeds that selected \(L_s\geq2\).}\label{fig:main5}
}
\end{figure}

In \(\mathcal D_{\mathrm{test}}^{\mathrm{post}}\) for Epileptor/iEEG, the \(R^2\) values for active 1, \(I_2\), and \(t_{\mathrm{F/S}}\) under the CNN--LSTM were \(0.957\pm0.006\), \(0.953\pm0.006\), and \(0.938\pm0.016\), respectively; under the dynamical features, the corresponding values were \(0.946\pm0.010\), \(0.943\pm0.007\), and \(0.915\pm0.013\). Waveform PCA retained only moderate point recovery for \(I_2\) and \(t_{\mathrm{F/S}}\). Across targets, \(E_{90}^{\mathrm{raw}}\) ranged from \(0.045\) to \(0.113\), and \(D_{\mathrm{SBC}}^{\mathrm{raw}}\) ranged from \(0.063\) to \(0.111\). In conjunction with the mismatch in the core waveform diagnostic space identified in Step 1, however, these results can be interpreted only as recoverability within the current simulation configuration and cannot be extrapolated as reliable derived inversion in real iEEG.

CMC/MMN likewise showed summary dependence. Under waveform PCA, the \(R^2\) values for \(t_{\mathrm{E/I}}\), \(g_{\mathrm{ii}}\), and \(g_{\mathrm{sp}}\) were \(0.788\pm0.014\), \(0.723\pm0.026\), and \(0.722\pm0.015\), respectively. Under the ERP features, those for active 1, \(t_{\mathrm{E/I}}\), and \(g_{\mathrm{sp}}\) were \(0.740\pm0.088\), \(0.698\pm0.019\), and \(0.675\pm0.015\), respectively. Under the CNN--LSTM, those for active 1 and \(g_{\mathrm{sp}}\) were only \(0.324\pm0.051\) and \(0.280\pm0.041\); \(g_{\mathrm{dp}}\) was below zero under all three summaries. Across targets, \(E_{90}^{\mathrm{raw}}\) ranged from \(0.020\) to \(0.067\), and \(D_{\mathrm{SBC}}^{\mathrm{raw}}\) ranged from \(0.042\) to \(0.088\). Good marginal calibration was therefore not equivalent to an informative target. The posterior for \(g_{\mathrm{dp}}\) could maintain approximately correct interval coverage and SBC performance, while its posterior mean still failed to track variation in the true value.

These results indicate that coverage of real observations, simulation-internal point recovery, and posterior calibration are three complementary but non-interchangeable questions. A summary can yield a well-calibrated but uninformative posterior, or it can accurately recover a target within the simulator yet be ineligible for real-mechanism interpretation because of real-data mismatch in Step 1. In conjunction with Step 1, targets with stronger recovery in CMC can proceed to subsequent joint interpretation conditional on the model; the Epileptor/iEEG results should remain restricted to evidence within the simulation configuration. Because this step audits only scalar posteriors, information overlap among multiple targets, compensation, and changes in joint performance when targets are inverted simultaneously are examined further in Step 3.

\hypertarget{sec-4-2-3}{%
\subsubsection{4.2.3 Within-track parameter synchrony and compensation}\label{sec-4-2-3}}

After completing the single-target audit, we separately trained joint posteriors within the Raw, Derived, and Active tracks. Posterior-mean synchrony measures whether posterior means vary synchronously across different simulated samples in \(\mathcal D_{\mathrm{test}}^{\mathrm{post}}\), whereas fixed-observation joint-sample correlation measures whether the joint posterior contains coupling in the same direction or negative compensation when the same posterior-test sample is held fixed.

Before interpreting parameter coupling, we first examined whether the within-track joint posterior altered the marginal recoverability of each target. Following the method in \hyperref[sec-2-4-1]{Section 2.4.1}, we report \(\Delta\mathrm{PSG}_k\) and the distribution, across different seeds, of whether the confidence interval from repeated resampling lay above zero, below zero, or spanned zero. The specific results from the two experiments are shown in \hyperref[tab:table6]{Tables 6} and \hyperref[tab:table7]{7}:

\clearpage
{\footnotesize
\begin{longtable}[]{@{}
  >{\raggedright\arraybackslash}p{(\columnwidth - 6\tabcolsep) * \real{0.2121}}
  >{\raggedleft\arraybackslash}p{(\columnwidth - 6\tabcolsep) * \real{0.2626}}
  >{\raggedleft\arraybackslash}p{(\columnwidth - 6\tabcolsep) * \real{0.2626}}
  >{\raggedleft\arraybackslash}p{(\columnwidth - 6\tabcolsep) * \real{0.2627}}@{}}
\caption{Change in marginal PSG for the within-track joint posterior of each parameter relative to the Step 2 single-target posterior in the Epileptor/iEEG experiment. Values are means \(\pm\) standard deviations across seeds; parentheses indicate the numbers of seeds for which the bootstrap \(95\%\) interval lay entirely above zero, entirely below zero, or spanned zero, respectively, denoted \(n_+/n_-/n_0\). Active 2 summarizes valid seeds only. \label{tab:table6}}\tabularnewline
\toprule
Parameter & CNN--LSTM & Waveform PCA & Dynamics feature \\
\midrule
\endfirsthead
\toprule
Parameter & CNN--LSTM & Waveform PCA & Dynamics feature \\
\midrule
\endhead
\(x_0\) & −0.024 ± 0.030 (0/6/4) & −0.016 ± 0.017 (0/4/6) & −0.022 ± 0.045 (2/5/3) \\
\(I_1\) & −0.020 ± 0.026 (0/5/5) & −0.015 ± 0.024 (0/3/7) & −0.013 ± 0.022 (1/4/5) \\
\(I_2\) & −0.009 ± 0.039 (1/1/8) & −0.053 ± 0.056 (0/7/3) & −0.005 ± 0.006 (1/1/8) \\
\(t_{\mathrm{eff}}\) & −0.077 ± 0.168 (0/7/3) & −0.017 ± 0.020 (0/4/6) & 0.002 ± 0.032 (3/1/6) \\
\(t_{F/S}\) & −0.011 ± 0.015 (1/6/3) & −0.060 ± 0.098 (0/6/4) & 0.005 ± 0.017 (4/1/5) \\
active 1 & 0.000 ± 0.000 (1/0/9) & 0.000 ± 0.007 (1/0/9) & −0.004 ± 0.007 (0/2/8) \\
active 2 & NaN & −0.003 ± 0.018 (0/0/2) & −0.018 ± 0.019 (1/4/5) \\
\bottomrule
\end{longtable}
}

{\footnotesize
\begin{longtable}[]{@{}
  >{\raggedright\arraybackslash}p{(\columnwidth - 6\tabcolsep) * \real{0.1386}}
  >{\raggedleft\arraybackslash}p{(\columnwidth - 6\tabcolsep) * \real{0.2871}}
  >{\raggedleft\arraybackslash}p{(\columnwidth - 6\tabcolsep) * \real{0.2871}}
  >{\raggedleft\arraybackslash}p{(\columnwidth - 6\tabcolsep) * \real{0.2872}}@{}}
\caption{Change in marginal PSG for the within-track joint posterior of each parameter relative to the Step 2 single-target posterior in the CMC/MMN experiment. Metric definitions and seed classifications are the same as in \hyperref[tab:table6]{Table 6} \label{tab:table7}}\tabularnewline
\toprule
Parameter & CNN--LSTM & Waveform PCA & Dynamics feature \\
\midrule
\endfirsthead
\toprule
Parameter & CNN--LSTM & Waveform PCA & Dynamics feature \\
\midrule
\endhead
\(g_{ss}\) & 0.008 ± 0.018 (3/1/6) & −0.007 ± 0.033 (1/2/7) & −0.001 ± 0.030 (2/1/7) \\
\(g_{sp}\) & 0.012 ± 0.026 (4/0/6) & 0.008 ± 0.011 (3/0/7) & 0.013 ± 0.009 (5/0/5) \\
\(g_{ii}\) & −0.002 ± 0.007 (0/1/9) & 0.026 ± 0.018 (8/0/2) & 0.003 ± 0.019 (1/1/8) \\
\(g_{dp}\) & 0.008 ± 0.013 (3/0/7) & 0.003 ± 0.012 (1/1/8) & 0.005 ± 0.007 (2/0/8) \\
\(\tau_i\) & 0.008 ± 0.010 (2/0/8) & 0.011 ± 0.018 (4/0/6) & −0.003 ± 0.039 (2/3/5) \\
\(t_{E/I}\) & −0.011 ± 0.014 (0/2/8) & −0.053 ± 0.173 (1/2/7) & −0.001 ± 0.010 (0/1/9) \\
\(t_{SP/DP}\) & 0.001 ± 0.017 (1/1/8) & 0.009 ± 0.013 (2/0/8) & 0.012 ± 0.015 (2/0/8) \\
\(t_{S/I}\) & 0.003 ± 0.008 (1/0/9) & 0.004 ± 0.011 (1/0/9) & 0.013 ± 0.028 (2/1/7) \\
active 1 & −0.001 ± 0.004 (1/0/9) & −0.010 ± 0.024 (0/2/8) & 0.003 ± 0.006 (2/0/8) \\
active 2 & 0.008 ± 0.008 (0/0/4) & −0.001 ± 0.034 (2/1/3) & 0.017 ± 0.001 (2/0/0) \\
\bottomrule
\end{longtable}
}

Overall, joint training did not systematically change marginal PSG. In Epileptor, negative intervals occurred more often for \(t_{\mathrm{eff}}\) under the CNN--LSTM and for \(I_2\) and \(t_{\mathrm{F/S}}\) under waveform PCA. In CMC, only \(g_{\mathrm{ii}}\) under waveform PCA improved in 8/10 seeds, while other targets showed no change that was consistent across summaries. The joint-minus-single-target differences in \(E_{90}^{\mathrm{raw}}\) and \(D_{\mathrm{SBC}}^{\mathrm{raw}}\) likewise showed no systematic direction.

\begin{figure}[H]
\hypertarget{fig:main6}{%
\centering
\includegraphics[keepaspectratio,width=1.00\textwidth]{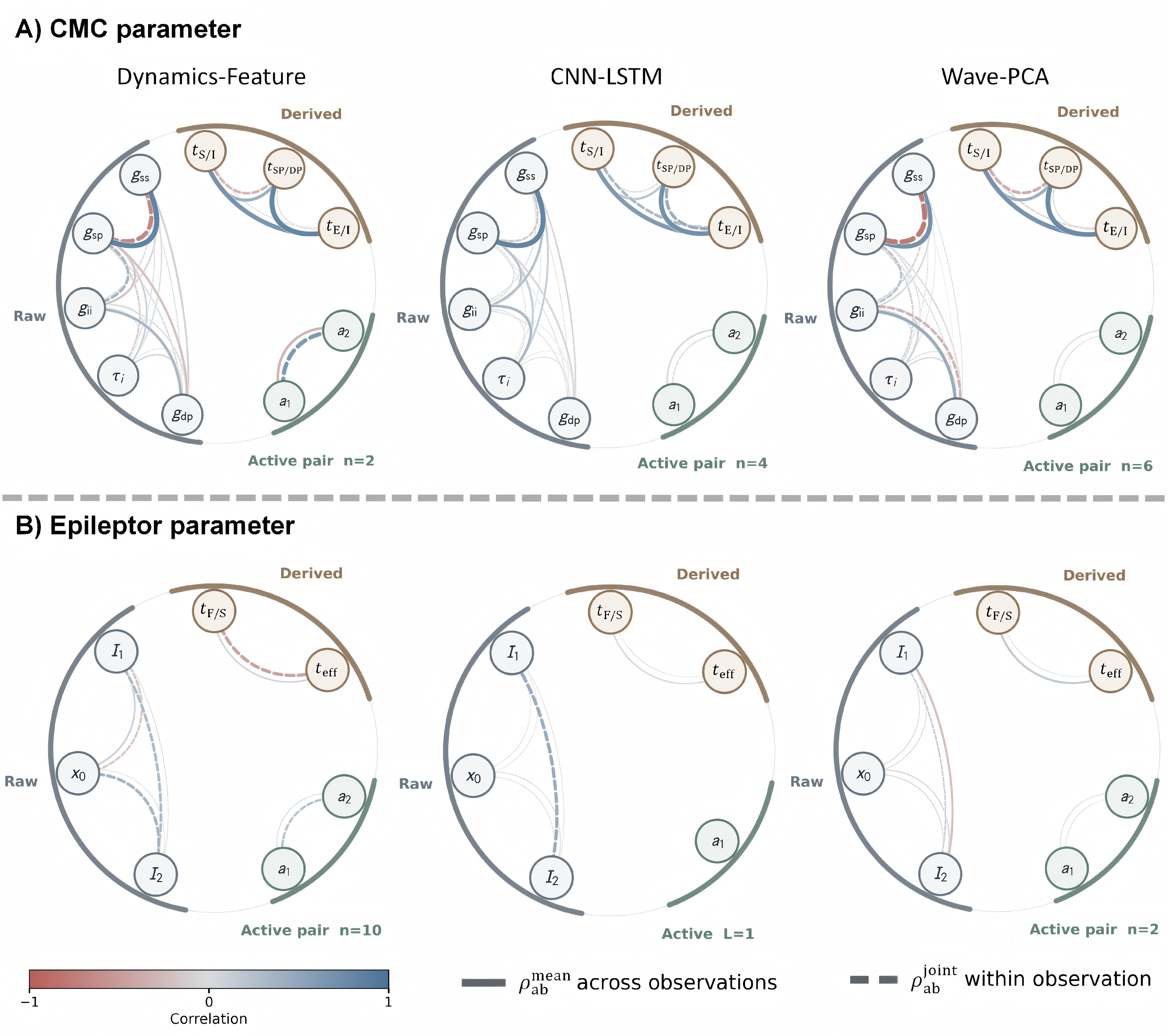}
\caption{Within-track posterior dependence structures in the CMC/MMN and Epileptor/iEEG experiments. (A) CMC/MMN; (B) Epileptor/iEEG. Each row shows the model-specific dynamical features, CNN--LSTM, and waveform PCA summaries, respectively, and the circular sectors correspond to the Raw, Derived, and Active tracks. Solid lines indicate posterior-mean correlations \(\rho_{ab}^{\mathrm{mean}}\) across samples in \(\mathcal D_{\mathrm{test}}^{\mathrm{post}}\), and dashed lines indicate joint-posterior correlations \(\rho_{ab}^{\mathrm{joint}}\) under a fixed posterior-test sample. Colour represents the direction and magnitude of the correlation, line width is proportional to the absolute correlation coefficient, and opacity indicates sign consistency across seeds. Active pair \(n\) denotes the number of valid seeds selecting \(L=2\).}\label{fig:main6}
}
\end{figure}

Within-track parameter-dependence structures are shown in \hyperref[fig:main6]{Fig. 6}, where solid lines indicate posterior-mean correlations \(\rho_{ab}^{\mathrm{mean}}\) across observations and dashed lines indicate joint-posterior correlations \(\rho_{ab}^{\mathrm{joint}}\) under a fixed observation. In the CMC/MMN experiment (\hyperref[fig:main6]{Fig. 6}A), the clearest relationship in the Raw track was between \(g_{\mathrm{ss}}\) and \(g_{\mathrm{sp}}\). Their \(\rho_{ab}^{\mathrm{mean}}\) values under the dynamical features, CNN--LSTM, and waveform PCA were \(0.853\pm0.055\), \(0.838\pm0.091\), and \(0.713\pm0.066\), respectively. Under the dynamical features and waveform PCA, their \(\rho_{ab}^{\mathrm{joint}}\) values were \(-0.689\pm0.036\) and \(-0.801\pm0.046\), respectively, indicating clear negative compensation under a fixed observation. This relationship was weaker under the CNN--LSTM (\(-0.208\pm0.036\)). Other relationships among raw parameters were mainly summary-dependent; for example, \(g_{\mathrm{sp}}\)--\(g_{\mathrm{ii}}\) under the dynamical features and \(g_{\mathrm{dp}}\)--\(g_{\mathrm{ii}}\) under waveform PCA showed only moderate joint correlations.

In the CMC Derived track, \(t_{\mathrm{E/I}}\)--\(t_{\mathrm{SP/DP}}\) and \(t_{\mathrm{E/I}}\)--\(t_{\mathrm{S/I}}\) had high \(\rho_{ab}^{\mathrm{mean}}\) values (\(0.655\)--\(0.821\)) under all three summaries, although this synchrony may have been affected by shared raw parameters. Under a fixed observation, the CNN--LSTM primarily showed moderate positive correlations, whereas the dynamical features and waveform PCA primarily showed moderate negative correlations between \(t_{\mathrm{SP/DP}}\) and \(t_{\mathrm{S/I}}\). In the Active track, only 2, 4, and 6 seeds selected \(L=2\), respectively, and the correlation directions were inconsistent; there was no stable two-dimensional active coupling.

The Epileptor/iEEG experiment (\hyperref[fig:main6]{Fig. 6}B) showed no coupling structure that was consistent across summaries. Under the dynamical features, the \(\rho_{ab}^{\mathrm{joint}}\) values for \(x_0\)--\(I_2\), \(I_1\)--\(I_2\), and \(t_{\mathrm{eff}}\)--\(t_{\mathrm{F/S}}\) were \(0.401\pm0.154\), \(0.358\pm0.108\), and \(-0.467\pm0.093\), respectively. Under the CNN--LSTM, only \(I_1\)--\(I_2\) showed a moderate joint correlation (\(0.477\pm0.101\)), whereas all relationships under waveform PCA were generally weak. In the Active track, the active 1--active 2 correlation under the dynamical features was only weak to moderate, the CNN--LSTM selected only \(L=1\), and waveform PCA selected \(L=2\) in only two seeds. Thus, parameter dependence observed in Epileptor was primarily summary-specific, and no stable cross-representation compensation structure was identified.

\hypertarget{sec-4-2-4}{%
\subsubsection{4.2.4 Cross-track parameter synchrony and compensation}\label{sec-4-2-4}}

\hypertarget{sec-3-3-4-1}{%
\paragraph{(1) Active--derived direction consistency}\label{sec-3-3-4-1}}

To determine whether the data-driven active subspace contained the linear proxy directions of the predefined derived coordinates, we followed the method in \hyperref[sec-2-4-2-a]{Section 2.4.2(a)} and performed comparisons within each seed using its independently selected active dimensionality \(L_s\). For \(L_s=1\), a single active direction was compared with the derived direction; for \(L_s=2\), the two-dimensional subspace spanned by active 1 and active 2 was compared. Because the projection proportion depends on \(L_s\), the results were summarized separately by active dimensionality. The framework reports the degree of misalignment \(s_h^{\mathrm{AD}}\), the empirical upper-tail fraction from the structure-matched permutation \(p_h^{\mathrm{AD}}\), and the proportion of seeds for which the derived direction achieved the highest alignment rank in the complete permutation reference set. The primary results are shown in \hyperref[tab:table8]{Tables 8} and \hyperref[tab:table9]{9}.

{\scriptsize
\begin{longtable}[]{@{}
  >{\centering\arraybackslash}p{(\columnwidth - 10\tabcolsep) * \real{0.1400}}
  >{\centering\arraybackslash}p{(\columnwidth - 10\tabcolsep) * \real{0.1300}}
  >{\centering\arraybackslash}p{(\columnwidth - 10\tabcolsep) * \real{0.0900}}
  >{\centering\arraybackslash}p{(\columnwidth - 10\tabcolsep) * \real{0.2750}}
  >{\centering\arraybackslash}p{(\columnwidth - 10\tabcolsep) * \real{0.2750}}
  >{\centering\arraybackslash}p{(\columnwidth - 10\tabcolsep) * \real{0.0900}}@{}}
\caption{Active--derived direction consistency in the Epileptor/iEEG experiment. Values are reported as median [min, max]. \label{tab:table8}}\tabularnewline
\toprule
Summary & Derived target & \(L_s\) (\(n_s\)) & \(s_h^{\mathrm{AD}}\) & \(p_h^{\mathrm{AD}}\) & Top-rank \\
\midrule
\endfirsthead
\toprule
Summary & Derived target & \(L_s\) (\(n_s\)) & \(s_h^{\mathrm{AD}}\) & \(p_h^{\mathrm{AD}}\) & Top-rank \\
\midrule
\endhead
\multirow[c]{2}{=}{Dynamics feature} & \(t_{\mathrm{eff}}\) & 2 (10) & 0.989 {[}0.916, 0.997{]} & 1.000 {[}1.000, 1.000{]} & 0/10 \\
& \(t_{\mathrm{F/S}}\) & 2 (10) & 0.001 {[}0.000, 0.002{]} & 0.167 {[}0.167, 0.167{]} & 10/10 \\
\midrule
\multirow[c]{2}{=}{CNN–LSTM} & \(t_{\mathrm{eff}}\) & 1 (10) & 0.997 {[}0.982, 0.999{]} & 1.000 {[}1.000, 1.000{]} & 0/10 \\
& \(t_{\mathrm{F/S}}\) & 1 (10) & 0.002 {[}0.000, 0.016{]} & 0.167 {[}0.167, 0.167{]} & 10/10 \\
\midrule
\multirow[c]{4}{=}{Waveform PCA} & \(t_{\mathrm{eff}}\) & 1 (8) & 0.947 {[}0.898, 0.998{]} & 0.917 {[}0.667, 1.000{]} & 0/8 \\
& \(t_{\mathrm{eff}}\) & 2 (2) & 0.837 {[}0.789, 0.884{]} & 0.917 {[}0.833, 1.000{]} & 0/2 \\
& \(t_{\mathrm{F/S}}\) & 1 (8) & 0.556 {[}0.243, 0.966{]} & 0.500 {[}0.333, 1.000{]} & 0/8 \\
& \(t_{\mathrm{F/S}}\) & 2 (2) & 0.029 {[}0.017, 0.042{]} & 0.500 {[}0.500, 0.500{]} & 0/2 \\
\bottomrule
\end{longtable}
}

\clearpage
{\scriptsize
\begin{longtable}[]{@{}
  >{\centering\arraybackslash}p{(\columnwidth - 10\tabcolsep) * \real{0.1400}}
  >{\centering\arraybackslash}p{(\columnwidth - 10\tabcolsep) * \real{0.1300}}
  >{\centering\arraybackslash}p{(\columnwidth - 10\tabcolsep) * \real{0.0900}}
  >{\centering\arraybackslash}p{(\columnwidth - 10\tabcolsep) * \real{0.2750}}
  >{\centering\arraybackslash}p{(\columnwidth - 10\tabcolsep) * \real{0.2750}}
  >{\centering\arraybackslash}p{(\columnwidth - 10\tabcolsep) * \real{0.0900}}@{}}
\caption{Active--derived direction consistency in the CMC/MMN experiment. Values are reported as median [min, max]. \label{tab:table9}}\tabularnewline
\toprule
Summary & Derived target & \(L_s\) (\(n_s\)) & \(s_h^{\mathrm{AD}}\) & \(p_h^{\mathrm{AD}}\) & Top-rank \\
\midrule
\endfirsthead
\toprule
Summary & Derived target & \(L_s\) (\(n_s\)) & \(s_h^{\mathrm{AD}}\) & \(p_h^{\mathrm{AD}}\) & Top-rank \\
\midrule
\endhead
\multirow[c]{6}{=}{Dynamics feature} & \(t_{\mathrm{E/I}}\) & 1 (8) & 0.765 {[}0.573, 0.957{]} & 0.300 {[}0.150, 0.600{]} & 0/8 \\
& \(t_{\mathrm{E/I}}\) & 2 (2) & 0.184 {[}0.097, 0.272{]} & 0.125 {[}0.050, 0.200{]} & 1/2 \\
& \(t_{\mathrm{SP/DP}}\) & 1 (8) & 0.533 {[}0.416, 0.724{]} & 0.100 {[}0.050, 0.150{]} & 2/8 \\
& \(t_{\mathrm{SP/DP}}\) & 2 (2) & 0.568 {[}0.558, 0.578{]} & 0.400 {[}0.350, 0.450{]} & 0/2 \\
& \(t_{\mathrm{S/I}}\) & 1 (8) & 0.978 {[}0.863, 1.000{]} & 0.817 {[}0.500, 0.967{]} & 0/8 \\
& \(t_{\mathrm{S/I}}\) & 2 (2) & 0.774 {[}0.724, 0.823{]} & 0.692 {[}0.667, 0.717{]} & 0/2 \\
\midrule
\multirow[c]{6}{=}{CNN–LSTM} & \(t_{\mathrm{E/I}}\) & 1 (6) & 0.862 {[}0.758, 0.999{]} & 0.475 {[}0.400, 0.850{]} & 0/6 \\
& \(t_{\mathrm{E/I}}\) & 2 (4) & 0.485 {[}0.235, 0.903{]} & 0.325 {[}0.100, 0.650{]} & 0/4 \\
& \(t_{\mathrm{SP/DP}}\) & 1 (6) & 0.538 {[}0.449, 0.827{]} & 0.075 {[}0.050, 0.150{]} & 3/6 \\
& \(t_{\mathrm{SP/DP}}\) & 2 (4) & 0.392 {[}0.350, 0.755{]} & 0.225 {[}0.150, 0.450{]} & 0/4 \\
& \(t_{\mathrm{S/I}}\) & 1 (6) & 0.960 {[}0.884, 0.999{]} & 0.717 {[}0.550, 0.950{]} & 0/6 \\
& \(t_{\mathrm{S/I}}\) & 2 (4) & 0.551 {[}0.394, 0.782{]} & 0.325 {[}0.217, 0.517{]} & 0/4 \\
\midrule
\multirow[c]{6}{=}{Waveform PCA} & \(t_{\mathrm{E/I}}\) & 1 (4) & 0.785 {[}0.351, 0.945{]} & 0.350 {[}0.050, 0.650{]} & 1/4 \\
& \(t_{\mathrm{E/I}}\) & 2 (6) & 0.293 {[}0.041, 0.844{]} & 0.150 {[}0.050, 0.750{]} & 3/6 \\
& \(t_{\mathrm{SP/DP}}\) & 1 (4) & 0.947 {[}0.797, 0.975{]} & 0.675 {[}0.500, 0.800{]} & 0/4 \\
& \(t_{\mathrm{SP/DP}}\) & 2 (6) & 0.509 {[}0.090, 0.841{]} & 0.325 {[}0.100, 0.700{]} & 0/6 \\
& \(t_{\mathrm{S/I}}\) & 1 (4) & 0.541 {[}0.353, 0.899{]} & 0.117 {[}0.050, 0.550{]} & 0/4 \\
& \(t_{\mathrm{S/I}}\) & 2 (6) & 0.615 {[}0.278, 0.886{]} & 0.558 {[}0.067, 0.867{]} & 0/6 \\
\bottomrule
\end{longtable}
}

In Epileptor/iEEG, \(t_{\mathrm{F/S}}\) lay almost entirely within the selected active subspace under both the dynamical features and the CNN--LSTM and ranked first in the structure-matched reference set for every seed. Because its permutation reference set contained only six directions, \(p_h^{\mathrm{AD}}=1/6=0.167\) was the minimum attainable value. Waveform PCA did not reproduce this structure. Even though its two \(L_s=2\) seeds had low absolute misalignment, the corresponding \(p_h^{\mathrm{AD}}=0.500\), indicating that the projection lacked structural specificity. \(t_{\mathrm{eff}}\) showed high misalignment under all three summaries.

In CMC/MMN, no derived direction was stably reproduced across summaries and active dimensionalities. Relatively stronger local evidence included the lower misalignment and upper-tail fraction of \(t_{\mathrm{E/I}}\) under waveform PCA with \(L_s=2\), and the lower permutation fraction of \(t_{\mathrm{SP/DP}}\) under the dynamical features and CNN--LSTM with \(L_s=1\). However, these results occurred only in some seeds or at specific dimensionalities, and neither absolute misalignment nor the top-rank success rate remained stable simultaneously.

Overall, Epileptor \(t_{\mathrm{F/S}}\) showed reproducible internal active--derived alignment under the dynamical features and CNN--LSTM, which was not detected under waveform PCA. In CMC, no derived coordinate was stably aligned across all summaries and seeds. These results support directional correspondence within specific configurations.

\hypertarget{sec-3-3-4-2}{%
\paragraph{(2) Raw--derived distributional consistency}\label{sec-3-3-4-2}}

\begin{figure}[p]
\centering
\includegraphics[keepaspectratio,width=1\textwidth,height=0.90\textheight]{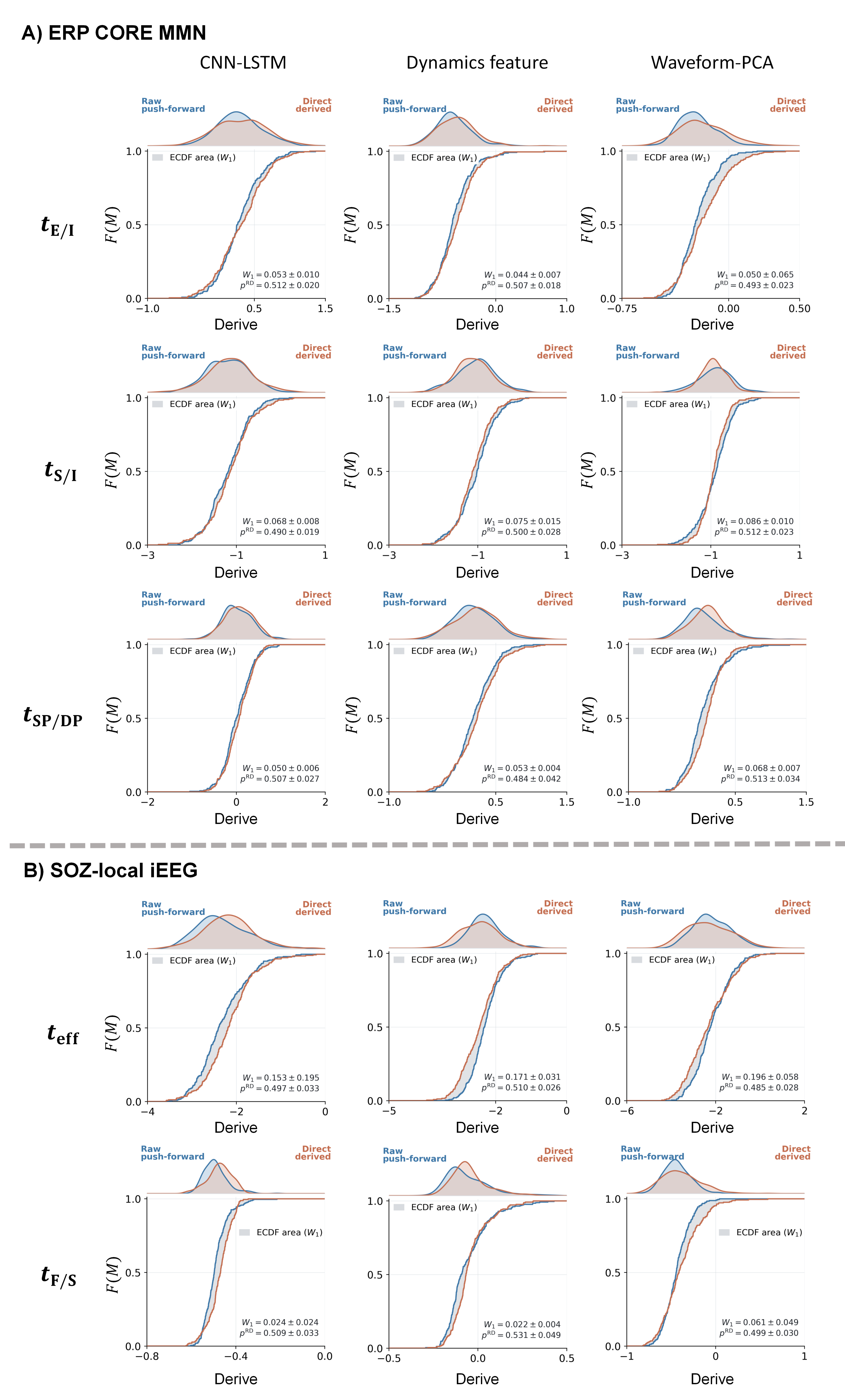}
\captionsetup{font=scriptsize,labelfont=bf,textfont=normalfont,labelsep=period,justification=justified,singlelinecheck=true,skip=2pt}
\caption{Distributional consistency between the raw push-forward and direct-derived posterior. Blue indicates the distribution obtained by pushing raw joint samples through the derived function, and orange indicates the direct derived posterior. KDEs are shown above, ECDFs below, and the grey area equals the one-dimensional \(W_1\). Each panel shows an actual observation whose \(W_1\) was closest to the corresponding summary--target centre across seeds. The lower-right corner reports the across-seed median \(\pm\) standard deviation of the 10 seed-level medians. (A). CMC/MMN experiment; (B). Epileptor/SOZ-local iEEG experiment}\label{fig:main7}
\end{figure}

Raw--derived distributional consistency assesses, for the same observation, whether the coordinate distribution obtained by pushing the Raw-track joint posterior through the derived function is consistent with the posterior estimated directly in the Derived track. The difference between the two paths is represented by the one-dimensional 1-Wasserstein distance \(W_1\), and the empirical upper-tail fraction \(p_i^{\mathrm{RD}}\) is calculated using \(\mathcal D_{\mathrm{cal}}^{\mathrm{post}}\), which is excluded from posterior fitting. A lower \(p_i^{\mathrm{RD}}\) indicates that the current discrepancy exceeds the normal error of the reference inference procedure; other values are not treated as a continuous grade of agreement. The results are shown in \hyperref[fig:main7]{Fig. 7}.

\clearpage
\clearpage

In CMC/MMN, no systematic raw--derived distributional mismatch was detected for any of the three derived coordinates under any summary. For \(t_{\mathrm{E/I}}\), \(W_1\) ranged from \(0.044\) to \(0.053\), with corresponding values of \(p^{\mathrm{RD}}=0.493\)--\(0.512\). For \(t_{\mathrm{S/I}}\) and \(t_{\mathrm{SP/DP}}\), \(W_1\) ranged from \(0.068\) to \(0.086\) and from \(0.050\) to \(0.068\), respectively, and their \(p^{\mathrm{RD}}\) values were all close to \(0.5\). Although the distance for \(t_{\mathrm{E/I}}\) under waveform PCA varied substantially across seeds, the calibrated fraction remained in the middle of the reference distribution.

Epileptor/iEEG yielded the same conclusion. For \(t_{\mathrm{eff}}\), \(W_1\) ranged from \(0.153\) to \(0.196\) across the three summaries, with \(p^{\mathrm{RD}}=0.485\)--\(0.510\). For \(t_{\mathrm{F/S}}\), \(W_1\) ranged from \(0.022\) to \(0.061\), with \(p^{\mathrm{RD}}=0.499\)--\(0.531\). Although some distances under the CNN--LSTM showed substantial variation across seeds, the calibrated fractions remained stable.

Because \(W_1\) retains the original scale of the derived coordinate, it should not be used for direct comparisons across coordinates. Consistency is therefore judged primarily from the calibrated \(p^{\mathrm{RD}}\). A \(p^{\mathrm{RD}}\) value near \(0.5\) indicates that the difference between the two posterior paths for real observations lies in the middle of the independent simulated calibration distribution, whereas only lower values indicate anomalous mismatch. Overall, no systematic inconsistency beyond the range of calibration error was detected between the raw posterior pushed forward through the derived function and the directly trained derived posterior.

\hypertarget{sec-3-3-4-3}{%
\paragraph{(3) Local--global active consistency audit}\label{sec-3-3-4-3}}

Local--global active consistency assesses whether the global active subspace determined for each seed on \(\mathcal D_{\mathrm{train}}^{\mathrm{post}}\) can represent the local sensitivity structure near real observations in \(\mathcal D^{\mathrm{obs}}_{\mathrm{test}}\). The Active dimensionality \(L_s\) was selected independently within each summary and seed; \(\mathcal D_{\mathrm{cal}}^{\mathrm{post}}\) provided a dimension-matched reference distribution of misalignment. The results are shown in \hyperref[tab:table10]{Table 10}.

{\scriptsize
\begin{longtable}[]{@{}cccccc@{}}
\caption{Local--global active subspace consistency in the Epileptor/iEEG and CMC/MMN experiments. Within each seed, the median was first taken across all real test observations, and the mean \(\pm\) standard deviation was then reported across seeds with the same summary and dimensionality. \label{tab:table10}}\tabularnewline
\toprule
Experiment & Summary & \(L_s\) & \(n_{\mathrm{seed}}\) & \(s^{\mathrm{LG}}\) & \(p^{\mathrm{LG}}\) \\
\midrule
\endfirsthead
\toprule
Experiment & Summary & \(L_s\) & \(n_{\mathrm{seed}}\) & \(s^{\mathrm{LG}}\) & \(p^{\mathrm{LG}}\) \\
\midrule
\endhead
\multirow[c]{4}{*}{Epileptor/iEEG} & CNN--LSTM & 1 & 10 & 0.043 ± 0.044 & 0.424 ± 0.167 \\
& Dynamics feature & 2 & 10 & 0.039 ± 0.052 & 0.544 ± 0.216 \\
\cmidrule(lr){2-6}
& \multirow[c]{2}{*}{Waveform PCA} & 1 & 8 & 0.327 ± 0.169 & 0.432 ± 0.119 \\
& & 2 & 2 & 0.100 ± 0.026 & 0.514 ± 0.068 \\
\midrule
\multirow[c]{6}{*}{CMC/MMN} & \multirow[c]{2}{*}{CNN–LSTM} & 1 & 6 & 0.511 ± 0.257 & 0.387 ± 0.178 \\
& & 2 & 4 & 0.389 ± 0.053 & 0.481 ± 0.151 \\
\cmidrule(lr){2-6}
& \multirow[c]{2}{*}{ERP core feature} & 1 & 8 & 0.273 ± 0.234 & 0.498 ± 0.168 \\
& & 2 & 2 & 0.361 ± 0.045 & 0.450 ± 0.039 \\
\cmidrule(lr){2-6}
& \multirow[c]{2}{*}{Waveform PCA} & 1 & 4 & 0.590 ± 0.246 & 0.482 ± 0.236 \\
& & 2 & 6 & 0.465 ± 0.055 & 0.507 ± 0.162 \\
\bottomrule
\end{longtable}
}

The Active dimensionality in Epileptor/iEEG showed clear summary dependence. Absolute misalignment was low for the CNN--LSTM and dynamical features but higher for waveform PCA when \(L_s=1\); \(p^{\mathrm{LG}}\) showed no systematic lower tail across seeds in any group. CMC/MMN showed greater summary and seed dependence in both dimensionality selection and absolute misalignment. For the CNN--LSTM, misalignment was \(0.511\pm0.257\) and \(0.389\pm0.053\) at \(L_s=1\) and \(L_s=2\), respectively. The mean \(p^{\mathrm{LG}}\) values for the CMC groups ranged from \(0.387\) to \(0.507\) and still showed no consistently low tail, although the proportion of low-tail observations and variation across seeds suggested that the local--global relationship was not equally stable across all initializations.

Because \(s^{\mathrm{LG}}\) is affected by both the dimensionality of the parameter space and the selected \(L_s\), its absolute value should not be compared directly across dimensionalities or experiments. The dimension-matched calibrated \(p^{\mathrm{LG}}\) and the proportion of low-tail observations are the primary criteria for determining whether a local shift is anomalous. Overall, neither experiment showed a local--global active subspace mismatch that was stable across seeds. The results nevertheless indicate that global active coordinates are only a low-dimensional summary of the overall sensitivity structure under a particular summary and seed.

\hypertarget{sec-5}{%
\section{5 Discussion}\label{sec-5}}

\hypertarget{sec-5-1}{%
\subsection{5.1 Discussion of the controlled experimental results}\label{sec-5-1}}

The controlled experiments assessed whether the proposed audit framework could distinguish different types of inference failure under known ground truth. To this end, we combined statistically controlled experiments with 200 independent replicates to estimate type I error rates and statistical power. In Step 1, the condition-matched pseudo-observed batch null controlled the type I error rates of the four audits at \(5.0\%\)--\(7.0\%\), and power for visible mismatch was \(100\%\) in every audit. When the shift was present only in dimensions removed by the summary, the summary-space rejection rate remained at the null-condition level, whereas waveform-space power remained \(100\%\). This finding supports the ability of the dual-space design to localize representational masking, while also indicating that it can test only for mismatch in the preregistered observation representations and cannot detect latent derived shifts that do not alter these representations.

Step 2 separated information content from raw-posterior calibration. PermWave retained the dimensionality, scale, marginal distribution, and network structure of Wave+ but removed the waveform--target correspondence. Its performance was similar to that of Original, whereas true Wave+ markedly improved PSG and \(R^2\), indicating that the detected gain arose from additional information rather than branch capacity. The underdispersed control retained the same posterior mean and therefore still had high PSG and \(R^2\), but \(E_{90}^{\mathrm{raw}}\) and \(D_{\mathrm{SBC}}^{\mathrm{raw}}\) increased to \(0.593\) and \(0.307\), respectively. Thus, PSG, point recovery, and raw-posterior calibration cannot substitute for one another.

The controlled experiments in Step 3 evaluated the joint and cross-track audits. The framework distinguished an independent joint posterior from prespecified strong negative compensation, identified an artificially shifted raw-derived posterior, and detected a pre-injected direction of active-derived alignment. Observed power reached \(100\%\) for all three diagnostics, with empirical type I error rates of \(0\%\)--\(3\%\). These negative controls indicate that raw-derived consistency and active-derived alignment do not necessarily arise from a shared simulator and training procedure. When the two inference paths or candidate directions were deliberately misspecified, the corresponding audits produced anomalous results.

Overall, the controlled experiments indicate that the framework can separately localize insufficient observation coverage, summary information loss, insufficient target information, posterior miscalibration, and anomalies in joint structure.

\hypertarget{sec-5-2}{%
\subsection{5.2 Discussion of the SOZ-local iEEG and Epileptor results}\label{sec-5-2}}

Using a simplified Epileptor dynamical model, we conducted a complete SBI recoverability analysis under the specified configuration on SOZ-local iEEG data. The experimental results indicate that parameter recoverability is constrained simultaneously by the coverage of the generative model, the summary representation, and the specific target definition. The different levels of the audit yielded clearly differentiated evidence.

In Step 1, waveform PCA did not show lower-tail anomalies within the formal Summary Space as strong as those under the other two summaries, whereas the CNN--LSTM and Epileptor dynamical features both indicated mismatch. More importantly, the results in the waveform diagnostic space shared across the three summaries all reached the permutation resolution limit. In contrast, Step 2 still identified several recoverable targets in the simulated blind test set. These findings are not contradictory: the former evaluates whether real observations fall within simulated support, whereas the latter evaluates whether parameters can be recovered from simulated signals conditional on the simulator being correct.

This real-data mismatch may arise first from insufficient model hierarchy. Epileptor is primarily intended to summarize general dynamics such as seizure initiation, termination, and transitions between fast and slow subsystems, rather than to reproduce patient-specific raw iEEG point by point \cite{ref54}. Real seizures exhibit multiple dynamotypes, and their waveforms are also affected by regional propagation, channel location, and patient-specific network structure \cite{ref66}. We used a simplified configuration with a single source, fixed coupling, and fixed readout, which therefore had difficulty covering seizure heterogeneity across patients. Observation-layer calibration can adjust gain, noise, and limited temporal shifts, but cannot supply propagation structures missing from the simulator.

The different performance of the summaries further indicates that observation coverage and parameter-information retention are separate questions. Waveform PCA preferentially preserves the linear directions with the greatest variance in the raw waveform. It may therefore bring simulations and observations close in these directions while disregarding parameter-related low-variance or nonlinear structure. Consistent with this interpretation, waveform PCA yielded low PSG and \(R^2\) values for most targets in Step 2. Although the CNN--LSTM and dynamical features did not pass the real-data coverage audit, they retained more simulation-internal parameter information. Studies of SBI for neural dynamics likewise indicate that time-series representations, prior ranges, and target parameterization can all substantially affect parameter recovery \cite{ref61}.

Within the current simulation configuration, recovery was strongest for \(I_2\), \(t_{\mathrm{F/S}}\), and active 1. This may be related to the direct inclusion of \(G_{\mathrm{SW}}x_2\) in the observation equation: the effect of \(I_2\) on the second spike-and-wave subsystem can produce stable time--frequency and morphological changes, and \(t_{\mathrm{F/S}}\), which contains \(I_2\), inherits this information. In contrast, \(x_0\) and \(I_1\) primarily affect the critical seizure state and onset process. Extracting all trajectories around the detected endogenous seizure onset may have attenuated the corresponding temporal information. The clear improvement of several targets under waveform PCA after Wave+ was added also indicates that some parameter information remained in the waveform descriptors but was not retained by the principal PCA directions.

Step 3 was generally consistent with this interpretation. The joint posterior did not systematically improve marginal recovery for weak targets, indicating that insufficient information cannot be resolved automatically through joint training. Some parameter correlations also did not form a compensation structure that was stable across summaries. The push-forward of the raw posterior and the direct derived posterior were largely consistent, indicating that the two paths were numerically self-consistent within the current simulator. The stable alignment of \(t_{\mathrm{F/S}}\) with the active subspace under the CNN--LSTM and dynamical features was consistent with its high PSG, but this relationship still depended on the current prior, summary, and observation mapping and cannot be assigned direct physiological meaning.

Previous patient-specific Epileptor studies have generally embedded multiple local models within structural-connectivity networks and focused on estimating regional epileptogenicity and spatiotemporal propagation \cite{ref63,ref64,ref65}. The present study therefore indicates only that the current single-source configuration is insufficient to explain continuous 30 s raw SOZ-local waveforms. Future work may incorporate multinode propagation, patient-level priors, channel-level observation mappings, and multiple seizures from the same patient to distinguish between-patient from within-patient variation.

Overall, this experiment illustrates the primary value of the framework. Even when some targets have high PSG, good point recovery, and posterior calibration within the simulator, real-data mismatch in Step 1 still prevents them from being interpreted as patient mechanisms. The current Step 2--3 results therefore serve only to localize simulation-internal sources of information, summary loss, and parameter geometry, rather than as evidence of physiological validity.

\hypertarget{sec-5-3}{%
\subsection{5.3 Discussion of the ERP CORE MMN and CMC results}\label{sec-5-3}}

In a separate five-node CMC auditory network model, we conducted a complete SBI inversion-validity analysis under the specified configuration using ERP CORE auditory mismatch negativity (MMN) data. Unlike the persistent mismatch in core dynamics in the Epileptor/iEEG experiment, the CMC/MMN experiment did not show simultaneous placement of all representation spaces in the extreme lower tail of the calibration reference. Some targets therefore had limited eligibility for interpretation using real data when the corresponding summary-coverage conditions were satisfied.

In Step 2, waveform PCA and the preregistered ERP features retained the strongest information for \(g_{\mathrm{sp}}\), \(g_{\mathrm{ii}}\), \(t_{\mathrm{E/I}}\), and active 1. This ranking was broadly consistent with the current model structure: the superficial pyramidal population participates in both local recurrent activity and the principal feedforward output of the auditory hierarchy, whereas the inhibitory population directly regulates the amplitude and duration of superficial responses. The relative-gain coordinate \(t_{\mathrm{E/I}}\) can therefore integrate their joint effects on the evoked waveform. Classical CMC theory likewise treats superficial pyramidal cells, inhibitory interneurons, and hierarchical feedforward and feedback pathways as important components of prediction-error signalling \cite{ref67}.

This result was also directionally consistent with previous dynamical studies of MMN. DCM studies have indicated that frequency-deviance responses are not generated by a single local process but involve local adaptation, intrinsic connections, and feedforward and feedback interactions across the auditory hierarchy \cite{ref68}. Studies of attention and prediction have further suggested that inhibitory-interneuron gain can modulate the mismatch response and sensory precision \cite{ref69}. Thus, the stronger simulation-internal recovery of \(g_{\mathrm{sp}}\) and \(g_{\mathrm{ii}}\) has some theoretical plausibility. In contrast, \(g_{\mathrm{dp}}\) was near the prior level under all three summaries. This result indicates that the independent contribution of deep pyramidal cells is difficult to separate under the current observation design. The deep state is simultaneously influenced by superficial input, local inhibition, and fixed feedback connections and enters the ERP readout jointly with the slow state through fixed sensor weights; its effect may therefore be absorbed by other pathways. In addition, the frontocentral virtual channel and difference wave retained stable MMN morphology but discarded spatial-topological information that might help distinguish superficial from deep sources.

The performance of the different summaries also revealed the importance of the mode of information compression. Waveform PCA retained the principal linear covariance of the standard stimulus and difference wave, whereas the ERP features directly encoded the N1, P2, and MMN windows and the recovery process. Both could therefore capture stable changes related to local gains. The waveform-reconstruction objective of the CNN--LSTM did not guarantee retention of target-specific information, and its original recovery was weak. Adding an independent waveform branch substantially improved most targets, indicating that some parameter-related ERP information had been compressed during encoding. Coverage and SBC were generally close across the posteriors, suggesting that these differences mainly reflected the degree of information retention rather than evident posterior underdispersion.

Step 3 further showed that parameters recoverable in isolation can nevertheless form substitution relationships in joint inversion. \(g_{\mathrm{ss}}\) and \(g_{\mathrm{sp}}\) showed stable negative joint correlations under the ERP features and waveform PCA, indicating that, for the same waveform, stronger sensory input can be compensated by weaker superficial gain. Correlations among some derived targets can also be explained by shared raw parameters or by propagation of this compensation structure. Raw-derived distributions were generally self-consistent, but the active dimensionality and directions changed with summary and seed, and no derived direction achieved stable, structurally specific alignment. The Active track should therefore remain exploratory.

The principal limitation of this experiment is that the current model is a fixed five-node CMC-inspired auditory network rather than the complete SPM DCM-CMC; regional connections, conduction delays, and virtual-sensor weights were all fixed. Averaging Fz, FCz, and Cz improved the signal-to-noise ratio but removed spatial information, and no ground-truth dynamical parameters were available in the real data. Future work may incorporate multichannel observation mappings and participant-level priors and allow some feedforward and feedback connections to vary, thereby assessing whether local gains and network connections remain distinguishable.

Overall, the CMC/MMN experiment provided more favourable evidence of observation fit than the Epileptor/iEEG experiment and supported conditional recovery of \(g_{\mathrm{sp}}\), \(g_{\mathrm{ii}}\), \(t_{\mathrm{E/I}}\), and active 1 within the simulator. At the same time, the framework identified insufficient information for \(g_{\mathrm{dp}}\), summary loss in the CNN--LSTM, and joint compensation between \(g_{\mathrm{ss}}\) and \(g_{\mathrm{sp}}\). The results therefore support limited parameter information and interpretive boundaries under the specified CMC configuration.

\hypertarget{sec-6}{%
\section{6. Conclusion}\label{sec-6}}

We propose a hierarchical validity-audit framework for simulation-based inference with neural mass models. Before reporting dynamical parameters or derived interpretations, the framework sequentially addresses three questions: 1. Can the candidate simulator--prior--observation--summary configuration cover the data to be analysed? 2. Under this configuration, which individual targets contain recoverable information, and does the current summary cause target-specific information loss? 3. When multiple recoverable targets are inverted jointly, do they exhibit synchronous compensation or cross-track interpretive inconsistency? The framework retains three parallel tracks, Raw, Derived, and Active, and provides continuous and consistency evidence.

Step 1 separately performs local-support and posterior-predictive audits in the summary space and preregistered waveform diagnostic space. Automatic observation-layer calibration estimates and freezes only nuisance factors such as gain, noise, and measurement scale; it does not force improved coverage by changing the morphology of the dynamical response. Thus, when mismatch remains in the waveform diagnostic space, the framework retains this risk rather than allowing superficial proximity in a high-dimensional summary to mask model insufficiency.

Step 2 jointly reports proper-score gain relative to the prior and single-value recovery capability for raw parameters, predefined coordinates, and automatically discovered active coordinates. An additional summary-loss branch compares posterior performance among the summary, the summary plus a null-waveform control, and the summary plus independent waveform descriptors. This comparison determines whether target-recovery failure primarily arises from insufficient information in the observations or from inadequate retention of existing waveform information by the current summary. The experiments indicate that the input-coverage capability of a summary is not equivalent to its parameter-recovery capability and that there is no universally valid performance ranking among the dynamical features, waveform PCA, and CNN--LSTM.

Step 3 extends single-target recoverability to the joint interpretability of multiple targets. Through derived push-forward of the raw posterior, the direct derived posterior, active--derived direction alignment, and local--global active consistency, the framework checks whether different inversion paths yield self-consistent interpretations. This step can identify posterior trade-offs formed by two parameters that are individually recoverable and can also prevent data-sensitive directions from being equated directly with predefined physiological mechanisms.

The two real-data applications demonstrate the framework's capacity to produce graded conclusions. In the SOZ-local iEEG experiment, the single-source Epileptor configuration showed systematic mismatch in the core waveform diagnostic space shared by all three summaries. Although \(I_2\), \(t_{\mathrm{F/S}}\), and active 1 showed strong evidence of recovery within the simulator, the Step 1 results prevented these targets from being interpreted as real patient mechanisms. This case illustrates that good simulation-internal recovery and posterior calibration cannot compensate for inadequate fit of the generative model to real observations.

In contrast, the fixed five-node auditory CMC configuration showed better conditional fit to the ERP CORE MMN data. Waveform PCA and the preregistered ERP features retained strong information for \(g_{\mathrm{sp}}\), \(g_{\mathrm{ii}}\), \(t_{\mathrm{E/I}}\), and active 1, whereas \(g_{\mathrm{dp}}\) was near the prior level. The joint posterior revealed a stable substitution relationship between \(g_{\mathrm{ss}}\) and \(g_{\mathrm{sp}}\), indicating that parameters recoverable in isolation may still be independently uninterpretable in joint inversion. The Raw-derived paths were generally self-consistent, but the active dimensionality and directions were summary-dependent, and none formed stable, structurally specific alignment with a predefined derived direction.

Overall, this study provides a hierarchical failure-localization procedure with evidence of controlled error rates. The framework explicitly distinguishes three different levels: no detected observation mismatch, simulation-internal recoverability, and eligibility for real-mechanism interpretation. It thereby constrains overinterpretation in NMM-SBI in which good parameter recovery is used to infer physiological mechanisms directly. Future work may extend the framework to other forms of dynamical data, including hierarchical priors and multinode dynamics.

\hypertarget{sec-appendix}{%
\section{Appendices}\label{sec-appendix}}

\setcounter{figure}{0}
\setcounter{table}{0}
\renewcommand{\theHfigure}{appendix.\arabic{figure}}
\renewcommand{\theHtable}{appendix.\arabic{table}}
\captionsetup[figure]{name=Appendix Fig.}
\captionsetup[table]{name=Appendix Table}

\hypertarget{app-1}{%
\subsection{1 Epileptor dynamics and observation mapping}\label{app-1}}

\hypertarget{app-1-1}{%
\subsubsection{1.1 Epileptor dynamical model}\label{app-1-1}}

Epileptor is a phenomenological neural dynamical model that describes transitions between seizure states. The model contains five core state variables, denoted \(x_1, y_1, z, x_2, y_2\), and an auxiliary low-pass state \(g\) representing the historical activity of the first subsystem. Here, \(x_1, y_1\) constitute the first fast subsystem and primarily represent fast ictal activity; \(z\) is a slow variable that regulates transitions between ictal and non-ictal states; \(x_2, y_2\) constitute the second subsystem and primarily describe activity at intermediate timescales, such as spike-and-wave discharges; and \(g\) represents the low-pass historical effect transmitted from the first fast subsystem to the second subsystem. The single-source, time-rescaled implementation used in this study is:\[\begin{aligned}\dot{x}_1&=\frac{s_f}{\tau_1}\left[y_1-f_1(x_1,x_2,z)-z+I_1\right]+\sigma_{\mathrm{dyn}}\xi_1(t) \\
\dot{y}_1&=\frac{s_f}{\tau_1}\left[y_0-5x_1^2-y_1\right] \\
\dot{z}&=\frac{4(x_1-x_0)-z}{\tau_0} \\
\dot{x}_2&=s_f\left[-y_2+x_2-x_2^3+I_2+b_g g-0.3(z-3.5)\right]+\frac{\sigma_{\mathrm{dyn}}}{2}\xi_2(t) \\
\dot{y}_2&=\frac{s_f}{\tau_2}\left[-y_2+f_2(x_2)\right] \\
\dot{g}&=-s_f\lambda_g(g-c_gx_1) \end{aligned} \tag{1}\label{eq:app-1}\]where \(s_f=12 \mathrm{s}^{-1}\) is the fixed timescale-conversion factor for the fast dynamics and maps the dimensionless fast timescale of the classical model to the current physical observation window; the slow variable \(z\) is controlled separately by the respecified \(\tau_0\). The model fixes \(y_0=1, \tau_1=1, \tau_2=10, b_g=2.0, \lambda_g=0.01, c_g=0.10\). \(\xi_1(t)\) and \(\xi_2(t)\) are mutually independent Gaussian white-noise processes, \(\sigma_{\mathrm{dyn}}\) controls the intensity of dynamical perturbations, \(I_1, I_2\) denote external input terms, and \(\tau_0, \tau_1, \tau_2\) control the timescales of the slow variable and the first and second subsystems. The functions \(f_1\) and \(f_2\) are piecewise nonlinear functions, where\[\begin{aligned}f_1(x_1, x_2, z) =\begin{cases} x_1^3 - 3x_1^2, & x_1 < 0 \\ \left[x_2 - 0.6(z - 4)^2\right]x_1, & x_1\ge 0 \end{cases} \\ f_2(x_2) = 
\begin{cases} 
0, & x_2 < -0.25 \\ 
6(x_2 + 0.25), & x_2 \ge -0.25 
\end{cases}\end{aligned}\tag{2}\label{eq:app-2}\]They generate nonlinear transitions between different dynamical states. The initial state of the slow variable is generated according to \(z(0)=\operatorname{clip}\left(I_1+0.80+\eta_z,\ 0.5,\ 7.5\right),\eta_z\sim\mathcal N(0,0.04^2)\). The remaining fast variables are initialized near the negative stable branch under the corresponding parameter condition, with \(g(0)=0.1x_1(0)\). The model is then integrated autonomously.

\hypertarget{app-1-2}{%
\subsubsection{1.2 Epileptor observation mapping}\label{app-1-2}}

In this study, we used a single-source setting without introducing multinode brain-network connections or interregional propagation terms. This choice restricted the focus to parameter inversion of local seizure dynamics. Model output was treated as local epileptic source activity. Its observed signal was represented by a linear combination of the state variables and was then mapped through the observation model to an iEEG signal.

In addition to the core dynamical parameters, we introduced the spike-wave readout gain \(G_{\mathrm{SW}}\) at the observation layer to control the contribution of the second-subsystem variable \(x_2\) to the simulated iEEG source signal. This observation-readout parameter maps the five-state Epileptor latent variables to an SOZ-local iEEG-like waveform. We linearly combined the first fast variable \(x_1(t)\) and the second-subsystem variable \(x_2(t)\) representing spike-and-wave activity to obtain the latent source signal:\[r(t)=w_{x_1}x_1(t)+G_{\mathrm{SW}}x_2(t) \tag{3}\label{eq:app-3}\]where \(w_{x_1}\) specifies the readout polarity of the first fast variable, and \(G_{\mathrm{SW}}\) controls the fixed observed contribution of the second spike-and-wave subsystem. The latent source signal is mapped through a stationary observation model to an unprocessed simulated iEEG signal:\[\widetilde y(t)=\alpha r(t)+\sigma_c b_c(t)+\sigma_\theta b_\theta(t)+\sigma_b b_b(t) +\sigma_{\mathrm{obs}}\varepsilon(t) \tag{4}\label{eq:app-4}\]where \(\alpha\) is the source gain, \(b_c(t)\) is coloured background activity, \(b_\theta(t)\) is a 4--7 Hz background rhythm, \(b_b(t)\) is stationary 8--35 Hz broadband background activity, and \(\varepsilon(t)\) is independent observation noise. None of these observation terms depends on seizure state or absolute time. Source gain and background intensity are estimated only as bounded observation nuisance parameters in the observation calibration set and are then frozen and shared across all summaries. Simulated iEEG signals are subsequently compared using the same time window, sampling rate, and preprocessing procedure to ensure that model-generated and observed data are comparable in their statistical features.

\hypertarget{app-2}{%
\subsection{2 CMC dynamics and observation mapping}\label{app-2}}

\hypertarget{app-2-1}{%
\subsubsection{2.1 CMC dynamical model}\label{app-2-1}}

The Canonical Microcircuit model is commonly used to describe hierarchical interactions among different neural populations within a cortical column. In the conventional DCM-CMC framework, cortical microcircuits generally contain multiple neural populations that jointly shape ERP/ERF responses through feedforward, feedback, and local inhibitory connections. Rather than reproducing the complete SPM DCM-CMC model parameter by parameter, we constructed a fixed five-node auditory network based on CMC local dynamics. The network included bilateral primary auditory cortices (A1), bilateral superior temporal gyri (STG), and the right inferior frontal gyrus (IFG). The basic topology, nominal connection weights, and nominal conduction delays of the network were fixed before inversion, but each simulated participant had a bounded common delay scale. Prespecified modulations of feedforward, feedback, inhibitory gain, and superficial-pyramidal gain were introduced under the deviant condition. These participant-level and condition-level variations were marginalized as nuisance random variables and were not inversion targets. Thus, we inverted local parameters shared by the five nodes.

Each node \(j\) contains an input-layer or spiny-stellate-cell population \(s_{\mathrm{ss},j}\), superficial pyramidal-cell population \(s_{\mathrm{sp},j}\), inhibitory-interneuron population \(s_{\mathrm{ii},j}\), deep pyramidal-cell population \(s_{\mathrm{dp},j}\), and an additional slow state \(s_{\mathrm{slow},j}\). For stimulus condition \(c\in\{S,D\}\), the four principal neural populations follow common second-order synaptic dynamics:\[\tau_q^2\ddot{s}_{q,j}^{(c)} + 2\tau_q\dot{s}_{q,j}^{(c)} + s_{q,j}^{(c)} = I_{q,j}^{(c)} + \eta_{q,j}^{(c)}, q \in \mathcal Q = \{\mathrm{ss}, \mathrm{sp}, \mathrm{ii}, \mathrm{dp}\} \tag{5}\label{eq:app-5}\]where \(I_{q,j}^{(c)}\) denotes the instantaneous effective drive received by population \(q\) in node \(j\) under condition \(c\), and \(\eta_{q,j}^{(c)}\) denotes state-driving noise. The effective inputs to the four populations are defined as\[\begin{aligned} I_{\mathrm{ss},j}^{(c)} ={}& I_{\mathrm{in},j}^{(c)} + 0.30g_{\mathrm B}\phi(s_{\mathrm{dp},j}) + 0.20I_{\mathrm{fb},j}^{(c)} - 0.22m_{\mathrm I}^{(c)}g_{\mathrm{ii}} \phi(s_{\mathrm{ii},j}) \\ 
I_{\mathrm{sp},j}^{(c)} ={}& m_{\mathrm{SP}}^{(c)}g_{\mathrm{sp}} \left[ 0.78\phi(s_{\mathrm{ss},j}) + 0.16I_{\mathrm{in},j}^{(c)} + 0.28I_{\mathrm{fb},j}^{(c)} \right] - 0.42m_{\mathrm I}^{(c)}g_{\mathrm{ii}} \phi(s_{\mathrm{ii},j}) \\ 
I_{\mathrm{ii},j}^{(c)} ={}& m_{\mathrm I}^{(c)}g_{\mathrm{ii}} \left[ 0.55\phi(s_{\mathrm{ss},j}) + 0.45\phi(s_{\mathrm{sp},j}) \right] \\ 
I_{\mathrm{dp},j}^{(c)} ={}& g_{\mathrm{dp}} \left[ 0.62g_{\mathrm F}\phi(s_{\mathrm{sp},j}) - 0.18\phi(s_{\mathrm{ii},j}) + 0.22I_{\mathrm{fb},j}^{(c)} \right] \end{aligned} \tag{6}\label{eq:app-6}\]Here, \(I_{\mathrm{ss},j}^{(c)}\) describes the integration by the input-layer population of direct thalamic input, delayed feedforward and lateral inputs, local deep feedback, interregional feedback, and local inhibition. \(I_{\mathrm{sp},j}^{(c)}\) describes the integration by the superficial pyramidal population of input-layer, sensory-network, and feedback inputs, with negative modulation by local inhibitory interneurons. \(I_{\mathrm{ii},j}^{(c)}\) describes recruitment of the inhibitory population by the input-layer and superficial pyramidal populations. \(I_{\mathrm{dp},j}^{(c)}\) describes the integration by the deep pyramidal population of superficial pyramidal output, local inhibition, and interregional feedback. Here, \(I_{\mathrm{in},j}^{(c)}\) includes direct thalamic input, delayed feedforward input, and lateral input; \(I_{\mathrm{fb},j}^{(c)}\) denotes delayed interregional feedback; \(g_{\mathrm F}\) and \(g_{\mathrm B}\) are the fixed baseline feedforward and feedback gains, respectively; and \(m_{\mathrm I}^{(c)}\) and \(m_{\mathrm{SP}}^{(c)}\) denote condition-dependent modulations of inhibitory gain and superficial-pyramidal gain, respectively.

State-driving noise is scaled according to the effective number of ERP trials under the corresponding condition as \(\eta_{q,j}^{(c)} = \frac{\sigma_q}{\sqrt{N_c}} \epsilon_{q,j}^{(c)}\). Here, \(\epsilon_{q,j}^{(c)}\) denotes mutually independent standard Gaussian perturbations across nodes, populations, and discrete integration times, and \(N_S=570\) and \(N_D=195\). The prescaling noise coefficients for the four population types are \((\sigma_{\mathrm{ss}}, \sigma_{\mathrm{sp}}, \sigma_{\mathrm{ii}}, \sigma_{\mathrm{dp}}) = (0.035, 0.040, 0.035, 0.030)\).

Neural-population states are transformed into effective outputs through a bounded nonlinear function:\[\phi(x)=\frac{2}{1+\exp(-1.4x)}-1\tag{7}\label{eq:app-7}\]In addition, for computational simplicity, the input-layer and superficial pyramidal populations share the same fast excitatory timescale \(\tau_{\mathrm{ss}}=\tau_{\mathrm{sp}}=\tau_e=12\ \mathrm{ms}\). The timescale of the inhibitory interneurons is the inversion parameter \(\tau_{\mathrm{ii}}=\tau_i\). The timescale of the deep pyramidal population is determined by its gain parameter \(g_{\mathrm{dp}}\) as \(\tau_{\mathrm{dp}}=\left\{55+15\tanh\!\left[0.7\left(g_{\mathrm{dp}}-1\right)\right]\right\}\mathrm{ms}\). The additional slow state follows \(\tau_{\mathrm{slow}}\dot{s}_{\mathrm{slow},j}=-s_{\mathrm{slow},j}+0.40s_{\mathrm{dp},j}+0.25s_{\mathrm{sp},j},\tau_{\mathrm{slow}}=180\ \mathrm{ms}\) and low-pass tracks a weighted combination of superficial and deep pyramidal activity with a \(180\ \mathrm{ms}\) timescale. Because it does not feed back into the effective inputs of the four principal populations, it primarily serves as a slowly derived state for constructing the low-frequency response rather than as an adaptive variable that actively changes local circuit dynamics.

Feedforward, feedback, and within-level lateral connections were defined among the five network nodes. Feedforward pathways included left A1 to left STG, right A1 to right STG, left STG to right IFG, and right STG to right IFG. Feedback pathways connected the opposite hierarchical directions, and lateral pathways connected bilateral A1 and bilateral STG. The specific connections are shown in Appendix \hyperref[fig:appendix1]{Fig. 1}:

\begin{figure}[htbp]
\hypertarget{fig:appendix1}{%
\centering
\includegraphics[keepaspectratio,width=1\textwidth,height=0.35\textheight]{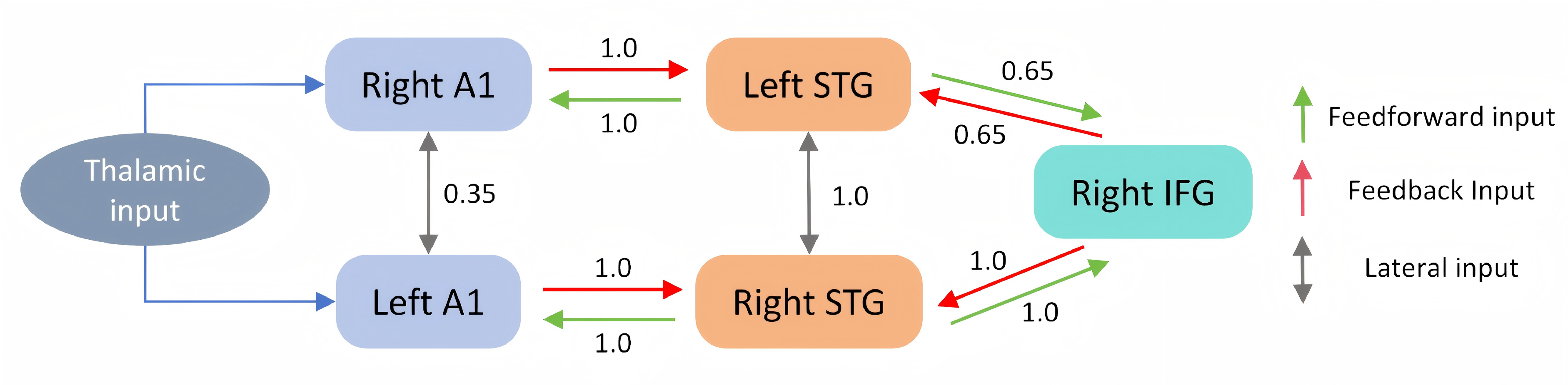}
\caption{Connection relationship of five nodes: left A1, left STG, right A1, right STG, and right IFG.}\label{fig:appendix1}
}
\end{figure}

In this study, let \(c \in \{S,D\}\) denote the standard and deviant conditions, respectively. The total input \(I_{\mathrm{in},j}\) received by each node \(j\) includes direct thalamic input \(I_{\mathrm{thal}}\), delayed feedforward input \(I_{\mathrm{ff}}\), and within-level lateral input \(I_{\mathrm{lat},j}\):\[\begin{aligned} 
I_{\mathrm{in},j}^{(c)}(t) &= I_{\mathrm{thal},j}^{(c)}(t) + I_{\mathrm{ff},j}^{(c)}(t) + I_{\mathrm{lat},j}^{(c)}(t)\\
I_{\mathrm{thal},j}^{(c)}(t) &= q_j g_{\mathrm{ss}} u_0(t)\\ 
I_{\mathrm{ff},j}^{(c)}(t) &= 0.55 g_{\mathrm F} m_{\mathrm F}^{(c)} \sum_{k=1}^{5} A_{jk}^{\mathrm{ff}} \phi\left[s_{\mathrm{sp},k}^{(c)}(t-d_{\mathrm F})\right]\\
I_{\mathrm{lat},j}^{(c)}(t) &= 0.10 \sum_{k=1}^{5} A_{jk}^{\mathrm{lat}} \phi\left[s_{\mathrm{sp},k}^{(c)}(t-d_{\mathrm L})\right] 
\end{aligned}\tag{8}\label{eq:app-8}\]The delayed feedback input from higher-order auditory regions is defined as\[I_{\mathrm{fb},j}^{(c)}(t) = 0.45 g_{\mathrm B} m_{\mathrm B}^{(c)} \sum_{k=1}^{5} A_{jk}^{\mathrm{fb}} \phi\left[s_{\mathrm{dp},k}^{(c)}(t-d_{\mathrm B})\right]. \tag{9}\label{eq:app-9}\]Thalamic stimulation acts directly only on bilateral A1; the node-input mask is therefore defined as \(\mathbf q=(q_1,q_2,q_3,q_4,q_5)^{\mathsf T}=(1,1,0,0,0)^{\mathsf T}\). Baseline connection gains are fixed at \(g_{\mathrm F}=0.85\) and \(g_{\mathrm B}=0.75\). All modulation coefficients are 1 under the standard condition. Under the deviant condition, the nominal values of feedforward, feedback, inhibitory-gain, and superficial-pyramidal-gain modulation are \(m_{\mathrm F}^{(D)}=1.15, m_{\mathrm B}^{(D)}=1.15, m_{\mathrm I}^{(D)}=1.15, m_{\mathrm{SP}}^{(D)}=0.89\), respectively, with bounded participant-level variation added around these nominal values. The feedforward and feedback condition multipliers are each applied only once in the corresponding interregional input.

The matrix element \(A_{jk}\) denotes a connection from node \(k\) to node \(j\). In the order left A1, right A1, left STG, right STG, and right IFG, the connection matrices are:\[A^{\mathrm{ff}}=\begin{pmatrix}
0&0&0&0&0\\
0&0&0&0&0\\
1&0&0&0&0\\
0&1&0&0&0\\
0&0&0.65&1&0
\end{pmatrix},
A^{\mathrm{fb}}=\begin{pmatrix}
0&0&1&0&0\\
0&0&0&1&0\\
0&0&0&0&0.65\\
0&0&0&0&1\\
0&0&0&0&0
\end{pmatrix},
A^{\mathrm{lat}}=\begin{pmatrix}
0&0.35&0&0&0\\
0.35&0&0&0&0\\
0&0&0&1&0\\
0&0&1&0&0\\
0&0&0&0&0
\end{pmatrix}\tag{10}\label{eq:app-10}\]The feedforward, feedback, and lateral delays are \(24\), \(48\), and \(12\ \mathrm{ms}\), respectively. To represent between-participant differences in conduction time, all three delays are multiplied by the participant-level scale \(\rho_d\): \((d_{\mathrm F}, d_{\mathrm B}, d_{\mathrm L}) = \rho_d (24, 48, 12)\ \mathrm{ms}, \rho_d \in [0.82, 1.22]\). Thalamic input is modelled as a causal Gaussian pulse peaking at approximately \(65\ \mathrm{ms}\) after stimulus onset:\[u_0(t)=A_{\mathrm{th}}\exp\!\left[-\frac{(t-\mu_{\mathrm{th}})^2}{2\sigma_{\mathrm{th}}^2}\right]H(t),\mu_{\mathrm{th}}\approx65\ \mathrm{ms}\tag{11}\label{eq:app-11}\]where \(A_{\mathrm{th}}\) is the thalamic-input amplitude, \(\mu_{\mathrm{th}}\) is the peak time, \(\sigma_{\mathrm{th}}\) controls pulse width, and \(H(t)\) is the Heaviside function. The complete model comprises external thalamic input, fixed five-node feedforward--feedback connections, local four-population dynamics, and the additional slow state.

\hypertarget{app-2-2}{%
\subsubsection{2.2 CMC observation mapping}\label{app-2-2}}

We used a fixed five-node auditory network. The observation mapping comprised three steps: first, neural-population states within each node were transformed into node-level source activity; second, the five nodes were mixed through virtual frontocentral sensor weights; and third, bounded EEG observation nuisance factors were added to construct the standard ERP and MMN difference wave. For condition \(c \in \{S,D\}\), local source activity at node \(j\) is defined as:\[r_j^{(c)}(t) = s_{\mathrm{sp},j}^{(c)}(t) - 0.90s_{\mathrm{dp},j}^{(c)}(t) + 0.45s_{\mathrm{slow},j}^{(c)}(t). \tag{12}\label{eq:app-12}\]Here, superficial and deep pyramidal activity enter the observation equation with opposite signs to approximate their opposing contributions to the laminar current dipole, and the slow state supplements responses at lower frequencies and longer timescales. The input-layer state \(s_{\mathrm{ss},j}\) does not directly enter the observation equation but still indirectly affects the final ERP through local dynamics. The baseline frontocentral weights of the five nodes are \(\boldsymbol{\ell}_0 = (0.038,\ 0.038,\ 0.377,\ 0.377,\ 0.170)^{\mathsf T}\), ordered as left A1, right A1, left STG, right STG, and right IFG. These baseline weights make the virtual sensor receive primarily contributions from bilateral STG and right IFG while retaining smaller bilateral A1 components.

To represent between-participant differences in source orientation and sensor projection, the weights of each simulated participant \(r\) are subjected to bounded random perturbations around their baseline values:\[\begin{aligned} \widetilde{\ell}_{rj} &= \ell_{0j}\exp(\zeta_{rj}), \zeta_{rj}\sim\mathcal N(0,0.18^2), \\ \ell_{rj} &= \frac{\widetilde{\ell}_{rj}}{\sum_{k=1}^{5}\vert{}\widetilde{\ell}_{rk}\vert{}} \end{aligned} \tag{13}\label{eq:app-13}\]The standard and deviant responses for the same participant share the same projection weights. The corresponding frontocentral source signal is:\[r_{\mathrm{FC},r}^{(c)}(t) = \sum_{j=1}^{5}\ell_{rj}r_j^{(c)}(t) \tag{14}\label{eq:app-14}\]The leadfield perturbation is part of a prespecified simulation nuisance distribution. At the sensor observation layer, the standard and deviant signals are expressed as\[\begin{aligned} \widetilde y_{S,r}(t) &= \alpha r_{\mathrm{FC},r}^{(S)}(t-\delta_r) + b_{\mathrm{shared},r}(t) + b_{S,r}(t) + \varepsilon_{S,r}(t), \\ \widetilde y_{D,r}(t) &= \alpha(1+\gamma_r)r_{\mathrm{FC},r}^{(D)}(t-\delta_r) + b_{\mathrm{shared},r}(t) + b_{D,r}(t) + \varepsilon_{D,r}(t). \end{aligned} \tag{15}\label{eq:app-15}\]where \(\alpha\) is the source gain from dimensionless network state to ERP amplitude; \(\delta_r\) is temporal jitter shared by the standard and deviant; \(\gamma_r\) represents participant-level amplitude fluctuation in the deviant response; \(b_{\mathrm{shared},r}\) is the broadband background and alpha rhythm shared by the two conditions; \(b_{S,r}\) and \(b_{D,r}\) are condition-specific residual backgrounds; and \(\varepsilon_{S,r}\) and \(\varepsilon_{D,r}\) are independent measurement noise. Standard residual noise is scaled by \(\sqrt{N_D/N_S}\) according to the effective trial counts \(N_S=570\) and \(N_D=195\), reflecting the noise reduction associated with averaging more standard-stimulus trials. The observation calibration parameters are written as \(\boldsymbol\psi = \left( \alpha, \sigma_\delta, \sigma_b, \sigma_\alpha, \rho_{\mathrm{res}}, \sigma_\varepsilon, \sigma_\gamma, \sigma_{\mathrm{smooth}} \right)\) and are estimated using only participants in the observed-calibration split. Each analysis replicate selects only one shared, target-neutral observation mapping \(\widehat{\boldsymbol\psi}\), which is then frozen and used jointly for all summaries. Observed-test participants do not participate in observation-parameter selection.

Let \(\mathcal P_{\mathrm{sim}}\) denote the preprocessing operator for simulated ERPs, including Gaussian smoothing, fourth-order zero-phase \(0.1\)--\(20\ \mathrm{Hz}\) Butterworth filtering, and mean baseline correction over the prestimulus interval \(-200\)--\(0\ \mathrm{ms}\). We finally construct:\[\begin{aligned} S_r(t) &= \mathcal P_{\mathrm{sim}}\left[\widetilde y_{S,r}(t)\right], D_r(t) = \mathcal P_{\mathrm{sim}}\left[\widetilde y_{D,r}(t)\right], \\ \Delta_r(t) &= D_r(t) - S_r(t),\mathbf Y_r = \begin{bmatrix} S_r \\ \Delta_r \end{bmatrix} \end{aligned} \tag{16}\label{eq:app-16}\]The two channels entering SBI are the standard ERP and the deviant-minus-standard MMN difference wave.

\hypertarget{app-3}{%
\subsection{3 Dimensionality selection for Active coordinates}\label{app-3}}

For a given summary, the free parameters and summary are first standardized using the means and standard deviations of the simulation-training split \(\mathcal D_{\mathrm{train}}^{\mathrm{post}}\), respectively:\[\mathbf z_i= \frac{\boldsymbol\theta_i-\boldsymbol\mu_\theta} {\boldsymbol\sigma_\theta+\epsilon}, \qquad \widetilde{\mathbf R}_i= \frac{\mathbf R_i-\boldsymbol\mu_R} {\boldsymbol\sigma_R+\epsilon}\tag{17}\label{eq:app-17}\]A total of \(N\) local centres are selected in the standardized parameter space. For the \(j\)th centre, after controlling for experimental condition \(c\), a local ridge regression is fitted using its parameter neighbourhood \(\mathcal N_j\):\[(\widehat{\mathbf a}_j,\widehat{\mathbf J}_j) = \arg\min_{\mathbf a,\mathbf J} \sum_{i\in\mathcal N_j} \left\Vert{} \widetilde{\mathbf R}_i-\mathbf a-\mathbf J\mathbf z_i \right\Vert{}_2^2 + \alpha\Vert{}\mathbf J\Vert{}_F^2\tag{18}\label{eq:app-18}\]Here, \(\widehat{\mathbf J}_j\) is the empirical local Jacobian of the summary with respect to the standardized free parameters. The local sensitivity matrix is then defined as \(\widehat{\mathbf F}_j = \widehat{\mathbf J}_j^\top \widehat{\mathbf J}_j\) and averaged over all local centres as \(\widehat{\mathbf F}_{\mathrm{global}} = \frac{1}{N_{\mathrm{center}}} \sum_{j=1}^{N_{\mathrm{center}}} \widehat{\mathbf F}_j\). This matrix describes the average gradient sensitivity under the current simulator--prior--summary configuration. After symmetrization, its nonnegative eigenvalues are arranged in descending order:\[\lambda_1\geq\lambda_2\geq\cdots\geq\lambda_d\geq0\tag{19}\label{eq:app-19}\]The corresponding orthogonal eigenvectors are denoted \(\mathbf V= [\mathbf v_1,\ldots,\mathbf v_d]\), and the \(k\)th candidate active target is defined as \(t_k^{\mathrm{active}} = \mathbf v_k^\top \mathbf z(\boldsymbol\theta)\). For \(d>1\), the framework sequentially examines \(L=1,\ldots,L_{\max}\), where \(L_{\max}=\min(L_{\mathrm{user}},d-1)\) and \(L_{\mathrm{user}}\) is the maximum number of active targets permitted by the user. \(L=d\) is excluded because the complete parameter space no longer constitutes low-dimensional reduction, and its cumulative sensitivity and subspace overlap degenerate to trivial results. When \(d=1\), the only candidate is \(L=1\).

For each candidate dimensionality, the cumulative sensitivity proportion is calculated as\[C_L= \frac{\sum_{j=1}^{L}\lambda_j} {\sum_{j=1}^{d}\lambda_j}\tag{20}\label{eq:app-20}\]and the relative eigengap normalized by the largest eigenvalue is calculated as:\[G_L= \frac{\lambda_L-\lambda_{L+1}} {\max(\lambda_1,\epsilon)}\tag{21}\label{eq:app-21}\]\(C_L\) measures the proportion of total sensitivity explained by the first \(L\) directions, and \(G_L\) measures the degree of spectral separation between the \(L\)th direction and the following direction. If \(\operatorname{tr} (\widehat{\mathbf F}_{\mathrm{global}}) \leq\epsilon_F\), the current summary is considered to contain no detectable stable sensitivity structure, and the framework returns no-detectable-active-structure rather than forcing the generation of an active target.

In addition, to evaluate the sensitivity of the active subspace to the sampling of local centres, the framework resamples the set of local matrices \(\{\widehat{\mathbf F}_j\}_{j=1}^{N_{\mathrm{center}}}\) with replacement in \(B_{\mathrm{AS}}\) bootstrap replicates. We used \(B_{\mathrm{AS}}=100\), which is user-adjustable within the framework. The \(b\)th resample yields \(\widehat{\mathbf F}_{\mathrm{global}}^{(b)} = \frac{1}{N_{\mathrm{center}}} \sum_{j\in\mathcal I_b} \widehat{\mathbf F}_j\), and its first \(L\) eigenvectors constitute the bootstrap active basis \(\mathbf V_L^{(b)}\). The basis corresponding to the complete training set \(\mathcal D_{\mathrm{train}}^{\mathrm{post}}\) is denoted \(\mathbf V_L\). Their subspace overlap is defined as\[S_L^{(b)} = \frac{1}{L} \left\Vert{} \mathbf V_L^\top\mathbf V_L^{(b)} \right\Vert{}_F^2 = \frac{1}{L}\sum_{r=1}^{L}\sigma_r^2\tag{22}\label{eq:app-22}\]where \(\sigma_r\) is a singular value of \(\mathbf V_L^\top\mathbf V_L^{(b)}\). The mean bootstrap stability is\[\overline S_L = \frac{1}{B_{\mathrm{AS}}} \sum_{b=1}^{B_{\mathrm{AS}}} S_L^{(b)}\tag{23}\label{eq:app-23}\]This metric ranges within \([0,1]\). A value closer to 1 indicates that local resampling within the same posterior-training split produces a more stable active subspace.

Finally, the primary analysis uses the prespecified reference conditions \(C_L\geq0.80, G_L\geq0.05, \overline S_L\geq0.80\). If multiple dimensionalities simultaneously satisfy all three conditions, the smallest \(L\) is selected to obtain the lowest-dimensional subspace that explains the principal sensitivity. If no candidate dimensionality satisfies all conditions but some candidates satisfy both the eigengap and bootstrap-stability conditions, the dimensionality with the largest \(G_L\) is selected; if there is a tie, the smaller \(L\) is selected. If there is still no candidate, the dimensionality that maximizes \(Q_L=\overline S_L\cdot\max(G_L,0)\) is selected.

For each summary, each random seed independently estimates the sensitivity matrix and selects an active dimensionality \(L_s\), with orthogonal basis:\[\mathbf V_s = [\mathbf v_{s,1},\ldots,\mathbf v_{s,L_s}] \in \mathbb R^{d\times L_s} \tag{24}\label{eq:app-24}\]The active targets for each random seed are independently defined within the framework as \(t_{k,s}^{\mathrm{active}} = \mathbf v_{s,k}^{\top}\mathbf z(\boldsymbol\theta), k=1,\ldots,L_s\). Here, \(\mathrm{active}_1\) denotes the first sensitivity direction independently discovered by each seed; \(\mathrm{active}_2\) summarizes only seeds satisfying \(L_s \ge 2\) and reports the number of valid seeds. The framework records the selection frequency of each dimensionality as \(n_l = \sum_{s=1}^{S}\mathbf 1(L_s=l)\) and calculates the pairwise dimensionality-matching proportion for any two seeds:\[P_{\mathrm{dim}} = \frac{2}{S(S-1)} \sum_{s<s'} \mathbf 1(L_s=L_{s'}) \tag{25}\label{eq:app-25}\]Second, for any pair of random seeds \((s,s')\), Active subspace overlap is defined as:\[O_{ss'} = \frac{2\left\Vert{} \mathbf V_s^\top\mathbf V_{s'} \right\Vert{}_F^2}{L_s+L_{s'}} \in [0,1] \tag{26}\label{eq:app-26}\]\(O_{ss'}\) applies uniformly to Active subspaces of the same or different dimensionalities and is strictly invariant to eigenvector sign flips, changes in ordering, or rotations within the subspace. A value closer to \(1\) indicates that the Active geometries captured by two random seeds are more consistent, whereas a value closer to \(0\) indicates orthogonal or divergent sensitivity directions. In particular, when \(L_s=L_{s'}=1\), this metric naturally reduces to the squared absolute cosine similarity between the two individual Active directions, \(O_{ss'} = \left\vert{} \mathbf v_s^\top\mathbf v_{s'} \right\vert{}^2\). The results of the two real-data experiments are shown in \hyperref[tab:appendix1]{Appendix Table 1}.

{\footnotesize
\begin{longtable}[]{@{}
  >{\centering\arraybackslash}p{(\columnwidth - 12\tabcolsep) * \real{0.1700}}
  >{\raggedright\arraybackslash}p{(\columnwidth - 12\tabcolsep) * \real{0.1600}}
  >{\raggedleft\arraybackslash}p{(\columnwidth - 12\tabcolsep) * \real{0.0800}}
  >{\raggedleft\arraybackslash}p{(\columnwidth - 12\tabcolsep) * \real{0.0800}}
  >{\raggedleft\arraybackslash}p{(\columnwidth - 12\tabcolsep) * \real{0.1300}}
  >{\raggedright\arraybackslash}p{(\columnwidth - 12\tabcolsep) * \real{0.1900}}
  >{\raggedright\arraybackslash}p{(\columnwidth - 12\tabcolsep) * \real{0.1900}}@{}}
\caption{Active dimensionality selection and subspace stability across seeds. Each summary includes 10 algorithmic random seeds and 45 seed pairs \label{tab:appendix1}}\tabularnewline
\toprule
Experiment & Summary & \(n_{L=1}\) & \(n_{L=2}\) & \(P_{\mathrm{dim}}\) & \(O_{\mathrm{ss}}\), mean \(\pm\) SD & \(O_{\mathrm{ss}}\), median {[}IQR{]} \\
\midrule
\endfirsthead
\toprule
Experiment & Summary & \(n_{L=1}\) & \(n_{L=2}\) & \(P_{\mathrm{dim}}\) & \(O_{\mathrm{ss}}\), mean \(\pm\) SD & \(O_{\mathrm{ss}}\), median {[}IQR{]} \\
\midrule
\endhead
\multirow[c]{3}{=}{Epileptor/iEEG} & CNN--LSTM & 10 & 0 & 1.000 & \(0.993\pm0.008\) & 0.995 {[}0.990, 0.998{]} \\
& Waveform PCA & 8 & 2 & 0.644 & \(0.685\pm0.171\) & 0.660 {[}0.590, 0.820{]} \\
& Dynamics features & 0 & 10 & 1.000 & \(0.979\pm0.025\) & 0.988 {[}0.973, 0.994{]} \\
\midrule
\multirow[c]{3}{=}{CMC/MMN} & CNN--LSTM & 6 & 4 & 0.467 & \(0.620\pm0.195\) & 0.591 {[}0.501, 0.763{]} \\
& Waveform PCA & 4 & 6 & 0.467 & \(0.377\pm0.194\) & 0.417 {[}0.264, 0.476{]} \\
& ERP features & 8 & 2 & 0.644 & \(0.783\pm0.134\) & 0.818 {[}0.644, 0.889{]} \\
\bottomrule
\end{longtable}
}

For Epileptor/iEEG, the CNN--LSTM and dynamical features stably selected \(L=1\) and \(L=2\), respectively, and both had across-seed subspace overlaps close to 1. Thus, the two summaries each discovered a highly stable sensitivity structure, but of different dimensionality. Waveform PCA had a relatively lower dimensionality-matching rate and subspace overlap. The Active structure in CMC/MMN was more dependent on the random seed overall: stability was highest for the ERP features, followed by the CNN--LSTM, whereas waveform PCA had the lowest dimensionality-matching rate and subspace overlap. Active targets under CMC waveform PCA should therefore remain exploratory.

\clearpage
\hypertarget{app-4}{%
\subsection{4 Hand-crafted feature table}\label{app-4}}

{\footnotesize
\begin{longtable}[]{@{}
  >{\centering\arraybackslash}p{(\columnwidth - 6\tabcolsep) * \real{0.1700}}
  >{\centering\arraybackslash}p{(\columnwidth - 6\tabcolsep) * \real{0.0600}}
  >{\raggedright\arraybackslash}p{(\columnwidth - 6\tabcolsep) * \real{0.2000}}
  >{\raggedright\arraybackslash}p{(\columnwidth - 6\tabcolsep) * \real{0.5700}}@{}}
\caption{Dynamical features used for Epileptor/iEEG and CMC/MMN-ERP; Epileptor used 15 dimensions and CMC used 39 dimensions \label{tab:appendix2}}\tabularnewline
\toprule
Experiment & Dim & Feature group & Features \\
\midrule
\endfirsthead
\toprule
Experiment & Dim & Feature group & Features \\
\midrule
\endhead
\multirow[c]{7}{*}[-2.5\baselineskip]{Epileptor/iEEG} & 1 & Baseline to ictal amplitude & log ratio of ictal RMS to baseline RMS \\
& 1 & Ictal peak timing & time fraction of the maximum absolute ictal peak \\
& 1 & Excess ictal envelope & time-normalized area of the ictal absolute envelope above the baseline absolute-amplitude level \\
& 3 & Ictal temporal thirds & early, middle, and late ictal log-RMS ratios relative to baseline RMS \\
& 2 & Boundary RMS contrasts & first-second ictal RMS minus baseline RMS; final-second ictal RMS minus preceding-second ictal RMS \\
& 4 & Low-frequency spectral shift & ictal delta fraction; ictal theta fraction; log ictal-to-baseline delta-fraction ratio; log ictal-to-baseline theta-fraction ratio \\
& 3 & Ictal autocorrelation & ictal ACF at 0.10 s, 0.50 s, and 1.00 s \\
\midrule
\multirow[c]{11}{*}[-4.5\baselineskip]{CMC/MMN-ERP} & 2 & Global MMN contrast & whole-epoch difference-wave mean; whole-epoch difference-wave RMS \\
& 3 & Poststimulus peak properties & maximum absolute difference-wave peak; signed difference-wave peak; poststimulus peak-time fraction \\
& 3 & Poststimulus temporal thirds & early, middle, and late difference-wave means \\
& 3 & N1-latency window, 70--130 ms & difference-wave mean, minimum, and maximum \\
& 3 & P2-latency window, 140--230 ms & difference-wave mean, minimum, and maximum \\
& 3 & MMN window, 120--250 ms & difference-wave mean, minimum, and maximum \\
& 3 & Late window, 250--400 ms & difference-wave mean, minimum, and maximum \\
& 3 & Difference-wave shape & poststimulus area; RMS first difference; zero-crossing rate \\
& 3 & Difference-wave autocorrelation & ACF at 20 ms, 50 ms, and 100 ms \\
& 10 & Difference-wave spectral features & log band power and band-power fraction in 0.5--4, 4--8, 8--13, and 13--30 Hz; spectral centroid; spectral entropy \\
& 3 & Deviant response level & early, middle, and late poststimulus deviant-ERP means \\
\bottomrule
\end{longtable}
}

\hypertarget{app-5}{%
\subsection{5 Dataset partitioning and parameter implementation}\label{app-5}}

To keep the main text focused on the statistical objects and interpretive boundaries, the implementation constants are listed here. The observation-layer loss is \(\mathcal L_{\mathrm{obs}}=\mathcal L_{\mathrm{median}}+0.20\mathcal L_{\mathrm{IQR}}+0.35\mathcal L_{\mathrm{tail}}\); each term is scale-standardized using the simulated IQR. The observed data are partitioned by participant at a 7:3 ratio into \(\mathcal D^{\mathrm{obs}}_{\mathrm{cal}}\) and \(\mathcal D^{\mathrm{obs}}_{\mathrm{test}}\), with no participant appearing in both sets.

Within each condition, Step 1 partitions the parent simulation library into \(\mathcal D_{\mathrm{ref}}^{\mathrm{sup}}\), \(\mathcal D_{\mathrm{cal}}^{\mathrm{sup}}\), and \(\mathcal D_{\mathrm{null}}^{\mathrm{sup}}\) (70\%:15\%:15\%). The Gaussian-kernel bandwidth for the Local predictive audit is the median neighbour distance. The batch test constructs \(B=2{,}000\) pseudo-observed batches from \(\mathcal D_{\mathrm{null}}^{\mathrm{sup}}\) with the same batch size and condition composition as \(\mathcal D_{\mathrm{obs,test}}\). Steps 2--3 repartition the same parent simulation library into \(\mathcal D_{\mathrm{train}}^{\mathrm{post}}\), \(\mathcal D_{\mathrm{val}}^{\mathrm{post}}\), \(\mathcal D_{\mathrm{cal}}^{\mathrm{post}}\), and \(\mathcal D_{\mathrm{test}}^{\mathrm{post}}\) (60\%:10\%:15\%:15\%). Thus, roles are mutually exclusive within each stage, but Step 1 and Steps 2--3 are not independent simulation replicates.

The raw-posterior \(E_{90}^{\mathrm{raw}}\) and randomized-PIT \(D_{\mathrm{SBC}}^{\mathrm{raw}}\) are both calculated directly on \(\mathcal D_{\mathrm{test}}^{\mathrm{post}}\). Summary-loss comparisons use a one-sided exact sign-flip randomization test within the same seed, with Holm correction applied within each target--summary combination to the Wave+--Original and Wave+--PermWave comparisons. The local--global active audit uses \(k=36\) posterior-training centres; the \(B_{\mathrm{AS}}=100\) bootstrap replicates and thresholds for Active dimensionality selection are provided in \hyperref[app-3]{Appendix 3}.

\begin{thebibliography}{69}

\bibitem{ref1} Nunez P L, Nunez M D, Srinivasan R. Multi-scale neural sources of EEG: genuine, equivalent, and representative. A tutorial review{[}J{]}. Brain Topography, 2019, 32(2): 193-214.

\bibitem{ref2} Chang E F. Towards large-scale, human-based, mesoscopic neurotechnologies{[}J{]}. Neuron, 2015, 86(1): 68-78.

\bibitem{ref3} Lu H Y, Lorenc E S, Zhu H, et al.~Multi-scale neural decoding and analysis{[}J{]}. Journal of neural engineering, 2021, 18(4): 045013.

\bibitem{ref4} Moran R J, Kiebel S J, Stephan K E, et al.~A neural mass model of spectral responses in electrophysiology{[}J{]}. NeuroImage, 2007, 37(3): 706-720.

\bibitem{ref5} Ahmadizadeh S, Karoly P J, Nešić D, et al.~Bifurcation analysis of two coupled Jansen-Rit neural mass models{[}J{]}. PloS one, 2018, 13(3): e0192842.

\bibitem{ref6} Douglas R J, Martin K A C, Whitteridge D. A canonical microcircuit for neocortex{[}J{]}. Neural computation, 1989, 1(4): 480-488.

\bibitem{ref7} Marreiros A C, Kiebel S J, Friston K J. A dynamic causal model study of neuronal population dynamics{[}J{]}. Neuroimage, 2010, 51(1): 91-101.

\bibitem{ref8} Moran R, Pinotsis D A, Friston K. Neural masses and fields in dynamic causal modeling{[}J{]}. Frontiers in computational neuroscience, 2013, 7: 57.

\bibitem{ref9} Friston K J, Harrison L, Penny W. Dynamic causal modelling{[}J{]}. NeuroImage, 2003, 19(4): 1273-1302.

\bibitem{ref10} Stephan K E, Penny W D, Moran R J, et al.~Ten simple rules for dynamic causal modeling{[}J{]}. NeuroImage, 2010, 49(4): 3099-3109.

\bibitem{ref11} Wang X, Kelly R P, Jenner A L, et al.~A comprehensive guide to simulation-based inference in computational biology{[}J{]}. arXiv preprint arXiv:2409.19675, 2024.

\bibitem{ref12} Deistler M, Boelts J, Steinbach P, et al.~Simulation-based inference: A practical guide{[}J{]}. arXiv preprint arXiv:2508.12939, 2025.

\bibitem{ref13} Lueckmann J M, Boelts J, Greenberg D, et al.~Benchmarking simulation-based inference{[}C{]}//International conference on artificial intelligence and statistics. PMLR, 2021: 343-351.

\bibitem{ref14} Hermans J, Delaunoy A, Rozet F, et al.~A crisis in simulation-based inference? Beware, your posterior approximations can be unfaithful{[}J{]}. Transactions on Machine Learning Research, 2022.

\bibitem{ref15} Gelman A, Meng X L, Stern H. Posterior predictive assessment of model fitness via realized discrepancies{[}J{]}. Statistica Sinica, 1996, 6(4): 733-807.

\bibitem{ref16} Talts S, Betancourt M, Simpson D, et al.~Validating Bayesian inference algorithms with simulation-based calibration{[}J{]}. arXiv preprint arXiv:1804.06788, 2018.

\bibitem{ref17} Säilynoja T, Schmitt M, Bürkner P C, et al.~Posterior SBC: simulation-based calibration checking conditional on data{[}J{]}. Statistics and Computing, 2026, 36(2): 78.

\bibitem{ref18} Pandeva T, Bakker T, Naesseth C A, et al.~E-valuating classifier two-sample tests{[}J{]}. arXiv preprint arXiv:2210.13027, 2022.

\bibitem{ref19} Lemos P, Coogan A, Hezaveh Y, et al.~Sampling-based accuracy testing of posterior estimators for general inference{[}C{]}//Proceedings of the 40th International Conference on Machine Learning. PMLR, 2023, 202: 19256-19273.

\bibitem{ref20} Linhart J, Gramfort A, Rodrigues P. L-c2st: Local diagnostics for posterior approximations in simulation-based inference{[}J{]}. Advances in Neural Information Processing Systems, 2023, 36: 56384-56410.

\bibitem{ref21} Cranmer K, Brehmer J, Louppe G. The frontier of simulation-based inference{[}J{]}. Proceedings of the National Academy of Sciences, 2020, 117(48): 30055-30062.

\bibitem{ref22} Hartoyo A, Cadusch P J, Liley D T J, et al.~Parameter estimation and identifiability in a neural population model for electro-cortical activity{[}J{]}. PLoS computational biology, 2019, 15(5): e1006694.

\bibitem{ref23} Gutenkunst R N, Waterfall J J, Casey F P, et al.~Universally sloppy parameter sensitivities in systems biology models{[}J{]}. PLoS computational biology, 2007, 3(10): e189.

\bibitem{ref24} Transtrum M K, Machta B B, Brown K S, et al.~Perspective: Sloppiness and emergent theories in physics, biology, and beyond{[}J{]}. The Journal of Chemical Physics, 2015, 143(1): 010901.

\bibitem{ref25} Constantine P G. Active subspaces: Emerging ideas for dimension reduction in parameter studies{[}M{]}. Society for Industrial and Applied Mathematics, 2015.

\bibitem{ref26} Lewis A L, Everett R A. Mitigating parameter identifiability issues through model calibration on the data-informed active subspace: An example in tumor growth{[}EB/OL{]}. SSRN, 2026{[}2026-08-14{]}. \url{https://doi.org/10.2139/ssrn.6055166}.

\bibitem{ref27} Daunizeau J, David O, Stephan K E. Dynamic causal modelling: A critical review of the biophysical and statistical foundations{[}J{]}. NeuroImage, 2011, 58(2): 312-322.

\bibitem{ref28} Friston K J, Preller K H, Mathys C, et al.~Dynamic causal modelling revisited{[}J{]}. Neuroimage, 2019, 199: 730-744.

\bibitem{ref29} Moran R J, Stephan K E, Kiebel S J, et al.~Bayesian estimation of synaptic physiology from the spectral responses of neural masses{[}J{]}. NeuroImage, 2008, 42(1): 272-284.

\bibitem{ref30} Fearnhead P, Prangle D. Constructing summary statistics for approximate Bayesian computation: Semi-automatic approximate Bayesian computation{[}J{]}. Journal of the Royal Statistical Society: Series B (Statistical Methodology), 2012, 74(3): 419-474.

\bibitem{ref31} Jiang B, Wu T Y, Zheng C, et al.~Learning summary statistic for approximate Bayesian computation via deep neural network{[}J{]}. Statistica Sinica, 2017, 27(4): 1595-1618.

\bibitem{ref32} Chen Y, Gutmann M U, Weller A. Is learning summary statistics necessary for likelihood-free inference?{[}C{]}//Proceedings of the 40th International Conference on Machine Learning. PMLR, 2023, 202: 4529-4544.

\bibitem{ref33} Schmitt M, Bürkner P C, Köthe U, et al.~Detecting model misspecification in amortized Bayesian inference with neural networks{[}C{]}//Dagm german conference on pattern recognition. Cham: Springer Nature Switzerland, 2023: 541-557.

\bibitem{ref34} Dignum E, Choudhary H, Lees M. Simulation-based inference in agent-based models using spatio-temporal summary statistics{[}C{]}//International Conference on Computational Science. Cham: Springer Nature Switzerland, 2025: 239-254.

\bibitem{ref35} Huang D, Bharti A, Souza A, et al.~Learning robust statistics for simulation-based inference under model misspecification{[}J{]}. Advances in Neural Information Processing Systems, 2023, 36: 7289-7310.

\bibitem{ref36} Anau Montel N, Alvey J, Weniger C. Tests for model misspecification in simulation-based inference: From local distortions to global model checks{[}J{]}. Physical Review D, 2025, 111(8): 083013.

\bibitem{ref37} Omar S M, Kimwele M, Olowolayemo A, et al.~Enhancing EEG signals classification using LSTM‐CNN architecture{[}J{]}. Engineering Reports, 2024, 6(9): e12827.

\bibitem{ref38} Wang G, Liu W, He Y, et al.~Eegpt: Pretrained transformer for universal and reliable representation of eeg signals{[}J{]}. Advances in Neural Information Processing Systems, 2024, 37: 39249-39280.

\bibitem{ref39} Kaur R, Jha S, Roy A, et al.~idecode: In-distribution equivariance for conformal out-of-distribution detection{[}C{]}//Proceedings of the AAAI conference on artificial intelligence. 2022, 36(7): 7104-7114.

\bibitem{ref40} Ishimtsev V, Bernstein A, Burnaev E, et al.~Conformal \(k\)-NN Anomaly Detector for Univariate Data Streams{[}C{]}//Conformal and Probabilistic Prediction and Applications. PMLR, 2017: 213-227.

\bibitem{ref41} Bates S, Candès E, Lei L, et al.~Testing for outliers with conformal p-values{[}J{]}. The Annals of Statistics, 2023, 51(1): 149-178.

\bibitem{ref42} Meshkat N, Kuo C E, DiStefano III J. On finding and using identifiable parameter combinations in nonlinear dynamic systems biology models and COMBOS: a novel web implementation{[}J{]}. PloS one, 2014, 9(10): e110261.

\bibitem{ref43} Constantine P G, Dow E, Wang Q. Active subspace methods in theory and practice: applications to kriging surfaces{[}J{]}. SIAM Journal on Scientific Computing, 2014, 36(4): A1500-A1524.

\bibitem{ref44} Gneiting T, Raftery A E. Strictly proper scoring rules, prediction, and estimation{[}J{]}. Journal of the American statistical Association, 2007, 102(477): 359-378.

\bibitem{ref45} Zamo M, Naveau P. Estimation of the continuous ranked probability score with limited information and applications to ensemble weather forecasts{[}J{]}. Mathematical Geosciences, 2018, 50(2): 209-234.

\bibitem{ref46} Combrisson E, Jerbi K. Exceeding chance level by chance: The caveat of theoretical chance levels in brain signal classification and statistical assessment of decoding accuracy{[}J{]}. Journal of neuroscience methods, 2015, 250: 126-136.

\bibitem{ref47} Ojala M, Garriga G C. Permutation tests for studying classifier performance{[}J{]}. Journal of Machine Learning Research, 2010, 11: 1833-1863.

\bibitem{ref48} Efron B, Tibshirani R J. An introduction to the bootstrap{[}M{]}. New York: Chapman \& Hall, 1993.

\bibitem{ref49} Cohen J. Statistical power analysis for the behavioral sciences{[}M{]}. 2nd ed.~Hillsdale: Lawrence Erlbaum Associates, 1988.

\bibitem{ref50} Chikuse Y. Statistics on special manifolds{[}M{]}. New York: Springer, 2003.

\bibitem{ref51} De Lara L, Deronzier M, González-Sanz A, et al.~On the nonconvexity of push-forward constraints and its consequences in machine learning{[}J{]}. SIAM Journal on Mathematics of Data Science, 2025, 7(2): 597-620.

\bibitem{ref52} Panaretos V M, Zemel Y. Statistical aspects of Wasserstein distances{[}J{]}. Annual review of statistics and its application, 2019, 6(1): 405-431.

\bibitem{ref53} Edelman A, Arias T A, Smith S T. The geometry of algorithms with orthogonality constraints{[}J{]}. SIAM journal on Matrix Analysis and Applications, 1998, 20(2): 303-353.

\bibitem{ref54} Jirsa V K, Stacey W C, Quilichini P P, et al.~On the nature of seizure dynamics{[}J{]}. Brain, 2014, 137(8): 2210-2230.

\bibitem{ref55} Bernabei J M, Li A, Revell A Y, et al.~HUP iEEG Epilepsy Dataset (Version v1.0.0){[}DS/OL{]}. NEMAR, 2026{[}2026-08-14{]}. \url{https://doi.org/10.82901/nemar.on004100}.

\bibitem{ref56} Pinotsis D A, Schwarzkopf D S, Litvak V, et al.~Dynamic causal modelling of lateral interactions in the visual cortex{[}J{]}. NeuroImage, 2013, 66: 563-576.

\bibitem{ref57} Kappenman E S, Farrens J L, Zhang W, et al.~ERP CORE: An open resource for human event-related potential research{[}J{]}. NeuroImage, 2021, 225: 117465.

\bibitem{ref58} Onton J, Westerfield M, Townsend J, et al.~Imaging human EEG dynamics using independent component analysis{[}J{]}. Neuroscience \& biobehavioral reviews, 2006, 30(6): 808-822.

\bibitem{ref59} Craik A, He Y, Contreras-Vidal J L. Deep learning for electroencephalogram (EEG) classification tasks: A review{[}J{]}. Journal of Neural Engineering, 2019, 16(3): 031001.

\bibitem{ref60} Greenberg D, Nonnenmacher M, Macke J. Automatic posterior transformation for likelihood-free inference{[}C{]}//International conference on machine learning. PMLR, 2019: 2404-2414.

\bibitem{ref61} Tolley N, Rodrigues P L C, Gramfort A, et al.~Methods and considerations for estimating parameters in biophysically detailed neural models with simulation based inference{[}J{]}. PLOS Computational Biology, 2024, 20(2): e1011108.

\bibitem{ref62} Blum M G B, Nunes M A, Prangle D, et al.~A comparative review of dimension reduction methods in approximate Bayesian computation{[}J{]}. Statistical Science, 2013, 28(2): 189-208.

\bibitem{ref63} Proix T, Bartolomei F, Guye M, et al.~Individual brain structure and modelling predict seizure propagation{[}J{]}. Brain, 2017, 140(3): 641-654.

\bibitem{ref64} Hashemi M, Vattikonda A N, Sip V, et al.~The Bayesian Virtual Epileptic Patient: A probabilistic framework designed to infer the spatial map of epileptogenicity in a personalized large-scale brain model of epilepsy spread{[}J{]}. NeuroImage, 2020, 217: 116839.

\bibitem{ref65} Vattikonda A N, Hashemi M, Sip V, et al.~Identifying spatio-temporal seizure propagation patterns in epilepsy using Bayesian inference{[}J{]}. Communications biology, 2021, 4(1): 1244.

\bibitem{ref66} Saggio M L, Crisp D, Scott J M, et al.~A taxonomy of seizure dynamotypes{[}J{]}. Elife, 2020, 9: e55632.

\bibitem{ref67} Bastos A M, Usrey W M, Adams R A, et al.~Canonical microcircuits for predictive coding{[}J{]}. Neuron, 2012, 76(4): 695-711.

\bibitem{ref68} Garrido M I, Kilner J M, Kiebel S J, et al.~Dynamic causal modeling of the response to frequency deviants{[}J{]}. Journal of neurophysiology, 2009, 101(5): 2620-2631.

\bibitem{ref69} Auksztulewicz R, Friston K. Attentional enhancement of auditory mismatch responses: a DCM/MEG study{[}J{]}. Cerebral cortex, 2015, 25(11): 4273-4283.

\end{thebibliography}
\end{document}